\documentclass{jfm}
\usepackage{tikz}
\usepackage[pdftex]{pict2e}
\usepackage{graphicx}
\usepackage{epstopdf,epsfig}
\usepackage{epstopdf}
\PrependGraphicsExtensions{.tif, .tiff}
\usepackage{newtxtext}
\usepackage{newtxmath}
\newcommand\tab[1][1cm]{\hspace*{#1}}
\usepackage{amsmath} 
\usepackage{makeidx} % Standard package for indexing
\usepackage{imakeidx} % Enhanced indexing package
\usepackage{multirow}
\usepackage{tabularx}
\usepackage{natbib}
\usepackage{hyperref}
\usepackage{xcolor}
\usepackage{setspace}
\hypersetup{
    colorlinks = true,
    urlcolor   = blue,
    citecolor  = blue,
}

\usepackage[ruled,vlined]{algorithm2e}
\SetKw{KwBy}{by}

\newcommand{\RomanNumeralCaps}[1]
\title{\vspace{-2.0cm}Transient segregation of different density granular mixtures \\ %Coupling rheology and segregation to predict transient flow of different granular mixtures
} 

 \author{Soniya Kumawat, Vishnu Kumar Sahu, \and Anurag Tripathi \corresp{\email{anuragt@iitk.ac.in}}}

 \affiliation{Department of Chemical Engineering, Indian Institute of Technology Kanpur 208016, India}

\begin{document}
\maketitle

\begin{abstract}
We study time-dependent density segregation of granular mixtures flowing over an inclined plane.
Discrete Element Method (DEM) simulations in a periodic box are performed for granular mixtures of same size and different density particles flowing under the influence of gravity. In addition, a continuum model is developed to solve the momentum balance equations along with species transport equation by accounting for the inter-coupling of segregation and rheology.
The particle force-based density segregation theory has been used along with the $\mu-I$ rheology to predict evolution of flow properties with time for binary and multicomponent mixtures. 
The effect of particle arrangements on the transient evolution of flow properties for three different initial configurations is investigated using both continuum and DEM simulations.
Continuum predictions for various flow properties of interest such as species concentration, velocity, pressure, and shear stress at different time instants are compared with DEM simulations. The results from the discrete and continuum models are found to be in good agreement with each other for well-mixed and heavy-near-base initial configurations.
However, the continuum model is unable to predict the flow evolution for the light-near-base initial configuration. 
\textcolor{black}{DEM simulations reveal the presence of an instability driven, quick segregation for this configuration which is not predicted by the one dimensional model and requires generalization to three dimensions.}
\end{abstract}


\section{Introduction}
\label{sec:intro}
\textcolor{black}{When a granular system is subjected to flow or vibration, separation of its components occurs, resulting in spatially distinct regions.} This phenomenon, known as segregation, is a key challenge in many solid-handling industries such as pharmaceutical, food processing, ceramic, metallurgical, and steel industries. Segregation also occurs in geophysical situations such as debris flows, snow avalanches, landslides, rock falls etc. Although surface properties and particle shapes can also cause segregation of granular materials, the differences in particle sizes, and/or densities are the primary factors that lead to the segregation. 
A large body of literature reporting the segregation of granular mixtures through experimental observations (\cite{gray1997pattern,HSIAU2002,Jain2005,YANG2006,jha2010percolation,asachi2018experimental,liao2019effect,liao2020behavior,BRANDAO20201,pillitteri2020size}) and Discrete Element Method simulations (\cite{pereira2011streak,pereira2011insights,chand2012discrete,combarros2014segregation,ayeni2015discrete,lim2016density,Pereira2017,Qiao2021}) is available. \textcolor{black}{However, these studies largely focus on reporting the observations for the segregating mixtures and lack theoretical formulation for predicting segregation.} %and don't provide any theoretical formulation to enable predictions for the segregation.  

Gray and coworkers (\cite{grayandthronton2005,grayandchugunov2006particle,gray_ancey_2011,GrayAncey2015,gray2018particle}) proposed and developed a theoretical framework to study the phenomena of segregation in free surface flows. In this approach, the advection-diffusion-segregation equation for each mixture species is solved to predict the segregation. The knowledge of the flow kinematics, segregation model, and diffusion coefficient is required to be able to predict the segregation. \textcolor{black}{An empirical segregation model for different size binary granular mixtures was proposed by \cite{fan2014modelling} and \cite{schlick2015granular} who used this framework to predict the segregation in heap and rotating cylinders by imposing the flow kinematics from the simulations/experiments.} This segregation model was extended by \cite{schlick2016} to accommodate polydisperse mixtures and was shown to be in qualitative agreement with DEM simulations. A similar approach was used to account for the combined effect of size and density segregation as well by \cite{Duan2021,duan2022designing}. 

\textcolor{black}{Relatively fewer studies have focused on segregation solely due to density differences.} 
\cite{xiao2016modelling} studied density segregation of binary granular mixtures in heap flow using both experiments and DEM simulations. The authors proposed a linear empirical segregation model based on their DEM data and solved the convection-diffusion-segregation equation by incorporating the flow kinematics from the DEM data. Notably, the theoretical predictions were in good agreement with DEM data at steady state. \cite{fry2018effect} modified the linear empirical model to account for the effect of pressure for size as well as density segregation of binary mixtures in the case of plane shear flow. Using this model, \cite{fry2019} predicted steady state concentration profiles that were found to be in good agreement with DEM data for a $50\% - 50\%$ mixture of different density grains in plane shear. %These studies also utilize flow kinematics obtained from the DEM simulations or experimental data.

%The linear empirical segregation model of (\cite{schlick2016}) has been used to predict the time-dependent evolution of the concentration of species in a size poly-disperse mixture in heap flow and rotating tumbler by 
\cite{deng2018continuum} generalized the segregation model for different density binary mixtures to multicomponent mixtures differing in density. %It is important to note that the model for solving transient species concentration profiles relies on time-dependent flow kinematics.
Following the previous studies, the authors utilized the velocity profiles obtained from DEM simulations to predict the evolution of segregation with time. This essential requirement of the knowledge of flow kinematics at each instant can be avoided by solving the momentum balance equations utilizing the granular rheology and intercoupling this with the advection-diffusion-segregation equation. This enhances the predictive capability of these approaches to systems where the flow kinematics may not be easy to measure.

Such an approach has been followed by \cite{barker2021OpenFoam} who implemented a $3D$ continuum model to predict the size segregation for multicomponent mixtures in free surface flows. 
The model was successfully able to predict qualitative behavior in flow geometries such as chute flow and rotating cylinder flow. This continuum model has recently been used to predict the size segregation in a triangular rotating drum as well (\cite{maguire2024particle}). 
\textcolor{black}{While \cite{barker2021OpenFoam} utilize the segregation model of \cite{trewhela2021experimental},  \cite{yang2021continuum} utilize the empirical segregation model of \cite{gajjar2014asymmetric} to predict transient size segregation in periodic shear cell, chute flow, and rotating drum.}
%A similar approach is used by \cite{yang2021continuum} to predict transient size segregation in periodic shear cell, chute flow, and rotating drum. While \cite{yang2021continuum} utilize the empirical segregation model of \cite{gajjar2014asymmetric} and \cite{barker2021OpenFoam} utilize the segregation model of \cite{trewhela2021experimental}. 
The predictions from the model were compared with DEM data and found to be in good agreement. Recently \cite{trewhela2024segregation} have utilized the segregation model of  \cite{trewhela2021experimental} along with the regularized $\mu-I$ rheology proposed by \cite{barker2017wellposedrheology} to predict the time-dependent size segregation in a plane shear flow using a $1D$ continuum model. %All these studies dealing with the prediction of transient segregation investigate segregation due to size differences. 

Transient segregation of granular mixtures differing in density has been studied by \cite{TIRAPELLE20211annular}. The authors predicted time-dependent behavior of binary mixture of different density particles using a segregation model based on the particle force-based theory. While the authors partially attempted to couple the rheology with the segregation model, they did not solve the momentum balance equations and used the flow kinematics obtained from the DEM simulations. The authors performed experiments in an annular cylinder in which the front and the back wall effects also play an important role along with the confinement at the top surface through which pressure was exerted on the material. They utilized an initial configuration where the light species was concentrated near the base. While the density ratios used in experiments ranged from  $1.4$ to $8.25$, the theoretical predictions of the model were compared with experimental data only up to density ratios $3.7$. A careful look at their results shows that the continuum simulations reported by the authors fails to qualitatively capture the concentration variation in an equal composition mixture even for low density ratios of $1.4$ and $2.59$. %These results indicate that a more detailed investigation of the time evolution of flow behavior of different density granular mixtures for different initial configurations is needed. 

\textcolor{black}{Segregation of different density mixtures by accounting for the intercoupling of the rheology and segregation was first studied by \cite{tripathi2013density}.} The authors derived a particle force-based model to predict the segregation of a binary mixture of different density particles in periodic chute flow at steady state. This segregation model has been generalized for multicomponent density mixture by \cite{sahu_kumawat_agrawal_tripathi_2023}. This particle force-based model stands out as the most promising segregation model for different density mixtures. The model does not require any empirically fitted parameter and is able to predict the steady state segregation of multicomponent mixtures for a wide range of density ratios, inclination angles, and composition of the species. 
%However, the applicability of the particle force-based model to accurately predict the time-dependent evolution of segregation and other flow properties has not been investigated yet. 

\textcolor{black}{In this work, we study the time-dependent behavior of flow and segregation of granular mixtures and explore the applicability of this particle force-based model to accurately predict the time-dependent evolution of segregation along with other flow properties.} We simultaneously solve the momentum balance equations by utilizing $\mu-I$ rheology along with the segregation-diffusion equation. Due to the periodic simulation box used in DEM simulations, we consider the variation of the flow properties only along the vertical direction and ignore the variations along the flow and vorticity directions.
We perform DEM simulations of binary and ternary mixtures differing in density for different mixture compositions and density ratios at two different inclination angles.
In addition, we explore the applicability of the proposed theoretical approach for different initial configurations of the species. Specifically, we consider the evolution of segregation for an initially well-mixed configuration along with two almost completely segregated configurations of light species near the base and heavy species near the base. \textcolor{black}{The results presented in this work show that transient segregation in a periodic chute can be very well captured by our $1D$ continuum model for binary, ternary, and quaternary mixtures for well-mixed and heavy-near-base configurations. However, the model predictions do not agree with DEM simulation results for the light-near-base initial configuration. More importantly, the failure of the one dimensional model for a short time duration for this configuration reveals the occurrence of an instability driven, quick segregation phenomenon that has received little attention in the literature.}

The organization of the paper is as follows: DEM simulation methodology is summarised in section \S \ref{sec:method}. The continuum model comprising of the time-dependent momentum balance and convection-diffusion equations inter-coupled with particle force-based segregation model and granular mixture rheology are detailed in section \S \ref{sec:theory}. We report the results of time-dependent properties in the case of initial well-mixed configuration for binary and multicomponent mixtures in section \S \ref{sec:binary} and \S \ref{sec:ternary}, respectively. Section \S \ref{sec:diff_initial_conf} presents the results for heavy-near-base and light-near-base initial configurations. We report the presence of Rayleigh-Taylor like instability in different density granular mixture flows in section \S \ref{sec:RT_instability} and provide a summary of our findings in section \S \ref{sec:summary}.

\section{DEM Simulation Methodology}
\label{sec:method}
\begin{figure}
    \centering
    \includegraphics[scale=0.305]{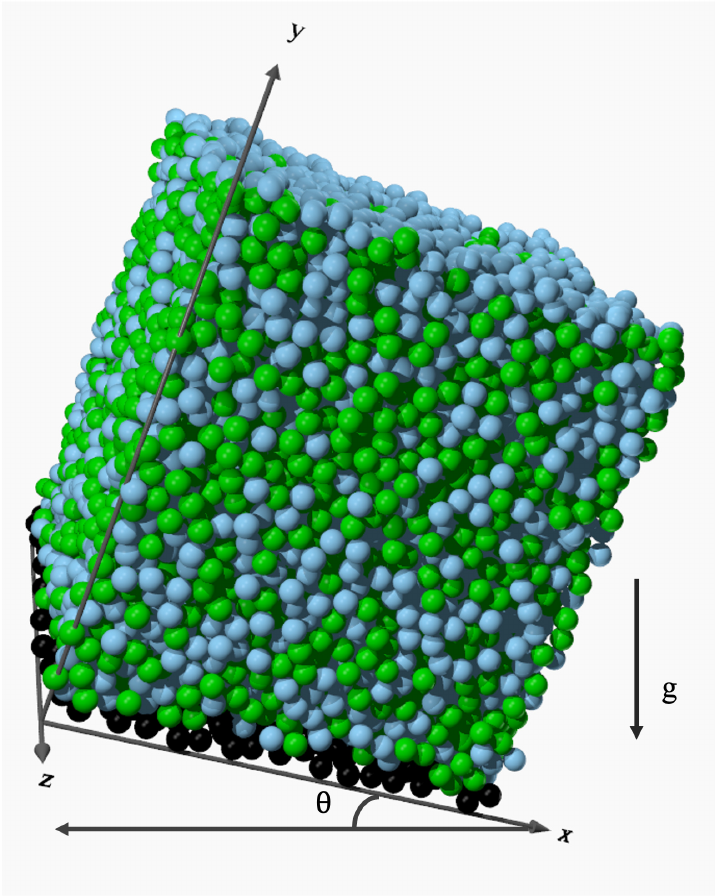}\put(-100,120){(a)}
    \includegraphics[scale=0.152]{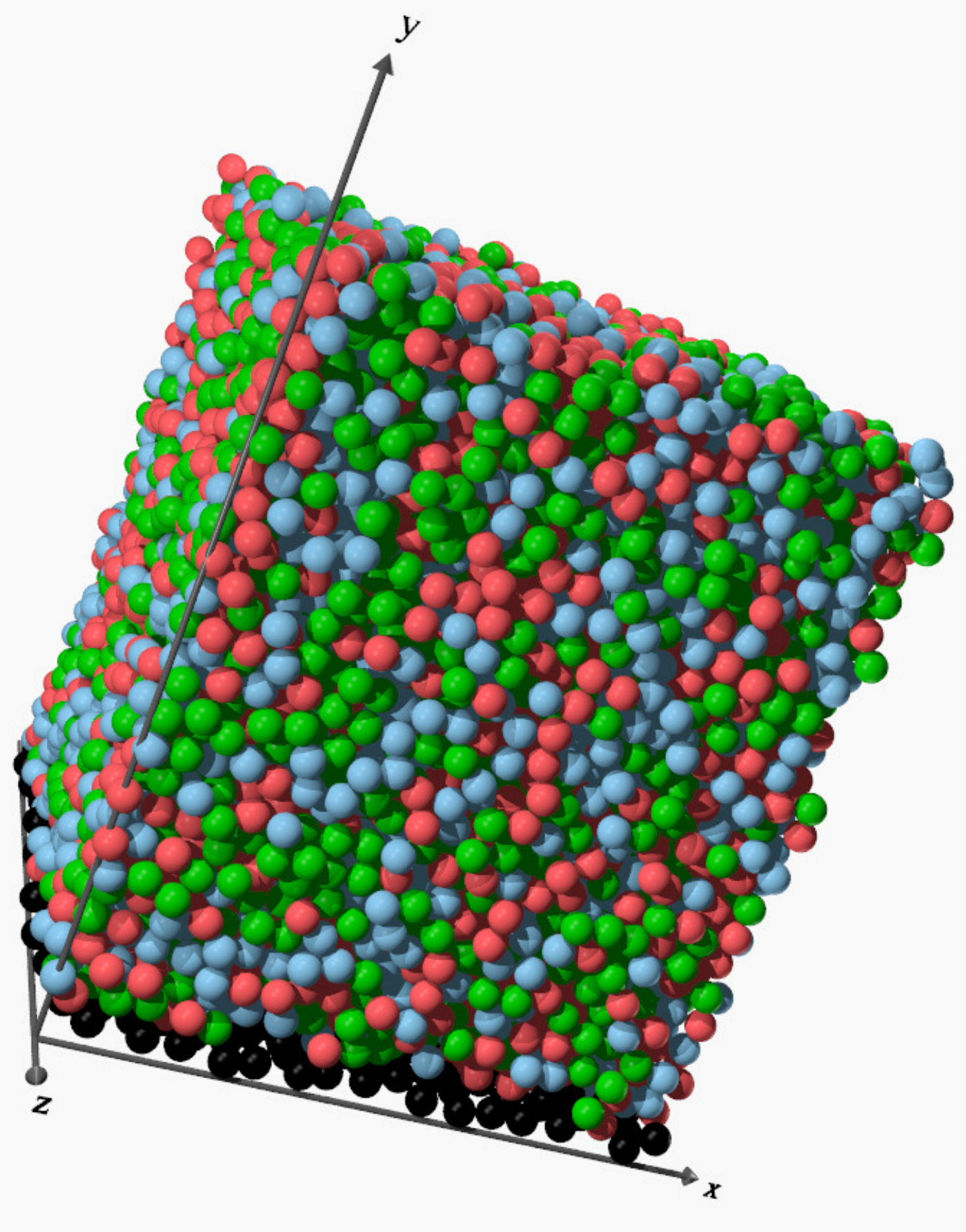}\put(-100,120){(b)}
    \caption{DEM simulation snapshots for equal composition mixture flowing over an inclined plane at an inclination angle of $\theta$. (a) Binary mixture having density ratio $\rho_H/\rho_L = 3.0$, and (b) Ternary mixture having $\rho_H: \rho_M : \rho_L$ = $3:2:1$. Green, red, and blue particles represent the heavy, medium, and light-density particles, respectively. Black particles represent the rough bumpy base of thickness $1.2d$.}
    \label{fig:DEM_snap_methodology}
\end{figure}
We perform simulations of spherical particles of identical sizes (diameter $d$) and different densities flowing over an inclined plane using the Discrete Element Method (DEM). The particles are assumed to be frictional and slightly inelastic. The inter-particle coefficient of normal restitution is $e = 0.88$ and the inter-particle friction coefficient of $\mu = 0.5$ is used in these simulations. 
We consider a simulation box that spans an area of $20d \times 20d$ in the streamwise ($x$) and vorticity ($z$) directions. A periodic boundary condition is imposed along the streamwise direction. To avoid the shearing along the vorticity direction, we use periodic boundary in the $z$ direction as well. \textcolor{black}{Use of the periodic boundaries in $x$ and $z$ directions facilitates detailed quantitative comparison of DEM simulation results with theoretical predictions of unidirectional density segregation. The height of the simulation box in the $y$ direction is $40d$ and total $10,000$ particles are used in the simulation.} At the start of the simulation, particles are generated in a cubic lattice so that two adjacent particles are not in contact. The particles are settled at inclination angle $\theta = 0^\circ$ under the influence of gravity until the average kinetic energy of particles in the layer becomes less than $10^{-6} mgd$.
The layer height after settling the particles turns out to be around $24d$. 
The inclination angle is then increased to the desired value so that particles start to flow. The simulation is performed for a sufficiently large time so that all the properties of interest achieve a steady value. This time duration ranges from $3000$ to $6000$ dimensionless time units. 
We employ the linear spring and dashpot model to calculate contact forces in both the normal and tangential directions. The normal and tangential spring stiffness is $k_{n} = 2 \times 10^{5}$ $mg/d$ and $k_{t} = 2k_{n}/7$, respectively. Further information regarding the contact force model can be found in \cite{tripathi2011rheology}.

% \textcolor{black}{At the start of the simulation, particles are generated in a cubic lattice so that two adjacent particles are not in contact and hence, the height of the simulation box in the $y$ direction is chosen/fixed to be $40d$. A total number of $10,000$ particles are used in the simulation so that the layer height after settling the particles at $0^\circ$ under the influence of gravity until the average kinetic energy of particles in the layer becomes less than $10^{-6} mgd$, turns out to be around $24d$.}

DEM snapshots for equal composition binary as well as ternary mixture, are shown in figure~\ref{fig:DEM_snap_methodology}a and \ref{fig:DEM_snap_methodology}b. Green, red, and blue particles represent the heavy, medium, and light-density particles, respectively. A rough bumpy base of thickness $1.2d$ is formed using black static particles to avoid slip at the base. In addition to investigating the well-mixed initial configuration, we explore the effect of two different initial configurations of heavy-near-base, and light-near-base for binary and ternary mixtures. 
In the heavy (light) near-base configuration particles are arranged in decreasing (increasing) order of density with the distance from the base. Thus the high density particles are situated near the base, and the low-density particles are located near the free surface in the heavy-near-base configuration while the reverse arrangement is used in the case of light-near-base configuration. 

Variation of the relevant flow properties is obtained by dividing the simulation box into strips of thickness $d$ covering the entire area of the simulation box in the $x-z$ plane. Thus we measure the variation of flow properties only along the normal ($y$) direction. 
In order to accurately measure the species concentration and other flow properties, the partial volume of the particles lying in the sampling volume is accounted. \textcolor{black}{The local concentration of $j^{th}$ species $f_j$ is calculated as the ratio of the volume of the $j^{th}$ species particles to the total volume of all the particles in the sampling volume.}
\begin{equation}
    f_j =  \frac{\sum_{i \epsilon j}^{N_j} w_i V_p}{\sum_i^N w_i V_p},
\end{equation}
where $N$ is the total number of all the particles, $N_j$ is the number of particles of $j^{th}$ species in the sampling volume and $V_p$ is the volume of the particle. $w_i$ is the fraction of the volume of $i^{th}$ particle in the sampling volume. The average velocity of all the $N$ particles partially or completely situated in the sampling volume $V_s$ is calculated as $\vec{v}=\sum_{i=1}^{N} w_i \vec{v}_{i}$, where $\vec{v}_{i}$ is the instantaneous velocity of particle $i$. 
The local shear rate ($\dot \gamma$) at each instant is calculated by numerically differentiating the velocity profile ($v_x$) using the forward difference method.
The stress tensor in each bin is calculated as
\begin{equation}
    \underline{\underline{\sigma}} = \frac{1}{V_s}\left[\sum_{i=1}^{N} w_i m_i(\vec{v}_{i} \vec{v}_{i} - \vec{v} \vec{v}) + \sum_{i=1,i\neq k}^{N_c} w_{ik}\vec{F}_{i k} \vec{x}_{i k} \right] \label{eq:sigma_tensor}.
\end{equation}
The first term in equation~\ref{eq:sigma_tensor} represents the kinetic contribution to the stress tensor due to velocity fluctuations. The second term represents the contribution due to the contact forces between the contacting particles summed over the $N_c$ pairs with $w_{ik}$ being the fraction of the branch vector (that joins the centres of contacting particles $i$ and $k$) lying within the volume $V_s$.
The pressure is calculated as one third of the trace of the stress tensor, i.e., $P = \left(\sigma_{x x} + \sigma_{y y} + \sigma_{z z} \right)/3 $ and the shear stress $\tau_{yx}  = \sigma_{yx}$. 
The results are reported in dimensionless units, employing the mass of the light species particles $m_L$ as the mass scale, particle diameter $d$ as the length scale, $(d/g)^{1/2}$ as the time scale, and $m_L g$ as the force scale. 
Consequently, velocity is normalized by $(gd)^{1/2}$, and the shear stress and pressure are normalized by $m_L g/d^2$. The instantaneous properties are calculated by averaging a total of  $20$ snapshots obtained after every $0.1$ time unit, thereby representing the average property over a span of $2$ time units.

\section{Continuum Model}
\label{sec:theory}
In the continuum framework, several models have been suggested to predict segregation phenomena (see for example \cite{bridgwater1985particle}, \cite{grayandthronton2005}, \cite{xiao2016modelling}, \cite{gray2018particle}). All of these models utilize the general form of the transport equation to account for the advection, segregation, and diffusion. The concentration of the $i^{th}$ species $f_i$ is governed by the following equation
\begin{equation}
   \frac{\partial f_i}{\partial t} +   \nabla \cdot (\textbf{v} f_i) +  \nabla \cdot (\textbf{J}_{i}^{S}) + \nabla \cdot (\textbf{J}_{i}^{D}) = 0,
   \label{eq:adv_diff_seg_vectoreqn}
\end{equation}
where, $\textbf{v}$ is the mixture velocity and $\textbf{J}_{i}^{S}$ and $\textbf {J}_{i}^{D}$ are segregation and diffusion fluxes of $i^{th}$ species, respectively. %\cite{barker2021OpenFoam} and \cite{maguire2024sizetriangularrotatingOpenFoam} studied the segregation phenomena for mixtures differing in sizes by solving the \ref{eq:adv_diff_seg_vectoreqn} along with the momentum balance equation, coupled with the segregation and rheological models. 
Using a similar approach, we present a continuum framework to describe the flow behavior and segregation phenomena for mixtures having different density particles for binary as well as multicomponent mixtures. Specifically, we consider the flow of granular mixtures over a surface inclined at an angle $\theta$ from the horizontal. \textcolor{black}{Experiments by \citet{Wiederseiner2011} have shown that chute length of few hundred particle diameters is required for complete segregation of mixtures.
Simulating such large systems using DEM is computationally expensive. 
In order to reduce the number of particles to be simulated, we have used periodic boundary in the flow ($x$) direction in our DEM simulations. To investigate segregation in a flow where shearing occurs only in the vertical direction, we utilize the periodic boundary condition in the vorticity ($z$) direction as well.} The usage of periodic boundary conditions in the flow and vorticity directions ensures the gradients in the $x $ and $z$ directions are practically zero and the time-dependent transport equation~\ref{eq:adv_diff_seg_vectoreqn} for $i^{th}$ species in the mixture reduces to,
\begin{equation}
   \frac{\partial f_i}{\partial t} +\frac{\partial (v_y f_i)}{\partial y} + \frac{\partial J_{iy}^{S}}{\partial y }  + \frac{\partial J_{iy}^{D}}{\partial y} =0.
   \label{eq:seg_diff1}
\end{equation}

We further assume that the flow is unidirectional, i.e., only $x$ component of the velocity is present, and \textcolor{black}{the $y$ component of the velocity is negligible, i.e., $v_y=0$. Using these assumptions, equation~\ref{eq:seg_diff1} reduces to}
\begin{equation}
   \frac{\partial f_i}{\partial t} + \frac{\partial J_{iy}^{S}}{\partial y }  + \frac{\partial J_{iy}^{D}}{\partial y} =0.
   \label{eq:seg_diff2}
\end{equation}
\textcolor{black}{This equation is identical to the form used in previous works of \cite{tripathi2013density} and \cite{sahu_kumawat_agrawal_tripathi_2023} for fully developed, unidirectional chute flow.} Previous studies by other researchers (\cite{fan2014modelling,xiao2016modelling,deng2018continuum,Duan2021}) report that segregation occurs mainly in the vertical direction (normal to the free surface) in geometries such as chute, heap, and rotating tumbler and the diffusion and segregation fluxes in the flow direction are ignored. However, unlike the present study, the presence of the convective flux term due to the mean flow is retained in these studies and hence, the variation of the species concentration along the flow direction is also captured in these studies. \textcolor{black}{However, for the periodic chute considered in this study, the convective flux term does not contribute to the species transport.}

In equation~\ref{eq:seg_diff2}, $J_{iy}^{S}$ is the segregation flux and $J_{iy}^{D}$ is the diffusion flux of $i^{th}$ species in the $y$-direction. \textcolor{black}{The expressions for these fluxes are given as \citep{tripathi2013density,sahu_kumawat_agrawal_tripathi_2023}}
\begin{equation}
    J_{iy}^{S} = v_{i} f_i ; \quad J_{iy}^{D} = -D\frac{\partial f_i}{\partial y}.
    \label{eq:fluxes}
\end{equation}
Here $v_{i}$ is the segregation velocity of the $i^{th}$ species in the $y$ direction, $f_i$ is the local volume concentration of $i^{th}$ species in the mixture and $D$ is the diffusivity. Substituting equation~\ref{eq:fluxes} in equation~\ref{eq:seg_diff2}, we obtain the following partial differential equation
\begin{equation}
   \frac{\partial f_i}{\partial t} + \frac{\partial }{\partial y } ( v_{i} f_i) = \frac{\partial}{\partial y} \left(D\frac{\partial f_i}{\partial y}\right).
   \label{eq:Conc_pde}
\end{equation}

The diffusivity $D$ has been shown to be dependent on the local shear rate ($\dot \gamma$) and volume average diameter ($d_{mix}$) in the case of granular shear flows \citep{bridgwater1985particle,utter2004self,tripathi2013density,fry2019,Duan2021,sahu_kumawat_agrawal_tripathi_2023}. Following these studies, we use the spatially varying diffusivity as $D = b \dot \gamma d_{mix}^2$. 
\textcolor{black}{Since the different density mixtures consist of particles of identical sizes of diameter $d$, local volume average diameter is equal to the particle diameter, i.e., $d_{mix} = d$. The value of the parameter $b = 0.041$ is used for binary as well as multicomponent mixtures as obtained from DEM simulations (\cite{tripathi2013density,sahu_kumawat_agrawal_tripathi_2023}).
Below, we first formulate the model equations for binary mixtures and subsequently extend it to multicomponent mixtures.}

\subsection{Binary mixtures}
\label{Subsec:theory_binarymix}
Let us consider a binary mixture having concentrations of light and heavy species in the mixture $f_L$ and $f_H$, respectively. The sum of concentrations of both the species must equal to $1$ i.e., $ f_H + f_L = 1$. We utilize particle force-based model of \cite{tripathi2013density} to determine the segregation velocity.
Considering the buoyancy, gravity and drag forces acting on a single heavy intruder particle in a mixture of light and heavy particles, the authors obtain the expression for the segregation velocity of the heavy species in case of a binary mixture as 
\begin{equation}
    v_{H} = \frac{g_y}{c \pi \eta d} (1-f_H)(m_H-m_L),
\end{equation}
where $m_L$ and $m_H$ are masses of the particles of light and heavy species in the mixture. $\eta$ is the mixture viscosity which is obtained as the ratio of the magnitude of the shear stress to the shear rate i.e., $|\tau_{yx}|/|\dot \gamma|$. The Stokes drag coefficient $c$ at different inclination angles is taken from the simulation study by \cite{tripathi2013density}. \textcolor{black}{The value of $c$ lies in the range of $2.5$ to $3.7$ for dense flows with solid fractions $\phi$ ranging between $0.5 - 0.59$.} Similarly, the segregation velocity for light species is given by 
\begin{equation}
    v_{L} =   -\frac{g_y}{c \pi \eta d} (1-f_L)(m_H-m_L).
\end{equation}
For the choice of axis as shown in figure~\ref{fig:DEM_snap_methodology}, $g_y=-g\cos\theta$. Hence, the segregation velocity of the light species is positive indicating that the light particles rise to the free surface while that of the heavy particles is negative indicating that they sink towards the base. 
The mixture velocity in $y$ direction for the binary mixture cases is calculated as 
\begin{equation}
v_y =  f_H v_H + f_L v_L. 
\end{equation}
Using the expressions for $v_H$ and $v_L$, the mixture velocity in the $y$ direction turns out to be zero, i.e., $v_y=0$ which is consistent with the unidirectional flow assumption made earlier. 
% \begin{equation}
%    v_y =  - \frac{g \cos\theta}{c \pi \eta d} (m_H-m_L) (1-f_H) f_H + \frac{g \cos\theta}{c \pi \eta d} (m_H-m_L) (1-f_L) f_L
% \end{equation}

% \begin{equation}
%    v_y =  -\frac{g \cos\theta}{c \pi \eta d} (m_H-m_L) ( - f_L f_H + f_H f_L ) = 0
%\end{equation}

%The Stokes drag coefficient $c$ at different inclination angles is taken from the simulation study by \cite{tripathi2013density}.
\textcolor{black}{Substituting the expressions of diffusivity and the heavy species segregation velocity in equation~\ref{eq:Conc_pde}, the species transport equation for heavy species takes the form}
% \begin{equation}
%    \frac{\partial f_H}{\partial t} + \frac{\partial }{\partial y } \left( -\frac{ g \cos\theta}{c \pi \eta d} (m_H-m_L) (1-f_H) f_H \right) =  \frac{\partial}{\partial y} \left(b \dot \gamma d^2 \frac{\partial f_i}{\partial y}\right)
%    \label{eq:final_Conc_pde}
% \end{equation}

\begin{equation}
   \frac{\partial f_H}{\partial t} = \frac{\partial }{\partial y } \left( \frac{g \cos\theta}{c \pi \eta d} (m_H-m_L) (1-f_H) f_H +  b \dot \gamma d^2 \frac{\partial f_H}{\partial y}\right). 
   \label{eq:final_Conc_pdepeform}
\end{equation}
Equation~\ref{eq:final_Conc_pdepeform} needs to be solved to obtain the concentration of heavy species $f_H$ (or light species $f_L$) with time $t$ and position $y$. \textcolor{black}{The initial and boundary conditions needed to solve equation~\ref{eq:final_Conc_pdepeform} are given as follows:} 
\begin{equation}
\begin{split}
     IC: \quad f_{H}(y,0) = f_{H,ini}(y),\\
     BC1: \quad J_{Hy}^{S} (0,t) + J_{Hy}^{D} (0,t)= 0,\\
     BC2: \quad J_{Hy}^{S}(h,t) + J_{Hy}^{D}(h,t) = 0.
    \end{split}
    \label{eq:IC_BC_concentration}
\end{equation}
% \begin{equation}
% \begin{split}
%      IC: t = 0, \quad f_{H}(y) = f_{H,int}(y).\\
%      BC1: y = 0, \quad J_{Hy}^{S} + J_{Hy}^{D} = 0.\\
%      BC2: y = h, \quad J_{Hy}^{S} + J_{Hy}^{D} = 0.
%     \end{split}
%     \label{eq:IC_BC_concentration}
% \end{equation}

Here, $f_{i,ini}(y)$ is the initial concentration profile of the $i^{th}$ species. In case of starting from a well-mixed configuration, $f_{i,ini}$ is equal to the total composition of that species in mixture i.e., $f^T_i$. \textcolor{black}{For the cases when the mixture is nearly segregated, $f_{i,ini}(y)$ is given by a step function. In order to enable comparison with our DEM simulation results, the initial concentration profile is approximated by fitting a sigmoid function to the concentration profile obtained from the DEM simulation data.} 
The boundary conditions correspond to no net mass flow in $y$ direction at the free surface ($y=h$) and the base ($y=0$) and state that the sum of segregation and diffusion fluxes will be zero at extreme values of $y$.
The concentration of light species in the binary mixture at any instant and location can be obtained as $f_L = 1 - f_H$. 

\subsection{Multicomponent mixtures}
We next consider a multicomponent mixture consisting of $N$ species differing only in density. The concentration of $i^{th}$ species in the mixture is $f_i$ with $\sum_{i=1}^N f_i = 1$. The expression for segregation velocity by considering the buoyancy, gravity and drag forces acting on a single intruder of $i^{th}$ species in a multicomponent mixture using the particle force-based theory of \cite{sahu_kumawat_agrawal_tripathi_2023} is given as 
\begin{equation}
   v_{i}  = -\dfrac{g\cos\theta}{c\pi\eta d}{\sum_{j=1}^{N} (m_{i} - m_{j}) f_{j}  },
  \label{eq:SegVel_multi}
\end{equation}
where $f_j$ is concentration of $j^{th}$ species in mixture. Equation~\ref{eq:SegVel_multi} states that the segregation velocity of any species at any location and time depends on the local concentration of all other species and the mass difference of the species with all other species. 
The mixture velocity in $y$ direction is calculated as $v_{mix} = \sum_{i=1}^{N} f_i v_i$. Using equation~\ref{eq:SegVel_multi} for $v_i$, we find that $\sum_{i = 1}^{N} f_i v_i = 0$. Thus, the mixture velocity in the vertical $y$ direction remains zero for multicomponent mixtures as well.

Substituting equation \ref{eq:SegVel_multi} in the segregation-diffusion equation \ref{eq:Conc_pde}, we get $N$ partial differential equations as follows, 
% \begin{equation}
%    \frac{\partial f_i}{\partial t} + \frac{\partial }{\partial y } \left(  -\dfrac{g\cos\theta}{c\pi\eta d}{\sum_{j=1}^{N} (m_{i} - m_{j}) f_{j} }  f_i\right) = \frac{\partial}{\partial y} \left(b \dot \gamma d^2 \frac{\partial f_i}{\partial y}\right),
% \label{eq:multi_seg_diff_pde}
% \end{equation}

% \begin{equation}
%    \frac{\partial f_i}{\partial t} = \frac{\partial }{\partial y } \left(  \dfrac{g\cos\theta}{c\pi\eta d}{\sum_{j=1}^{N} (m_{i} - m_{j}) f_{j} }  f_i  \right) + \frac{\partial}{\partial y} \left(b \dot \gamma d^2 \frac{\partial f_i}{\partial y}\right); ~~~ i\in [1:N].
%    \label{eq:multi_seg_pde_final}
% \end{equation}

\begin{equation}
   \frac{\partial f_i}{\partial t} = \frac{\partial }{\partial y } \left(  \dfrac{g\cos\theta}{c\pi\eta d}{\sum_{j=1}^{N} (m_{i} - m_{j}) f_{j} }  f_i + b \dot \gamma d^2 \frac{\partial f_i}{\partial y}\right); ~~~ i\in [1:N].
   \label{eq:multi_seg_pde_final}
\end{equation}
In order to get the concentration profiles of $N$ species in mixture, we need to solve $N-1$ partial differential equations simultaneously. The concentration of the $N^{th}$ species can be obtained using the identity $\sum_{i=1}^{N}f_i=1$.
For example, in case of ternary mixtures, we need to solve only two partial differential equations for any two (e.g., heavy and light density) species. The concentration of the third species (e.g., medium density species) can be obtained as $f_M = 1 - f_L - f_H $. 
The initial and boundary conditions are similar to equation~\ref{eq:IC_BC_concentration} and are given as 
\begin{equation}
\begin{split}
     f_{i}(y,0) = f_{i,ini}(y); ~~~ i\in [1:N -1],\\
     J_{iy}^{S}(0,t) + J_{iy}^{D}(0,t) = 0; ~~~ i\in [1:N - 1],\\
     J_{iy}^{S}(h,t) + J_{iy}^{D}(h,t) = 0; ~~~ i\in [1:N - 1].
    \end{split}
\label{eq:IC_BC_concentration_multi}
\end{equation}
Note that the initial concentration profiles of all the $N-1$ species need to be provided at $t=0$. Similarly, the boundary conditions (zero flux at the free surface and base) also need to be provided for all the $N-1$ species. 
%The simplified segregation-diffusion equations for light and heavy density species are given as, respectively, 

% \begin{equation}
%    \frac{\partial f_L}{\partial t}  = \frac{\partial }{\partial y } \left( \frac{g \cos\theta}{c \pi \eta d_L }  [(m_L - m_M) f_M + ( m_L - m_H) f_H] f_L  + b \dot \gamma d^2 \frac{\partial f_L}{\partial y}\right),
%   %\label{eq:light_pde_ternary_final}
% \end{equation}

% \begin{equation}
%    \frac{\partial f_H}{\partial t} = \frac{\partial }{\partial y } \left(  \frac{g \cos\theta}{c \pi \eta d_H}  [(m_H - m_M) f_M + ( m_H -m_L ) f_L] f_H + b \dot \gamma d^2 \frac{\partial f_H}{\partial y} \right).
% %\label{eq:heavy_pde_ternary_final}
% \end{equation}

% \textcolor{black}{The initial and boundary conditions required for solving equation~\ref{eq:multi_seg_pde_final} for different species in multicomponent mixtures are similar to equation~\ref{eq:IC_BC_concentration}.}

\subsection{Momentum balance equations for flow kinematics}
The convection-diffusion-segregation equations for binary (equation \ref{eq:final_Conc_pdepeform}) as well as multicomponent (equations \ref{eq:multi_seg_pde_final}) mixtures incorporate the influence of the local shear rate and local viscosity of mixtures on the concentration profile. The viscosity of the flowing granular medium depends on the local shear rate and the local pressure.
Hence, the knowledge of the flow kinematics and pressure is needed to predict the species concentrations. For this, we solve the momentum balance equations for the unidirectional, fully developed flow (in the $x$ direction) over an inclined surface at an inclination angle $\theta$. The simplified momentum balance equations in the $x$ and $y$ directions for this case reduce to
\begin{equation}
   \rho_b \frac{\partial v_x }{\partial  t} =   \rho_b g \sin\theta - \frac{\partial \tau_{yx}}{\partial y},
   \label{eq:mombal_x}
\end{equation}
\begin{equation}
    0 = - \frac{\partial \sigma_{yy}}{\partial y} - \rho_b g \cos\theta.
   % P(y,t) = g \cos \theta (1 - a \tan \theta )\int_{y}^{H}  \rho_b(y,t) dy.
    \label{eq:mombal_y_ differnetial}
\end{equation}
Here, $v_x$ is the velocity in the $x$ direction, $\rho_b$ is the bulk density, $\tau_{yx}$ is the shear stress.
\textcolor{black}{The normal stress $\sigma_{yy}$ at any location $y$ is obtained by integrating equation~\ref{eq:mombal_y_ differnetial} with the boundary condition $\sigma_{yy} = 0$ at $y = h$. We also account for the anisotropy of stress tensor due to non-zero normal stress differences in the case of granular flow \citep{tripathi2011rheology} by means of the parameter $a$ and obtain the expression for the pressure}
\begin{equation}
   P(y,t) = g \cos \theta (1 - a \tan \theta )\int_{y}^{h}  \rho_b(y,t) dy.
    \label{eq:mombal_y}
\end{equation}
We follow the approach of \cite{sahu_kumawat_agrawal_tripathi_2023} and account for the dependence of the stress anisotropy parameter upon the inertial number as $a = (0.35 - 0.27 I_{mix})/3$. In agreement with \cite{sahu_kumawat_agrawal_tripathi_2023}, we also find that using this relation leads to better predictions of the mixture velocity. However, the concentration profiles and the center of mass variation with time are found to be practically the same as that obtained by using a constant value of the stress anisotropy parameter used by \cite{tripathi2011rheology}.

The bulk density of the mixture is given by $\rho_b = \phi \rho_{mix}$, where $\rho_{mix}(y,t)=\sum_{i=1}^{N} \rho_i f_i(y,t)$ is the local volume average density of the mixture. 
Here, $\rho_i$ is the density and $f_i$ is the local concentration of species $i$ in the mixture.
The local packing fraction $\phi$ is obtained using the dilatancy law with the generalized inertial number for the granular mixtures (\cite{tripathi2011rheology}) 
\begin{equation}
    \phi(y,t)=\phi_{max}- \beta I_{mix}. 
    \label{eq:packing_fraction}
\end{equation}
$\phi_{max}$ and $\beta$ are the dilatancy law parameters obtained from DEM simulation data.
The expression of inertial number for granular mixture having identical size ($d_{mix} = d$) and different density particles is given by
\begin{equation}
    I_{mix}=\frac{|\dot\gamma| d }{\sqrt {P/\rho_{mix}}},
    \label{eq:inertial_no}
\end{equation}
where $\dot \gamma =dv_x/dy$ is the local shear rate. The shear stress ($\tau_{yx}$) and pressure ($P$) relate to each other by means of the inertial number dependent effective friction coefficient $\mu(I_{mix})$, 
\begin{equation}
  |\tau_{yx}| =   \mu(I_{mix}) P.
  \label{eq:shear_stress}
\end{equation}
\textcolor{black}{While a linear variation of $\mu$ with inertial number has been suggested by some researchers \citep{PlaneshearLinearmu_I}, and has been utilized by others \citep{parez2016unsteady}, 
%to obtain analytical expressions for the time evolution of granular flows, 
a nonlinear variation of $\mu$ with $I$ for dense granular flows proposed by \cite{jop2006constitutivenature} is commonly used for the dense flow regime. The JFP model by \cite{jop2006constitutivenature} empirically relates the variation of $\mu$ with $I$ as} 
\begin{equation}
\mu(I_{mix}) = \mu_{s}+\frac{\mu_{m}-\mu_{s}}{1+I_{0}/I_{mix}}, 
\label{eq_5:mu_I_JFP}
\end{equation}
where, $\mu_s$, $\mu_m$, and $I_0$ are the rheological model parameters. The values of these parameters are taken from \cite{tripathi2013density,sahu_kumawat_agrawal_tripathi_2023} and are reported in table~\ref{tab:rheopar}. \textcolor{black}{ Note that \cite{patro2021} have shown that the JFP model fails to capture the flow rheology at inertial numbers greater than $0.6$. Since, the focus of this study is limited to dense granular flow regime for inertial numbers up to $0.5$, equation~\ref{eq_5:mu_I_JFP} can be used to describe the variation of $\mu$ with $I$.} Using the expression for the bulk density $\rho_b$ and shear stress $\tau_{yx}$, the momentum balance equation~\ref{eq:mombal_x} reduces to
\begin{equation}
   \phi \rho_{mix} \frac{\partial v_x }{\partial  t} =  \phi \rho_{mix} g \sin \theta   + \frac{\partial }{\partial y} (\mu(I_{mix}) P).
   \label{eq:mombal_x_final}
\end{equation}

In our simulations, the flow is started from an initial condition of zero velocity and we incorporate a rough, bumpy base to ensure the no-slip boundary condition at the base. 
%($y = 0$) and $0$ shear stress at $y = H$ in the case of free surface flow. 
Hence we use the following initial and boundary conditions to solve the partial differential equation \ref{eq:mombal_x_final}
\begin{equation}
   \begin{split}
        IC: v_{x}(y,0) = 0,\\
        BC1: v_{x}(0,t) = 0,\\
        BC2: \tau_{yx}(h,t) =0.
   \end{split} 
    \label{eq:IC_BC_momentum}
\end{equation}
We note that \cite{patro2023unsteady} have solved the time-dependent momentum balance equation for the unidirectional flow of monodisperse grains using nearly identical approach and have shown that the approach is able to capture the time-dependent properties well. We follow a similar approach for mixtures and solve the intercoupled set of equations of advection-diffusion-segregation, momentum balance and the inertial number based constitutive equation as detailed below.
 
 \begin{table}
\centering
  \begin{center}
    \begin{tabular}{cccccc}  
    $\mu_{s}$ & $\mu_{m}$ & $I_{o}$  &$\phi_{max}$ & $\beta$ \\
    $\tan(20.16^{\circ})$ & $\tan(37.65^{\circ})$ & 0.434 & 0.59 & 0.16 \\
\end{tabular}
    \caption{Values of the rheological model parameters (same as in \cite{tripathi2011rheology}) used in this study.}
    \label{tab:rheopar}
  \end{center}
\end{table}

\subsection{Numerical solution}
%We use the similar approach utilized by 
\cite{patro2023unsteady} 
solved the momentum balance equations along with the rheological model of \cite{patro2021} using PDEPE solver in MATLAB for periodic chute flow geometry and
predicted the time evolution of a granular layer having identical size and density particles.
%Following studies (\cite{gray_ancey_2011,deng2018continuum,deng2019modeling,gao2021modeling}) also utilize the MATLAB $pdepe$ solver to obtain the concentration profile in heap flow. 
The standard representation of the equation in PDEPE solver is given as follows: 
\begin{equation}
C \left(y,t,u,\frac{\partial u}{\partial y}\right) \frac{\partial u}{\partial t} = S \left(y,t,u,\frac{\partial u}{\partial y}\right) + y^{-m} \frac{\partial}{\partial y}\left(y^m L\left(y,t,u,\frac{\partial u}{\partial y}\right)\right).
\label{eq:matlab_pdepeform}
\end{equation}
To solve the momentum balance equation using MATLAB PDEPE solver, we choose 
the variables 
\begin{equation}
  u = v_x, \quad m = 0, \quad C = \phi \rho_{mix}, \quad S = \phi \rho_{mix} g \sin\theta,  \quad L =  \mu(I_{mix}) P,  
\end{equation}
\textcolor{black}{so that the PDEPE equation~\ref{eq:matlab_pdepeform} becomes identical to the momentum balance equation~\ref{eq:mombal_x_final}. Note that the choices of parameter $m$, variable $u$, and relevant functions ($C,L,S$) are almost identical} to those used by \cite{patro2023unsteady}, the only difference being the usage of mixture density $\rho_{mix}$ and mixture inertial number $I_{mix}$ instead of pure component density $\rho$ and inertial number $I$.

\begin{algorithm}
\SetAlgoLined
\textbf{Initialize}:\\
\tab Layer height $h = h_{min}$ \\
 \tab Time $t = 0$, time step $\Delta t = 0.1$ \\
 \tab Number of spatial grids in $y$ direction $N_{bin} = 200$\\ 
 \tab Grid size $\Delta y = h/N_{bin}$.\\
 \tab Species concentration $f_{i}(y)$ $\forall$ $i\in 1:N-1$\\
\tab Packing fraction $\phi(y) = \phi_{max}$ \\
\tab Velocity $v_{x}(y) = 0$ \\
   % \For{$i\gets1$ \KwTo $N-1$}{
   %    \tab Species Concentration $f_i(y)$ \\ %$\forall$ $i\in 1:N-1$\\
   %    \tab Packing fraction $\phi(y) = \phi_{max}$ \\
   %    \tab Velocity $v_{x}(y) = 0$ 
   %    \tab }
    \While{$t < t_{final}$}
    {
    Compute: \\
  %  \tab Layer height \textbf{$H$} with accounting \textbf{$\phi$} \\
  %  Calculate the following properties at time $t$ \\
    \tab Velocity gradient $|\dot \gamma | = d v_{x}/dy$ numerically\\
    \tab Local mixture density $\rho_{mix} = \sum_{i=1}^{N} \rho_i f_{i}$ \\
    \tab Pressure $P$ by equation~\ref{eq:mombal_y} numerically \\
    \tab Inertial number $I$ using equation \ref{eq:inertial_no} \\
   % \tab Effective friction coefficient $\mu(I)$ using equation \ref{eq_5:mu_I_JFP}\\
    \tab Packing fraction $\phi(I)$ using equation \ref{eq:packing_fraction}\\
    Invoke MATLAB PDEPE Solver:\\
     \tab Solve momentum balance equation \ref{eq:mombal_x_final} with $ICs$, \& $BCs$ (eq.~\ref{eq:IC_BC_momentum}). \\
    \tab Solve segregation-diffusion equations~\ref{eq:multi_seg_pde_final} with $ICs$, \& $BCs$  (eq.~\ref{eq:IC_BC_concentration_multi}).\\
     \tab Obtain velocity $v_x(y)$ and species concentration $f_{i}(y)$ at $t + \Delta t$.\\
     
     Compute the following at $t + \Delta t$: \\
     \tab Inertial number $I$ using equation~\ref{eq:inertial_no} \\ \tab Packing fraction $\phi$ using equation~\ref{eq:packing_fraction}\\ \tab Effective friction coefficient $\mu$ using equation~\ref{eq_5:mu_I_JFP} \\ \tab Pressure $P$ using equation~\ref{eq:mombal_y} \\  \tab Shear stress $\tau_{yx}$ using equation~\ref{eq:shear_stress} \\
     \tab Average packing fraction $\phi_{avg}(t)$ \\
     \tab Layer height $h(t) = h_{min} \phi_{max}/\phi_{avg}(t)$ \\
     \tab Grid size $\Delta y = h(t)/N_{bin}$. \\
     Update time $t \rightarrow t + \Delta t$. \\
  %   \tab Velocity $v_{x}(y) = v_{x}(y) $. \\
  %   \tab Species concentration $f_{i}(y) = f_{i}(y)$ $\forall$ $i=1:N-1$.\\
    }
 \caption{Algorithm for predicting the concentration profiles and flow properties}
 \label{algo1}
\end{algorithm}

For the mixtures case, we also solve the simplified segregation-diffusion equation (equation~\ref{eq:final_Conc_pdepeform} for binary mixtures and equations~\ref{eq:multi_seg_pde_final} for multicomponent mixtures) along with the simplified momentum balance equation \ref{eq:mombal_x_final}. 
By comparing the diffusion-segregation equation~\ref{eq:multi_seg_pde_final} with MATLAB PDEPE equation~\ref{eq:matlab_pdepeform}, we obtain the parameter ($m$), variable ($u$) and functions ($C,L,S$) as follows:
\begin{equation}
u = f_i, ~~~ m = 0, ~~~ 
C = 1, ~~~ S = 0, ~~~ 
L = \dfrac{g\cos\theta}{c\pi\eta d}{\sum_{j=1}^{N} (m_{i} - m_{j}) f_{j} }  f_i + b \dot \gamma d^2 \frac{\partial f_i}{\partial y}.
\end{equation}

%The $PDEPE$ solver uses numerical discretization to approximate the derivatives by dividing the domain into finite spatial and temporal grids. \textcolor{black}{A total of $N = 200$ spatial grids along the $y$ direction are used which corresponds to a grid size of approximately one-eighth of the particle diameter.} We choose the time step $\Delta t = 0.1$ dimensionless time units. This time step corresponds to approximately $10^{-3}s$ for 1 mm particle size for earth's gravitational acceleration. \textcolor{black}{Similar to DEM simulations, instantaneous properties are obtained by averaging $20$ snapshots over a time period of $2$ time units.}
%In order to obtain species concentration and velocity profiles, the segregation-diffusion equation and momentum balance equations are solved simultaneously along with inter-coupled rheological constitutive correlations, shown in algorithm \ref{algo1}.  

\textcolor{black}{In the beginning, we provide the initial guess for velocity and species concentration with layer height $h = h_{min}$ assuming that the layer is at rest with the maximum solids fraction $\phi_{max}$ at time $t = 0$ across the entire layer. Next, we compute the shear rate by numerically differentiating the velocity and the mixture density as the density weighted average of the species concentrations, i.e., $\rho_{mix} = \sum_{i=1}^{N} \rho_i f_i$. Subsequently, we calculate the pressure, inertial number, and packing fraction. Further, the species concentration and velocity fields are obtained by solving the segregation-diffusion equation~\ref{eq:multi_seg_pde_final} and momentum balance equation~\ref{eq:mombal_x_final} simultaneously using the PDEPE solver along with the initial and boundary conditions given in equation~\ref{eq:IC_BC_concentration_multi} and equation~\ref{eq:IC_BC_momentum}, respectively. 
The PDEPE solver uses numerical discretization to approximate the derivatives by dividing the domain into finite spatial and temporal grids. A total of $N = 200$ spatial grids along the $y$ direction are used which corresponds to a grid size of approximately one-eighth of the particle diameter. We choose the time step $\Delta t = 0.1$ dimensionless time units. This time step corresponds to approximately $10^{-3}s$ for 1 mm particle size for earth's gravitational acceleration. In order to facilitate comparison with DEM simulations, the instantaneous properties from the theoretical predictions are also obtained by averaging $20$ snapshots over a time period of $2$ time units. The detailed description for solving these equations using MATLAB PDEPE solver is given in algorithm \ref{algo1}.}

\section{Evolution of flow properties of binary mixtures}
\label{sec:binary}

\begin{figure}
    \centering
    \includegraphics[scale=0.33]{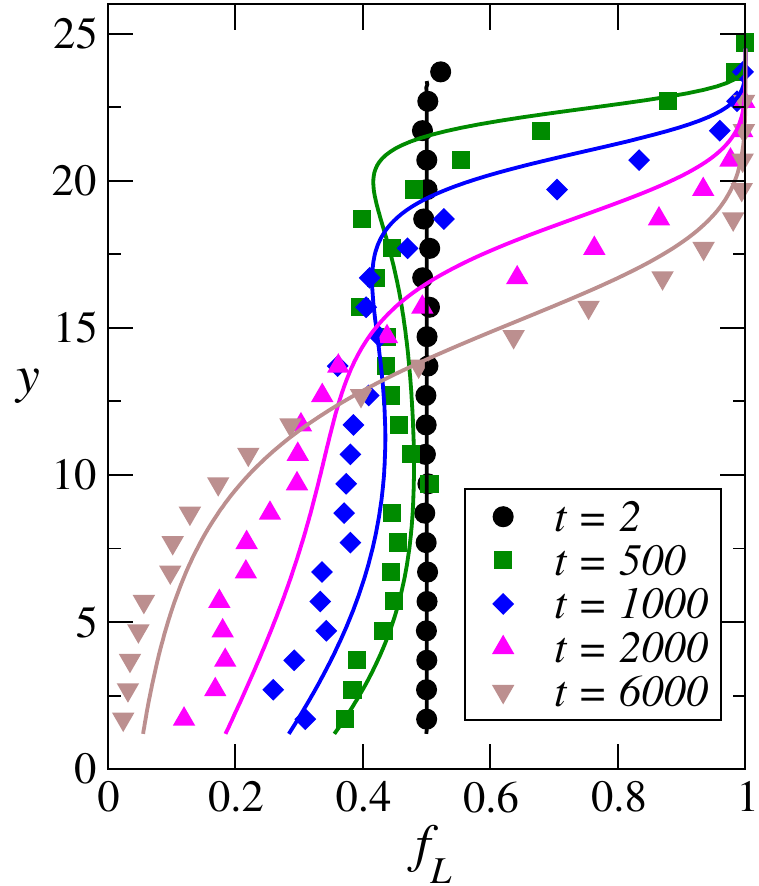}\put(-120,145){(a)}\hfill
    \includegraphics[scale=0.33]{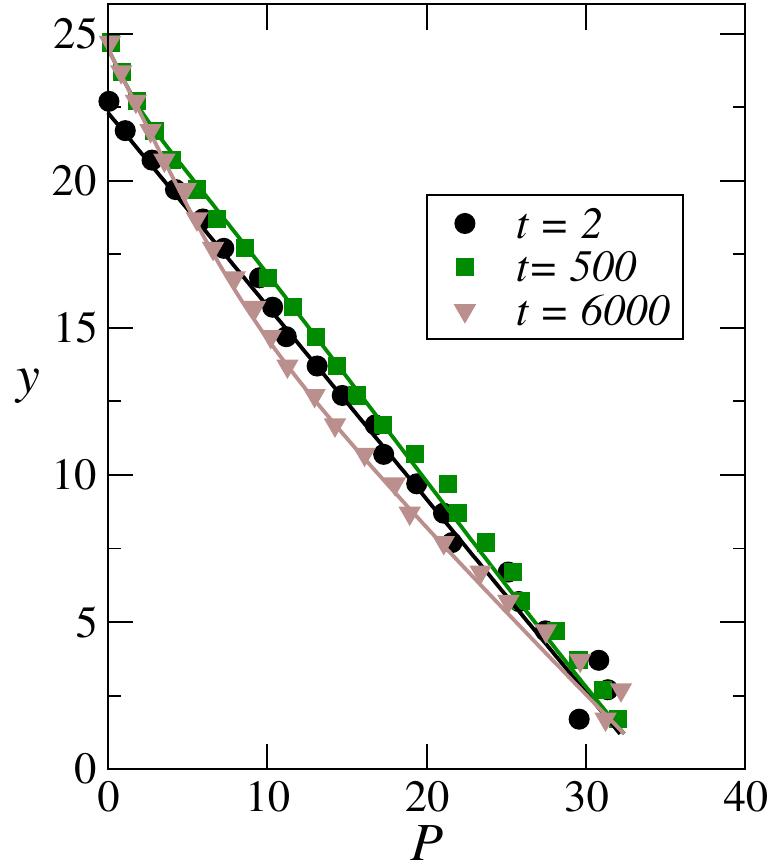}\put(-120,145){(b)}\hfill
     \includegraphics[scale=0.33]{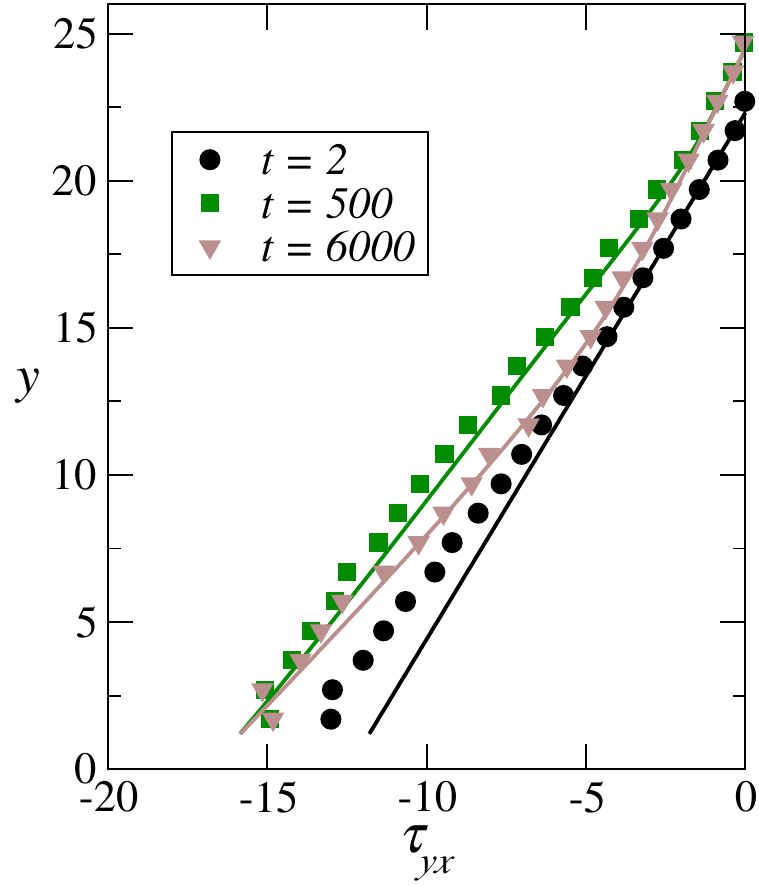}\put(-120,145){(c)}\hfill
   \vspace{0.5cm}
    \includegraphics[scale=0.33]{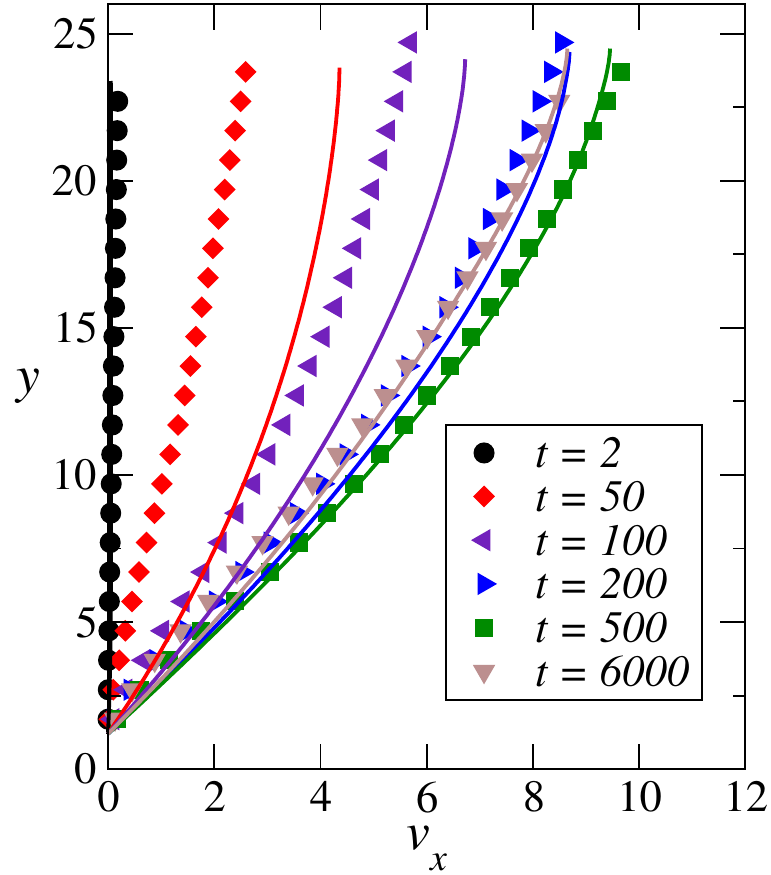}\put(-120,145){(d)}\hfill
     \includegraphics[scale=0.33]{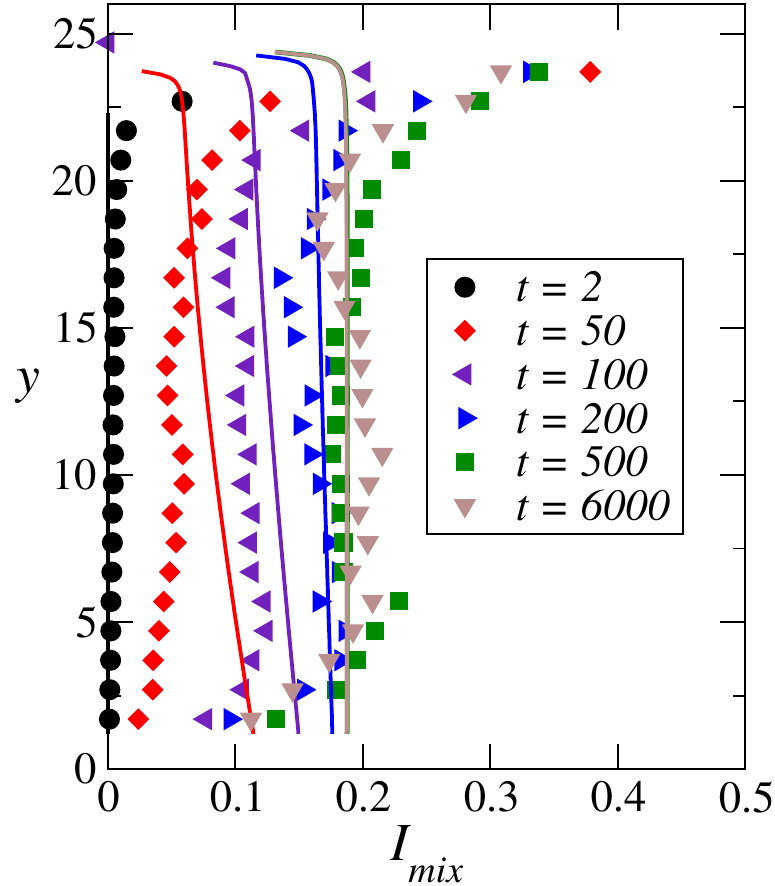}\put(-120,145){(e)}\hfill
      \includegraphics[scale=0.33]{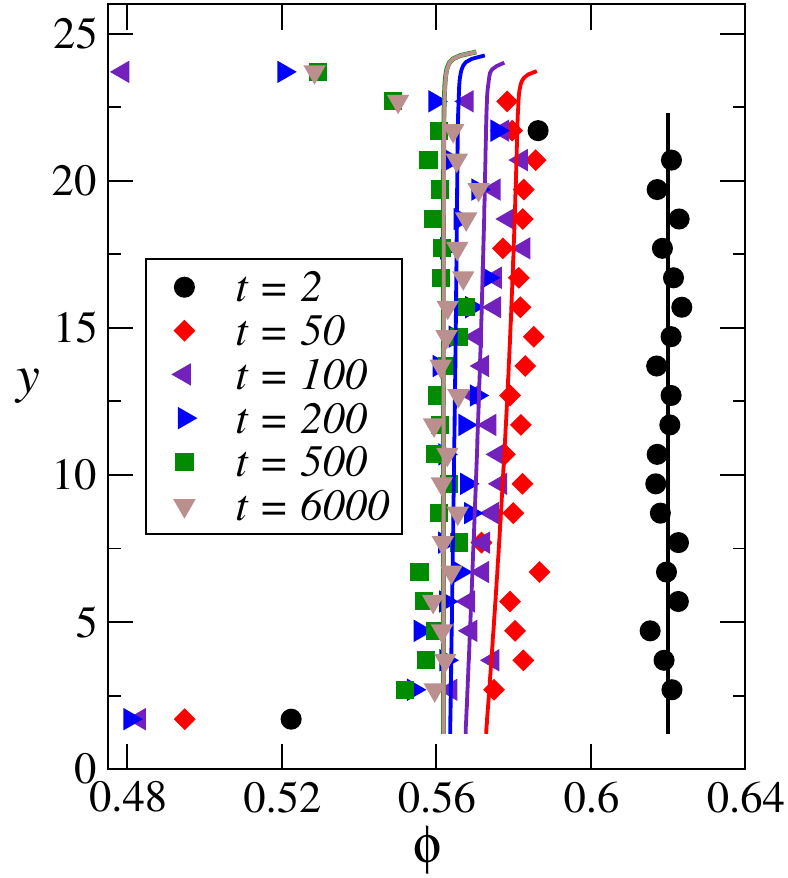}\put(-120,145){(f)}\hfill
    \caption{(a) Comparison of the instantaneous profiles for an equal composition binary mixture having density ratio $\rho = 2.0$. (a) Concentration $f_L$, (b) pressure $P$, (c) shear stress $\tau_{yx}$, (d) mixture velocity $v_x$, (e) inertial number $I_{mix}$, and (f) solids fraction $\phi$. Symbols represent the DEM data and solid lines represent the continuum predictions with each color corresponding to a different time instant. The flow starts from the initial well-mixed configuration at inclination angle $\theta = 25^\circ$.}
    \label{fig:Ins_rho_2_flt_50}
\end{figure}

We first report the results for an equal composition binary mixture with density ratio $\rho = 2.0$. The flow is started from an initially well-mixed state at an inclination angle of $\theta = 25^o$. 
Figure~\ref{fig:Ins_rho_2_flt_50} shows the instantaneous concentration profiles along with other flow properties of the binary mixture at different time instants. \textcolor{black}{Different symbols represent the DEM data at different times while solid lines represent the continuum model predictions. The DEM data as well as continuum predictions at a particular instant are represented by a specific color for the ease of comparison.} 
%such as velocity, pressure, shear stress, inertial no, and packing fraction. 
%The mixture has species density ratio $\rho = 2.0$ and is flowing at an inclination angle of $\theta = 25^o$. 
Figure~\ref{fig:Ins_rho_2_flt_50}a shows the concentration profile of light species $f_L$ in the mixture. Since the flow begins from an initially well-mixed state i.e., $f_L(y,0) = 0.5$, the theoretical prediction for the concentration profile of light species at time $t = 2.0$ remains almost constant (black line) with $f_L \approx 0.50$ for all $y$ values following the DEM data (black circles) closely. 
%This is because the initial state in the DEM simulation involves a nearly mixed state generated by randomly distributing both species particles. 
As time proceeds, the light particles move upwards and heavy particles move downwards leading to increase (decrease) in the $f_L$ values near free surface (base) 
which is clearly observable from the DEM data at $t = 500,1000,2000$ and $6000$. The predicted concentration profiles also depict the same trend and follow the DEM data very well. We observe that the concentration profiles become steady after $t=6000$ indicating that \textcolor{black}{steady state segregation is achieved.} 
%After $t = 6000$ time units, the concentration profile of light species doesn't change and steady state is obtained. We observe a region rich in light particles near the free surface and rich in heavy particles near the bottom.
%Predicted concentration profiles (lines) at different times are in good agreement with corresponding DEM data (symbols). 
Figures \ref{fig:Ins_rho_2_flt_50}b - \ref{fig:Ins_rho_2_flt_50}f show the pressure, shear stress, velocity, inertial number, and packing fraction profiles at different time instants. 
Figures \ref{fig:Ins_rho_2_flt_50}b and \ref{fig:Ins_rho_2_flt_50}c demonstrate that the pressure and shear stress profiles exhibit relatively small change over time due to segregation. The slight non-linear variation of the pressure and shear stress at $t=6000$ is observable due to the segregation of the mixture. The high density particles occupy the lower part of the layer giving rise to larger pressure and shear stress gradient along $y$ compared to that in the upper part of the layer which has larger fractions of low density particles.

\begin{figure}
    \centering 
\includegraphics[scale=0.255]{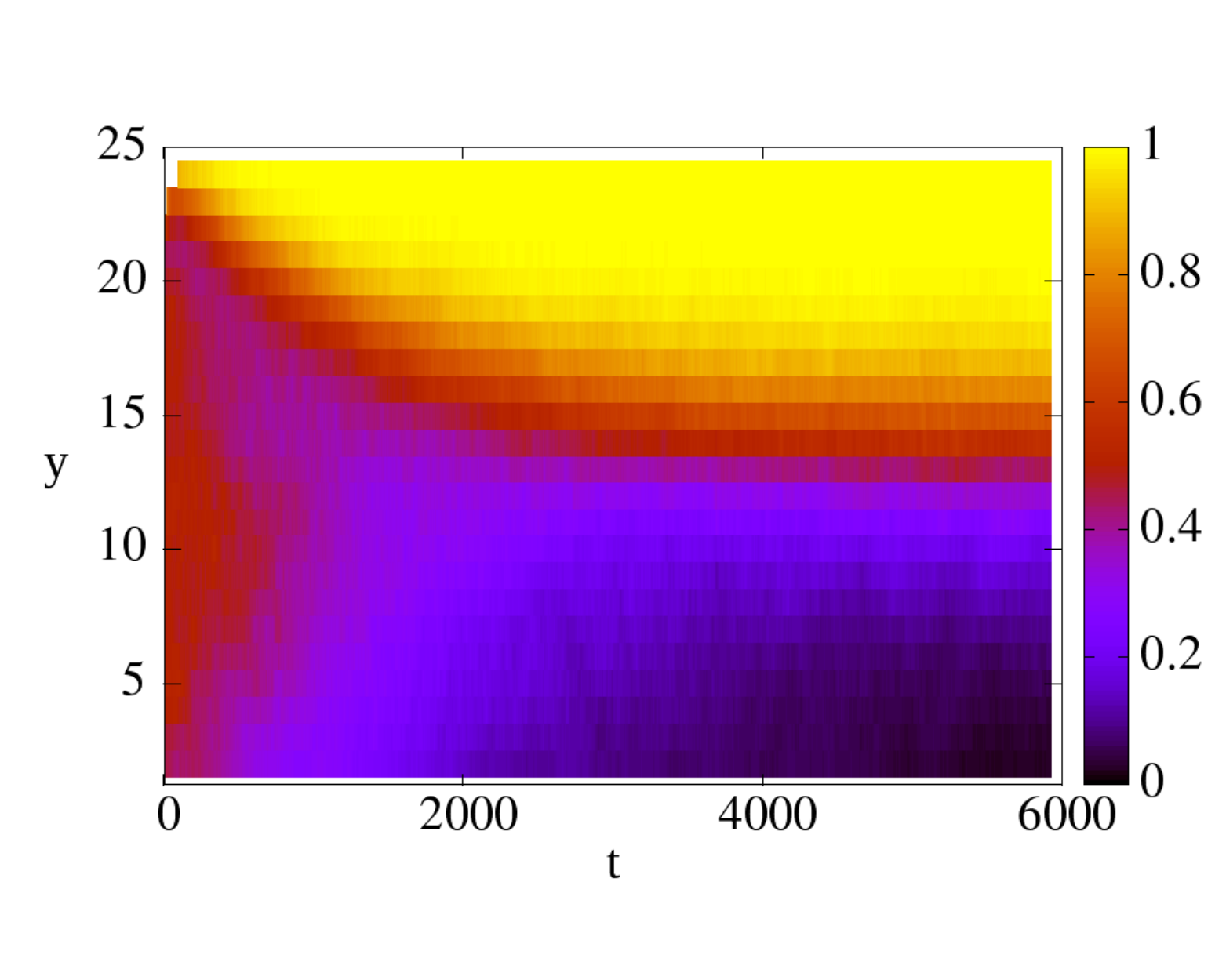}\put(-190,137){(a)} \put(-22,137){$f_L$}\hfill   
\includegraphics[scale=0.255]{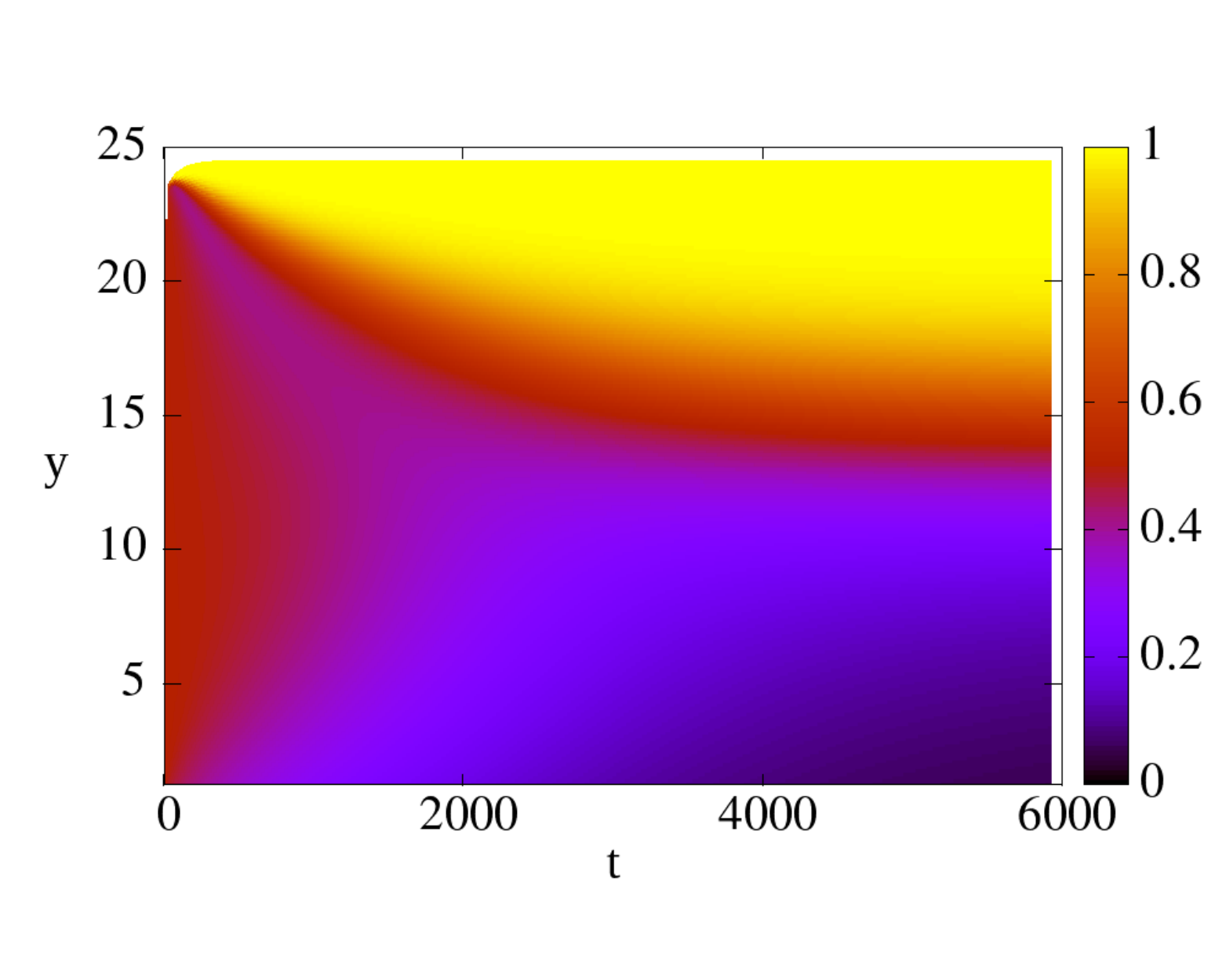}\put(-190,137){(b)}\put(-22,137){$f_L$}\hfill
\includegraphics[scale=0.35]{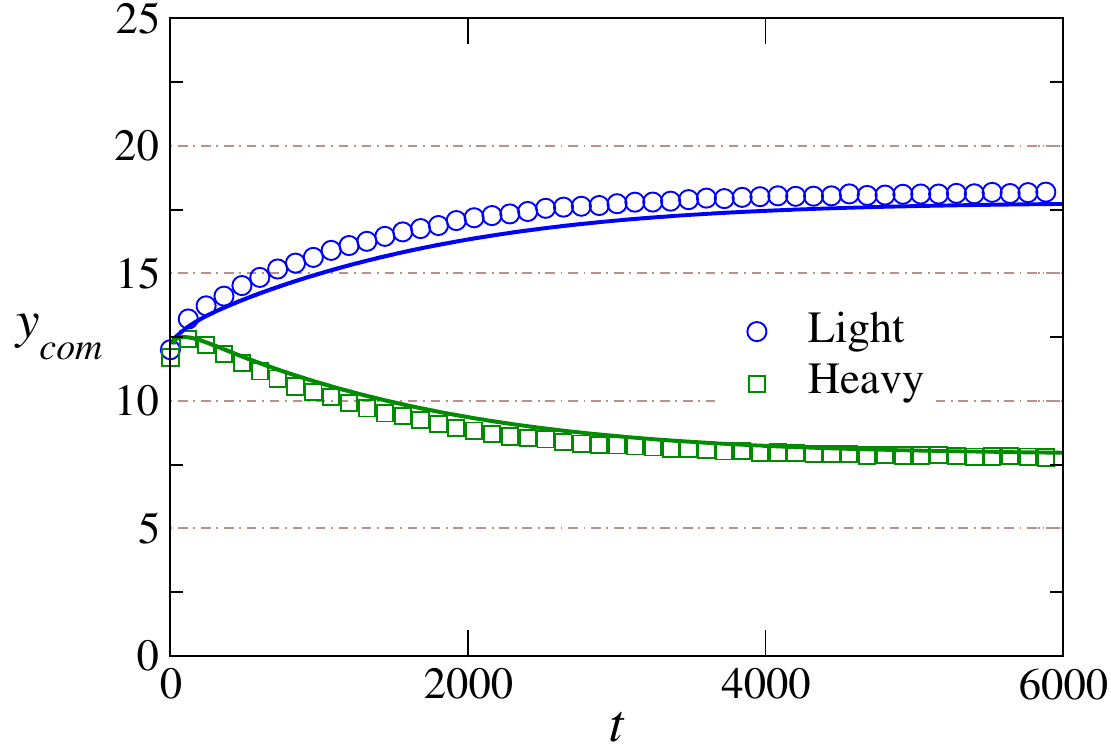}\put(-180,125){(c)}\quad 
\includegraphics[scale=0.35]{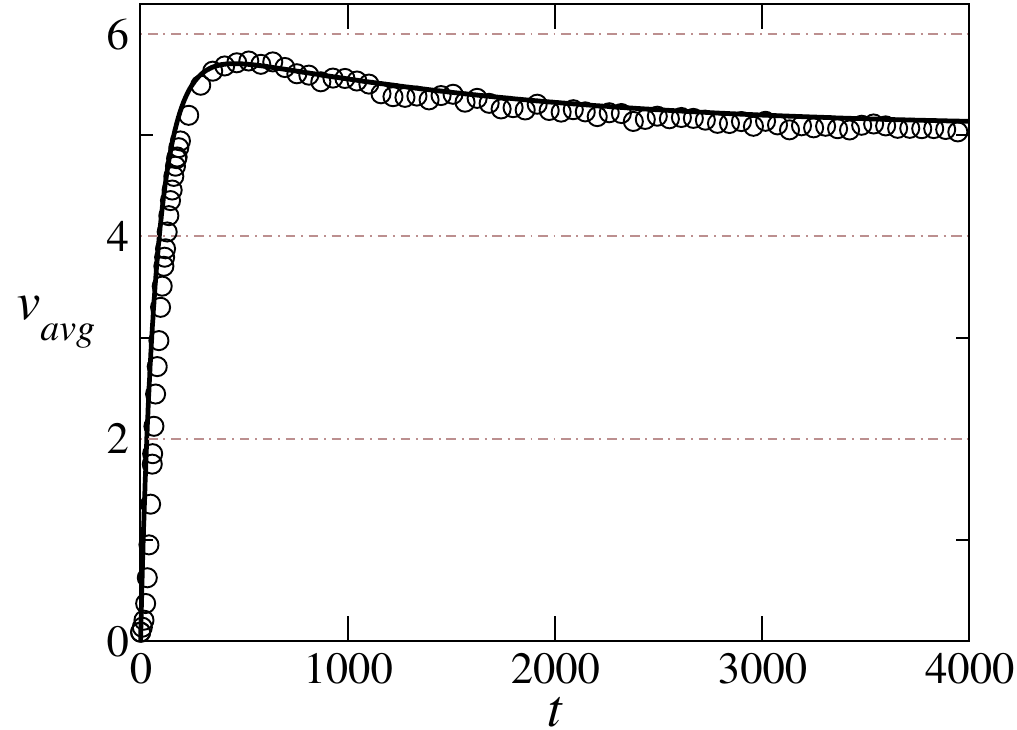}\put(-175,125){(d)}\hfill
\caption{Comparison of evolution of concentration of light species ($f_L$) with time for an equal composition binary mixture with density ratio $ \rho = 2.0$. (a) DEM data, and (b) Continuum simulation. Variation of (c) the species centre of mass and (d) average mixture velocity with time. Symbols represent the DEM data while solid lines represent the theoretical predictions.}
    \label{fig:Avg_rho_2_flt_50}
\end{figure}

Figure \ref{fig:Ins_rho_2_flt_50}d shows the variation of velocity in the flow direction ($v_x$) with the distance from the base. The predicted velocity profiles for $50$ and $100$ time units are found to be reasonably higher compared to the DEM data. After this initial time period, the predicted velocity profiles match very well with the DEM data. As expected, the instantaneous velocity at $t=500$ time units is found to be higher compared to that at $t=200$ as the flow evolves. However, the steady state velocity at $t=6000$ is found to be lower than the instantaneous velocity at $t=500$ time units. This very interesting feature is accurately predicted by the continuum model as well. We discuss this in more detail later while discussing the results for figure~\ref{fig:Avg_rho_2_flt_50}.
The deviation in the theoretical predictions at initial times can also be observed in the variation of the inertial number,  shown in figure \ref{fig:Ins_rho_2_flt_50}e. However the prediction for later times agree reasonably well with DEM simulations. The variation of solids fraction ($\phi$) at different times is shown in figure~\ref{fig:Ins_rho_2_flt_50}f and the predicted profiles match well with the DEM data. 
%The results and discussion in Appendix~\ref{sec:low_I_mu_phi} 

In order to understand the reason for the deviation of the early time predictions, we also performed DEM simulations of monodisperse systems having only one species. We found that \textcolor{black}{similar} differences in the early times predictions are observed in the case of monodisperse granular materials \textcolor{black}{as well}. Hence we conclude that the origin of these differences lies in the inertial number based $\mu-I$ rheological model.
We also note that the validity of the inertial number based rheological description becomes questionable at very low values of the inertial numbers~\citep{andreotti2013granular}. Our DEM simulations for monodisperse materials do show that the $\mu-I$ relation for $I<0.05$ deviates from the JFP model expression \textcolor{black}{when plotted} using the parameters reported in table~\ref{tab:rheopar}. 
However, our predictions at later times are expected to be relatively less affected by these deviations \textcolor{black}{observed for} the early time predictions. This is because the flow velocity varies only by a small amount for $t > 200$ (see figure~\ref{fig:Ins_rho_2_flt_50}d) while the species concentration evolution due to segregation occurs over a much larger timescale and reaches to near steady state at $t\approx 6000$. 

\begin{figure}
    \centering
  %  \includegraphics[scale=0.35]{Mono_plus_binary_velocity_fig4.pdf}
  %  \hfill
 %   \includegraphics[scale=0.35]{Mono_plus_binary_velocity_fig4b.pdf}
  % \includegraphics[scale=0.35]{Mono_plus_binary_velocity_fig4_shifted_v1.pdf} 
  % \includegraphics[scale=0.35]{Mono_plus_binary_velocity_fig4_shifted_withbinary_predictionsv1.pdf}
  \includegraphics[scale=0.35]{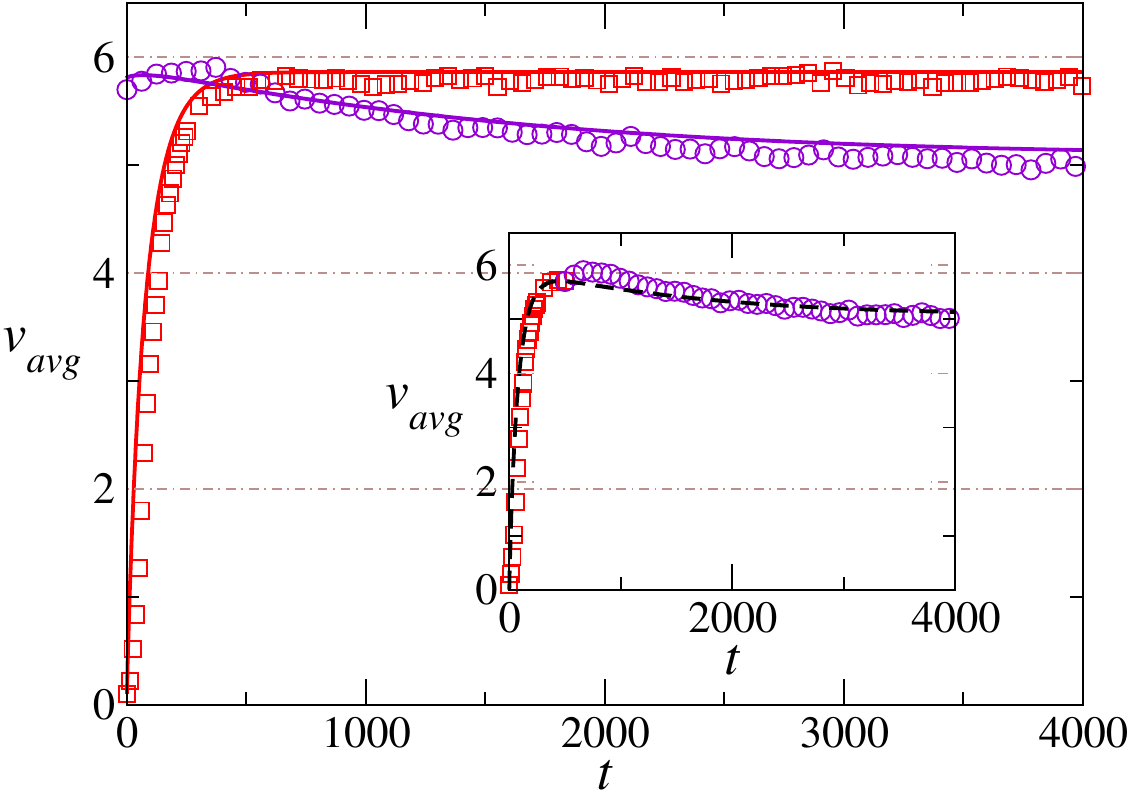} 
    \caption{Variation of average velocity with time for monodisperse mixture. Red squares represent DEM data while the red line represents theory data for monodisperse mixture. Purple circles/lines represent the data for the case where we increased the mass of randomly selected $50\%$ particles by a factor of two, starting from the steady state positions and velocities of the particles corresponding to monodisperse case after $500$ time units. \textcolor{black}{Inset shows the data for monodisperse case up to $500$ time units and the sudden mass change case for $t \geq 500$. Black dashed line represents model predictions for binary mixture shown by solid line in figure~\ref{fig:Avg_rho_2_flt_50}d.}}
\label{fig:Mono_plus_binary_avg_velocity}
\end{figure}

Variation of the light species concentration across the layer in figure \ref{fig:Ins_rho_2_flt_50}a could be shown only for few specific time instants. In order to get a better understanding of evolution of the concentration of light particles across both spatial as well as temporal coordinates, we plot the light species concentration using a color map across the space-time plane in figure~\ref{fig:Avg_rho_2_flt_50}a and \ref{fig:Avg_rho_2_flt_50}b. We report the data for the case shown in figure~\ref{fig:Ins_rho_2_flt_50}, i.e., density ratio $\rho = 2.0$ and inclination angle $\theta = 25^o$.
Figure \ref{fig:Avg_rho_2_flt_50}a shows the concentration color map of light species obtained from the DEM simulations while the predictions from the theory are shown in figure \ref{fig:Avg_rho_2_flt_50}b. In this representation, yellow denotes $f_L = 1$, i.e., pure light species, whereas black corresponds to $f_L = 0$, i.e., pure heavy species.
The spatial distribution of light and heavy species across flowing layer height using the DEM data (figure \ref{fig:Avg_rho_2_flt_50}a) matches well with the predictions (figure \ref{fig:Avg_rho_2_flt_50}b) throughout the space-time plane. \textcolor{black}{The concentration using DEM data in figure~\ref{fig:Avg_rho_2_flt_50}a exhibits more discrete variations compared to the theoretical model, which can be attributed to the large grid size (equal to $d$) used in the case of DEM.} The white region near the top represents the region beyond the free surface. 

For a more quantitative comparison of the theory with DEM simulations, we report the variation of the center of mass of the species $y_{com}$ with time $t$.
Figure \ref{fig:Avg_rho_2_flt_50}c shows the variation of $y_{com}$ for both light (blue) and heavy (green) species in the mixture. Symbols represent the DEM data while solid lines represent the theoretical predictions. 
Initially at time $t = 0$, $y_{com}$ is identical for both species since the flow is started from a well-mixed configuration. \textcolor{black}{The center of mass of the heavy species decreases while that of light species starts to increase until becoming constant for $t \geq 5000$ and the continuum predictions describe the DEM data very well. }
Figure~\ref{fig:Avg_rho_2_flt_50}d shows the variation of average mixture velocity along the flow direction ($v_{avg}$) with time. 
While a rapid increase in $v_{avg}$ is observable at early times, the flow velocity reaches a maximum \textcolor{black}{around $500$ time units} and then slowly decreases \textcolor{black}{afterwards}. 
This non-monotonic variation of the average flow velocity with time is very well captured by the \textcolor{black}{continuum model} as well. \textcolor{black}{This demonstrates the capability of the one dimensional continuum model which is able to capture the intricate details} and subtle nuances of the flow and segregation features observed in DEM simulations. The non-monotonic variation of $v_{avg}$ with time occurs due to the inter-coupling of the rheology and segregation.
\textcolor{black}{This peculiar behavior can be better understood by considering two different flow situations that minimize the effects of the inter-coupling between rheology and segregation.} In the first case, we remove the segregation altogether and consider the average velocity variation with time for a monodisperse granular material starting from rest. We plot $v_{avg}$ for monodisperse material in figure~\ref{fig:Mono_plus_binary_avg_velocity} using red squares (DEM data) and solid red lines (model). For this case, $v_{avg}$ increases with time and reaches a plateau around $500$ time units. \textcolor{black}{This rising behavior at early times} is similar to the early time behavior observed in figure~\ref{fig:Avg_rho_2_flt_50}d, confirming that the early rise of $v_{avg}$ with time occurs due to the acceleration of the layer. 
To simulate a situation where the inter-coupling of segregation with rheology is minimal, we use the following protocol. Starting from the steady state positions and velocities of the particles corresponding to monodisperse case, we randomly selected $50\%$ particles in the layer and increased their mass by a factor of two. \textcolor{black}{This protocol provides us a well-mixed mixture with an initial velocity equal to that for the steady flow of monodisperse granular material of same density. Using this combination of the initial species composition and velocity of the particles, we performed the DEM simulations as well as continuum simulations and observed the variation of the average flow velocity and species center of mass. 
The results from these DEM simulations are shown using purple circles and the theoretical predictions are shown using solid lines of the same color in figure~\ref{fig:Mono_plus_binary_avg_velocity}. The $v_{avg}$ for this case keeps decreasing with time. The inset in figure~\ref{fig:Mono_plus_binary_avg_velocity} shows that figure~\ref{fig:Avg_rho_2_flt_50}d can be approximately reconstructed by combining the monodisperse data (red symbols) of figure~\ref{fig:Mono_plus_binary_avg_velocity} up to $500$ time units with the data for the sudden change in the particle density case for the fully developed flow (purple symbols). The black dashed line in the inset corresponds to the model predictions shown in figure~\ref{fig:Avg_rho_2_flt_50}d. The excellent agreement of black dashed line with the data confirms that the initial rise in the average velocity occurs primarily due to acceleration of the layer. Further, the slow decreasing behavior of $v_{avg}$ in figure~\ref{fig:Avg_rho_2_flt_50}d occurs due to the slow evolution of segregation during the flow.} Due to the dependence of the mixture viscosity on local concentration in the inertial number based rheological model, a complex interplay of the segregation and rheology during the flow occurs. Since the segregation occurs over a large time scale, the viscosity of the granular mixture also keeps changing and leads to slow decrease of the average velocity. \textcolor{black}{The proposed continuum model accounts for this inter-coupled nature of rheological and segregation model, and hence, this interesting non-monotonic behavior of average velocity is captured very accurately by the model.} It is important to note that in absence of the inter-coupling of the rheology with segregation, such flow features cannot be captured.
%This decrease in the average velocity for the mixture case can be understood as follows. As the segregation proceeds, the heavy species concentrates towards the base while the 
%However, the segregation in this case is decoupled from the initial evolution of the flow due to rheology. 
%The inset of figure \ref{fig:Avg_rho_2_flt_50}d shows the significant differences from the DEM data are observable at initial times. Such differences are also observable in case of monodisperse mixture. 

\begin{figure}
    \centering
    \includegraphics[scale=0.33]{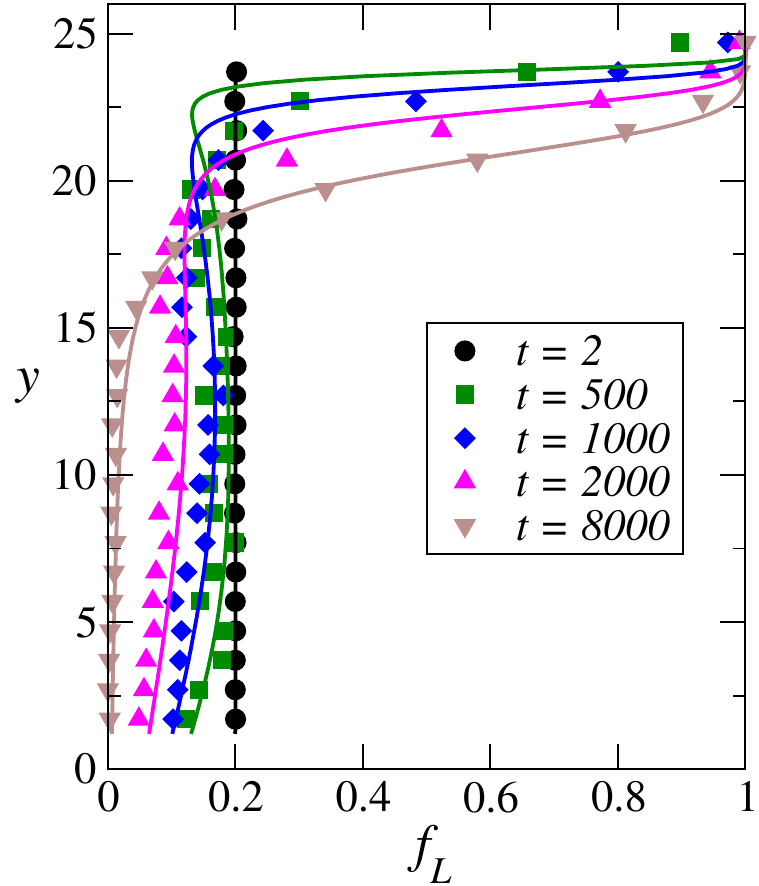} \put(-130,140){(a)}\quad \quad 
    \includegraphics[scale=0.37]{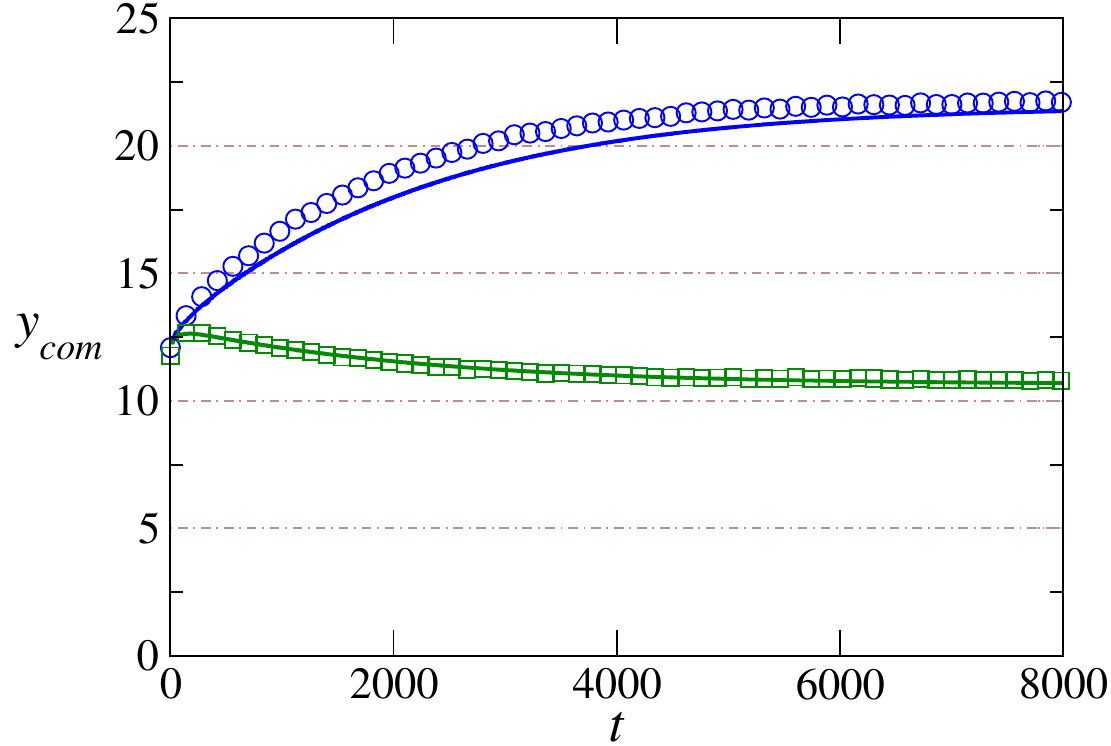}\put(-180,140){(b)}
    \vspace{0.1cm}
    \includegraphics[scale=0.33]{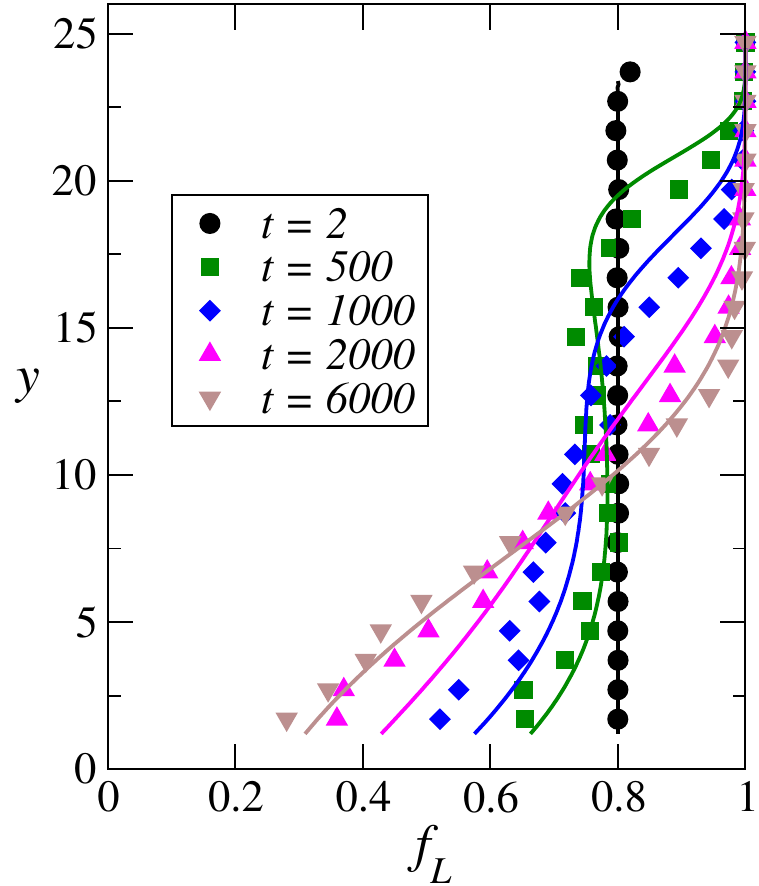}\put(-130,140){(c)}\quad \quad
    \includegraphics[scale=0.37]{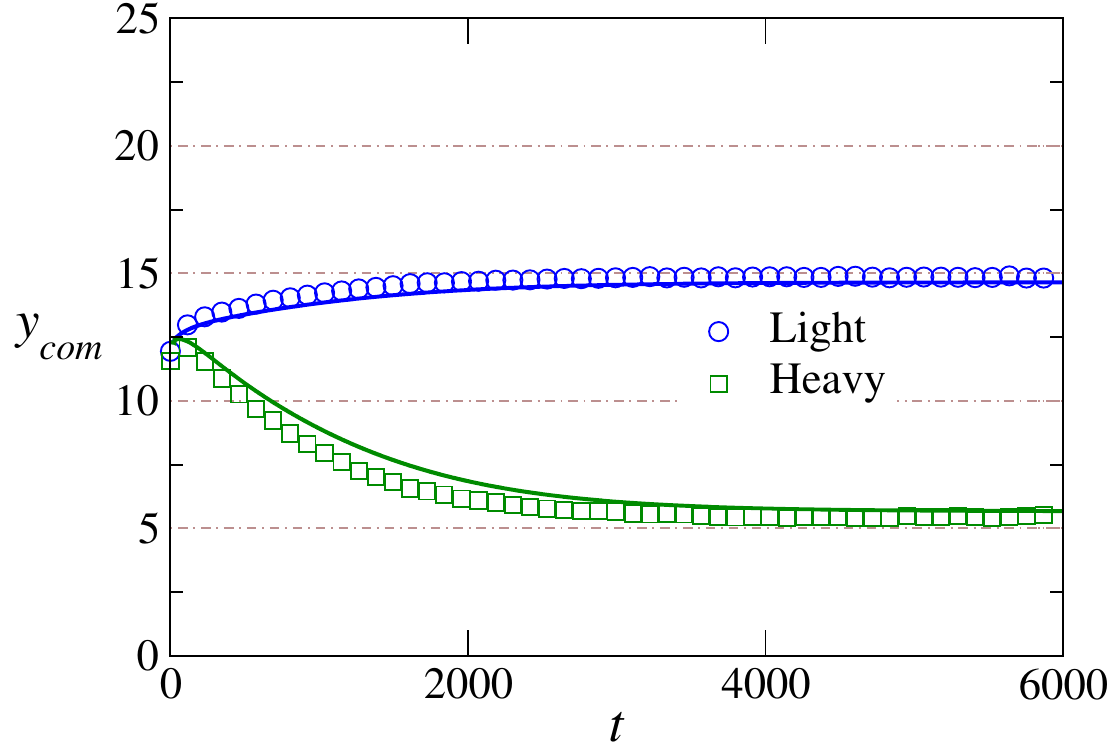}\put(-180,140){(d)}
    \caption{\textcolor{black}{(a) Instantaneous concentration profiles of light species ($f_L$) and (b) time evolution of center of mass ($y_{com}$) for binary mixture with $20\%$ light species ($f^T_L = 0.2$) for density ratio $\rho = 2.0$ at inclination angle $\theta = 25^\circ$. (c) and (d) show the same data for $f^T_L = 0.8$ case.
    Symbols represent the DEM data while solid lines represent the model predictions.}}
    \label{fig:flt_80_20}
\end{figure}
Figures~\ref{fig:Ins_rho_2_flt_50} and \ref{fig:Avg_rho_2_flt_50} confirm that the particle force-based theory is successfully able to predict the time-dependent segregation for equal composition binary mixtures. We next report the results for two different mixture compositions. As before, the flow is started from an initial well-mixed state at an inclination angle $\theta = 25^o$ and the density ratio of the heavy to light species is $\rho = 2.0$. 
Figure~\ref{fig:flt_80_20}a shows the instantaneous concentration profiles at different times and figure~\ref{fig:flt_80_20}b shows the time variation of the centre of mass position of light and heavy species for a mixture having $20\%$ light particles ($f^T_L = 0.20$). Figures~\ref{fig:flt_80_20}c and \ref{fig:flt_80_20}d show the same data for mixture having $80\%$ light particles ($f^T_L = 0.80$). 
Starting from a uniform concentration of $f_L = f^T_L$, the concentration of light species keeps decreasing near the base and keeps  increasing near the top as time progresses. %For $f^T_L = 20$, the pure region of light particles near the top spans about $2.5d$ at steady state. However, for $f^T_L = 0.80$, the pure light particle region extends to approximately $15d$ at steady state.
For the case of $f^T_L = 0.20$, the center of mass for light species (blue color) rises from $y_{com}\approx 12$ to $y_{com}\approx 22$ whereas for $f^T_L = 0.80$, it reaches only up to to $y_{com}\approx 15$. 
Figures~\ref{fig:flt_80_20}a - \ref{fig:flt_80_20}d show that the model predictions (solid lines) are in reasonably good agreement with DEM data (symbols) at all the times. 

\begin{figure}
    \centering
    \includegraphics[scale=0.35]{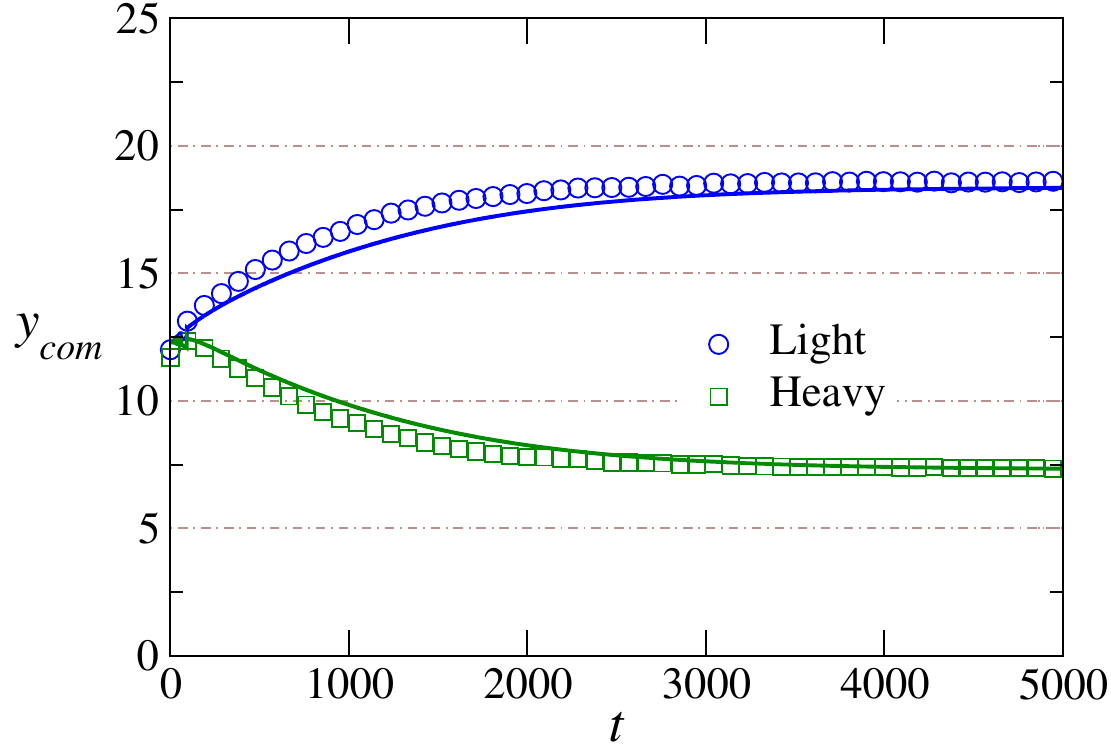}\put(-180,120){(a)} 
    \includegraphics[scale=0.35]{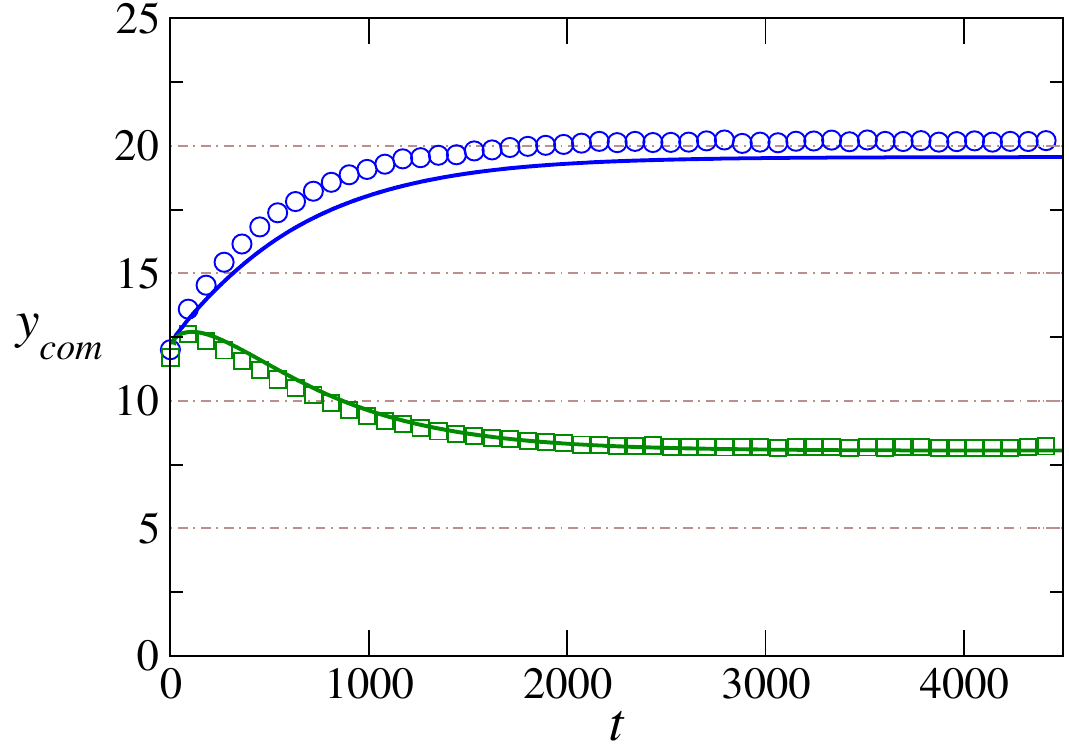}\put(-175,120){(c)}\hfill
    \vspace{0.4cm}
     \includegraphics[scale=0.37]{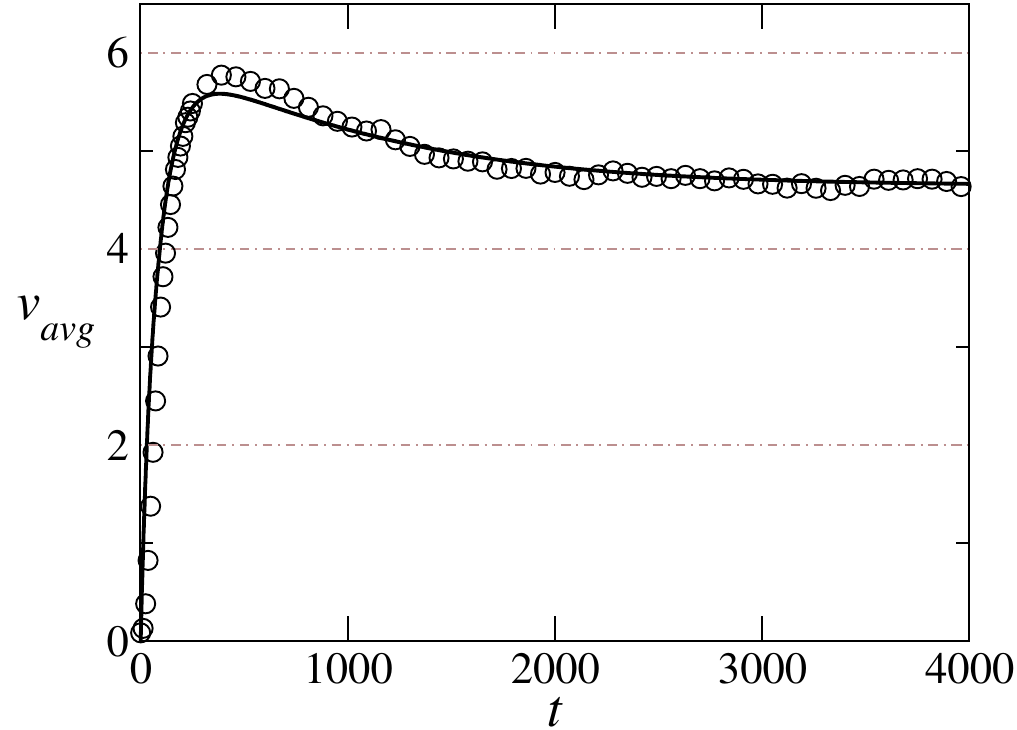}\put(-180,120){(b)} 
   \includegraphics[scale=0.37]{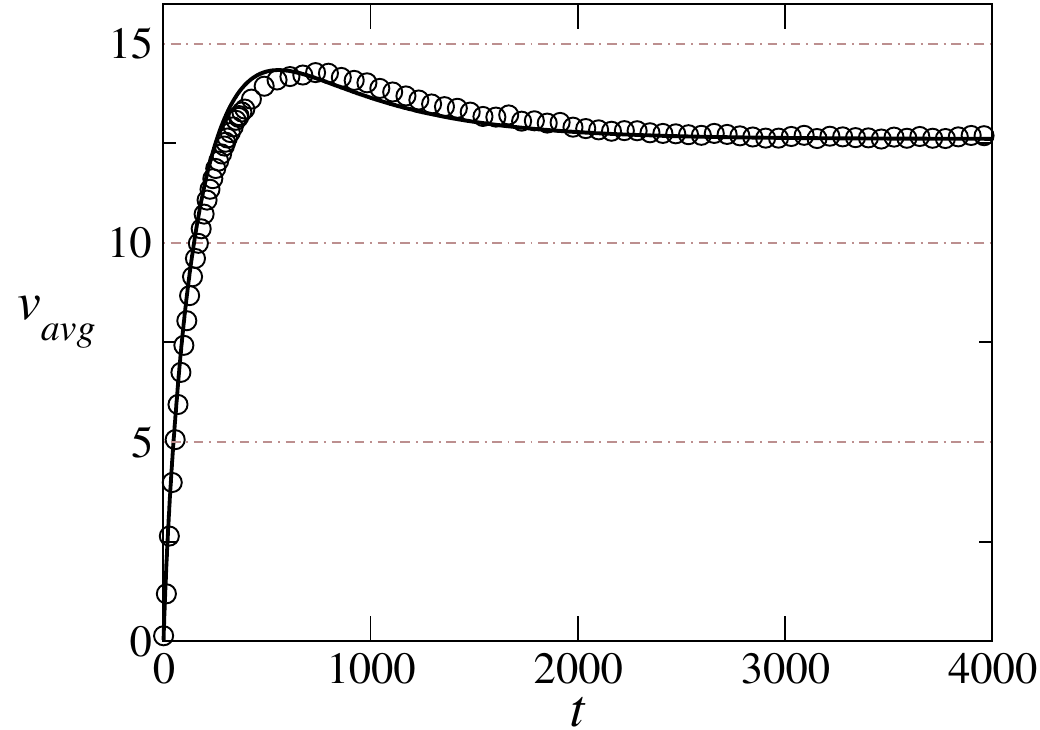}\put(-180,120){(d)}
    \caption{\textcolor{black}{Time variation of (a) centre of mass ($y_{com}$) and (b) average velocity ($v_{avg}$) for an equal composition binary mixture with density ratio $\rho = 3.0$ flowing at $\theta = 25^\circ$. (c) and (d) show the same data for $\theta = 29^\circ$.}}
    \label{fig:rho_3.0_theta_25_29}
\end{figure}

Next, we explore whether the proposed theory is able to capture the influence of the inclination angle on the mixture flow and segregation. 
Figure \ref{fig:rho_3.0_theta_25_29} shows the results for an equal composition binary mixture with density ratio $\rho = 3.0$. As before, the data is reported for a mixture starting from well-mixed initial configuration. Figure~\ref{fig:rho_3.0_theta_25_29}a shows time variation of the species centre of mass for both heavy (green) and light (blue) species \textcolor{black}{for that} mixture flowing at inclination angle $\theta = 25^o$ and the corresponding average mixture velocity is shown in figure~\ref{fig:rho_3.0_theta_25_29}b. Similarly, figures~\ref{fig:rho_3.0_theta_25_29}c and \ref{fig:rho_3.0_theta_25_29}d show the same results for inclination angle $\theta = 29^o$. 
%At steady state, the vertical position of light species at inclination angle $\theta = 25^o$ (shown in figure \ref{fig:rho_3.0_theta_25_29}a) is approximately $y_{com} = 18$ which is $20$ at inclination  $\theta = 29^o$ (shown in figure \ref{fig:rho_3.0_theta_25_29}b). Hence, the height of the flowing layer increases with the increase in inclination angle which is captured well by the theoretical approach. 
As before, the average velocity for both inclinations rises quickly, achieves a maximum value and then starts to slowly decrease due to the inter-coupling of rheology and segregation. The steady state average mixture velocity at inclination angle $\theta = 29^o$ is more than twice of the steady state average velocity at $\theta = 25^o$. 
%We don't observe the differences in the predicted velocity profile with DEM data at initial times at inclination angle $\theta = 29^o$. 
Theoretical predictions (shown using solid line) are in very good agreement with DEM simulation data (symbols) for both these inclination angles. We have also performed simulations for different binary mixtures with density ratios $\rho = 1.5, 2.0, 3.0$, and $5.0$ over a wide range of compositions and inclination angles. While these results are not reported here to avoid repetition, we find that the theoretical predictions for all these different cases are found to be in very good agreement with DEM simulations. \textcolor{black}{Hence, we conclude that the particle force-based theory for binary mixture is successfully able to predict the evolution of segregation over a wide range of concentrations, inclination angles, and density ratios.}

\section{Evolution of flow properties of multicomponent mixtures}
\label{sec:ternary}
In this section, we report the results for mixtures having more than two species. %For this, we utilize the generalized particle force-based density segregation model of \cite{sahu_kumawat_agrawal_tripathi_2023} and predict the time-dependent concentration profiles and flow properties and compare them with the DEM results. 
\begin{figure}
    \centering
     \includegraphics[scale=0.33]{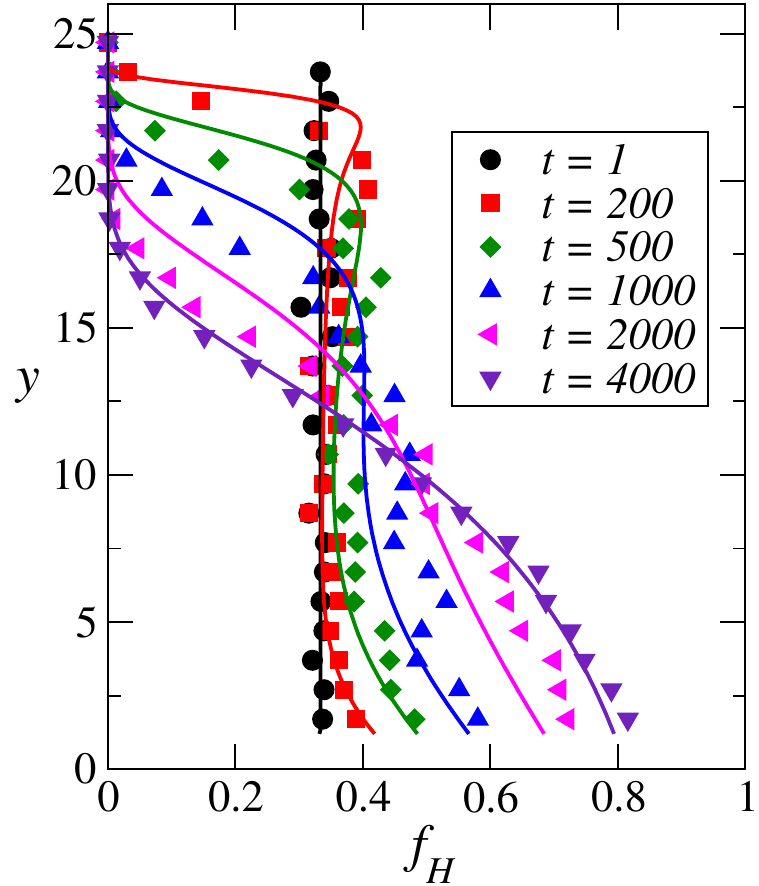}\put(-125,135){(a)}\hfill
    \includegraphics[scale=0.33]{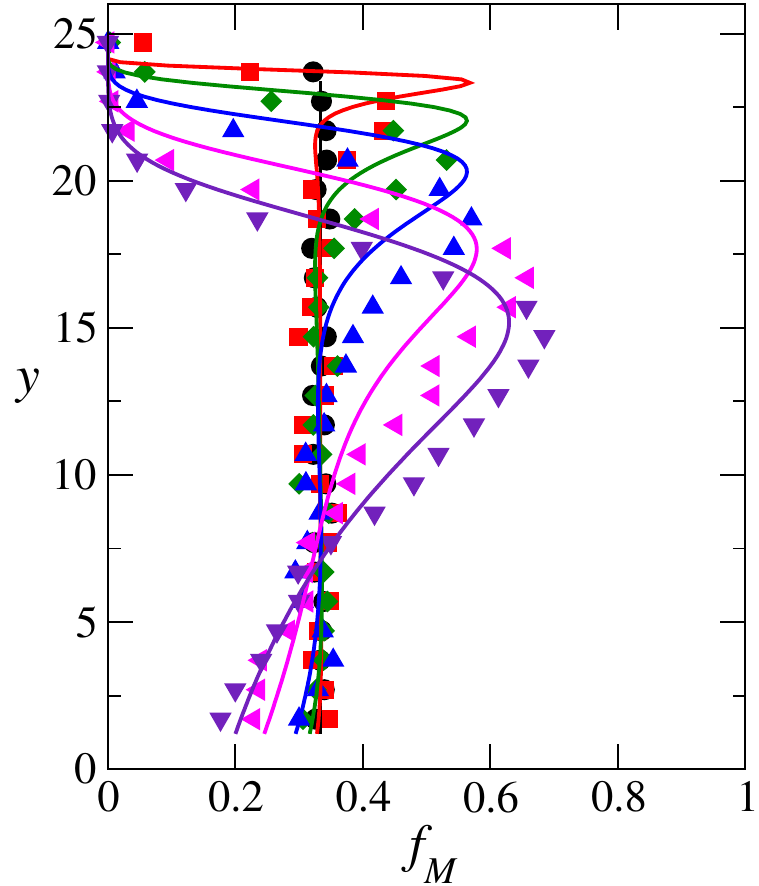}\put(-125,135){(b)}\hfill
    \includegraphics[scale=0.33]{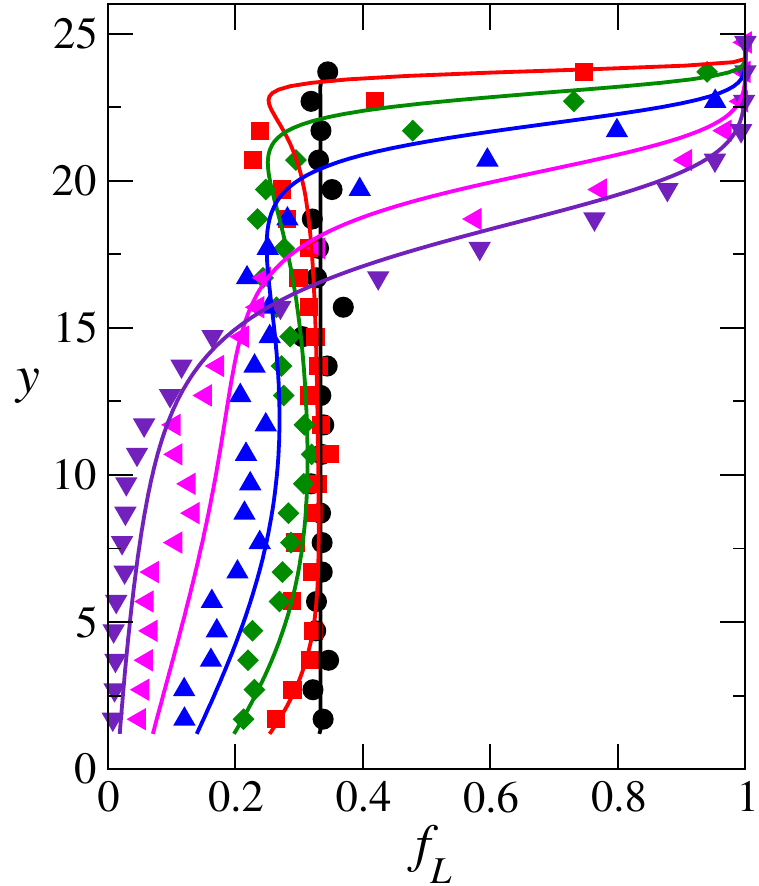}\put(-125,135){(c)}\hfill
    \caption{Instantaneous concentration profiles of (a) high density, (b) medium density, and (c) low density species in an equal composition ternary density mixture having density ratios $\rho_H: \rho_M : \rho_L$ = $3:2:1$. Flow is started from an initial mixed configuration at inclination angle $\theta = 25^\circ$.}
    \label{fig:ternary_equal_comp}
\end{figure}
Figure~\ref{fig:ternary_equal_comp} shows the results for a ternary mixture having equal composition of each of the three species $f^T_L = f^T_M = f^T_H = 1/3$. Species density ratios are $\rho_H: \rho_M : \rho_L$ = $3:2:1$ and the flow starts from a well-mixed initial configuration at inclination angle $\theta = 25^o$. Figure \ref{fig:ternary_equal_comp}a shows the instantaneous concentration profiles of heavy species at different time instants. The concentration profiles for medium and light species are shown in figure~\ref{fig:ternary_equal_comp}b and \ref{fig:ternary_equal_comp}c, respectively. Symbols represent the DEM data while solid lines represent continuum model predictions. Different colors correspond to \textcolor{black}{model predictions} at different times. Initially, at time unit $t = 2$, the concentration of heavy, medium, and light species are $1/3$ throughout the layer height as the flow starts from a uniformly mixed state. As the flow evolves, the concentration of heavy species increases near the base and decreases near the free surface due to segregation. At $t = 4000$, the concentration of heavy species (purple line and symbols) is found to be zero near the free surface (see figure \ref{fig:ternary_equal_comp}a). 
Similarly, the light particles experience net upward force in the presence of medium and heavy-density particles. Hence, the concentration of light species decreases gradually toward the base with time, %. At $t = 4000$, it is almost zero at $40\%$ of height near the base and almost $1$ at $20\%$ of height near the top,
as shown in figure~\ref{fig:ternary_equal_comp}c. 
The medium-density particles, however, experience an upward force in the basal region rich in heavy-density particles and a downward force in the free surface region rich in low-density particles. Hence, the concentration of medium-density species decreases near the base as well as near the surface with time. More and more medium density particles concentrate in the middle \textcolor{black}{part of the layer with time} and hence, the concentration of medium species at $t = 4000$ is larger in the middle region compared to the free surface and base, as shown in figure~\ref{fig:ternary_equal_comp}b. 

\begin{figure}
    \centering
     \includegraphics[scale=0.35]{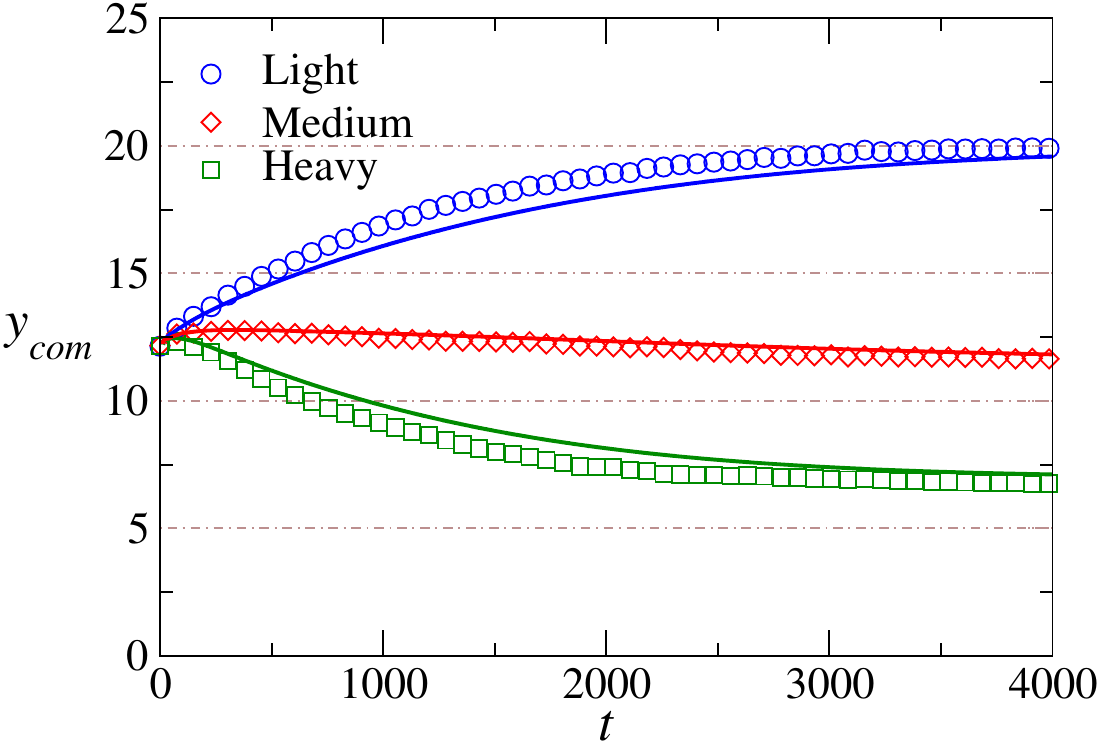}\put(-180,125){(a)}
     \includegraphics[scale=0.35]{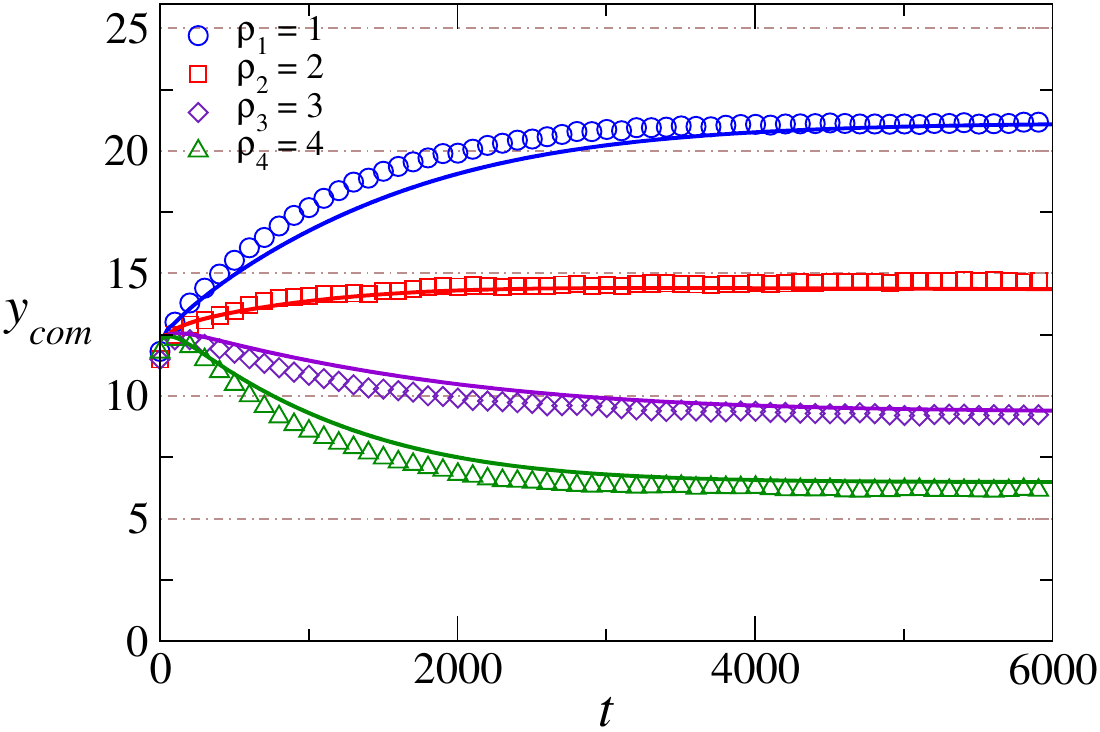}\put(-175,125){(b)}
    \caption{Time evolution of centre of mass of all species in equal composition mixtures flowing at inclination angle $\theta = 25^\circ$. (a) Ternary mixture having density ratios $\rho_H: \rho_M : \rho_L$ = $3:2:1$, and (b) quaternary mixture having density ratios $\rho_4: \rho_3 : \rho_2 : \rho_1 = 4:3:2:1$.
    }
\label{fig:ternary_equal_comp_3_4_comp}
\end{figure}

Figure~\ref{fig:ternary_equal_comp_3_4_comp}a shows the variation of centre of mass $y_{com}$ with time for this equal composition ternary mixture. Evidently, theoretical predictions (solid lines) using the continuum model match well with the DEM data (symbols) for all the three species for the case of ternary mixture as well. 
Figure~\ref{fig:ternary_equal_comp_3_4_comp}b shows the variation of the center of mass of the species for a quaternary mixture having density ratios $\rho_4: \rho_3 : \rho_2 : \rho_1 = 4:3:2:1$ starting from a well-mixed initial configuration at inclination angle $\theta = 25^\circ$. Again, excellent match is observed between the DEM simulations and theoretical predictions for this case as well, without using any fitting parameter in the theoretical formulation. These results confirm that \textcolor{black}{the proposed continuum model} is able to capture the time evolution of the flow and segregation for multicomponent mixtures as well.

\section{Effect of initial configuration of species}
\label{sec:diff_initial_conf}
All the results reported above considered mixtures having a well-mixed initial configuration. %We conclude that the particle force-based theory is able to predict successfully the segregation for transient states. 
We now report results exploring the applicability of our one-dimensional segregation model to capture the flow evolution for different initial configurations for binary and ternary density mixtures. Specifically, we consider the cases for binary and ternary mixtures with an initial configuration that is nearly completely segregated. 
\subsection{Heavy-Near-Base configuration}
\begin{figure}
    \centering
   \includegraphics[scale=0.35]{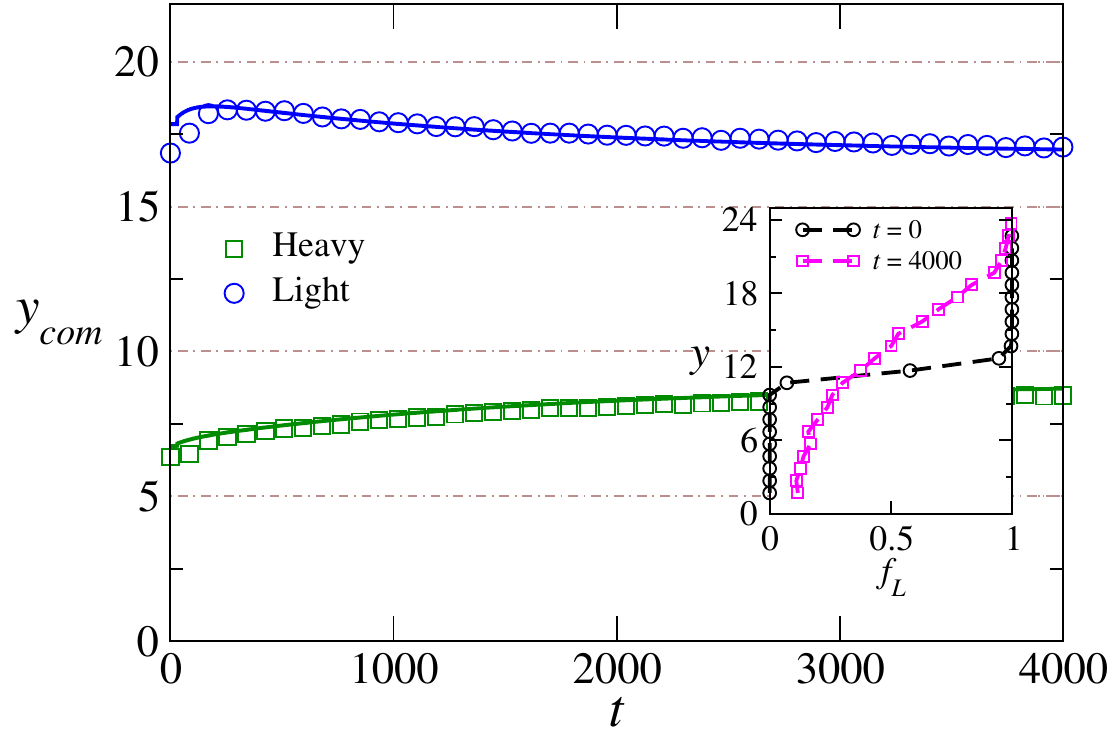}\put(-180,125){(a)} \put(-125,140){Binary mixture}\hfill
    \includegraphics[scale=0.35]{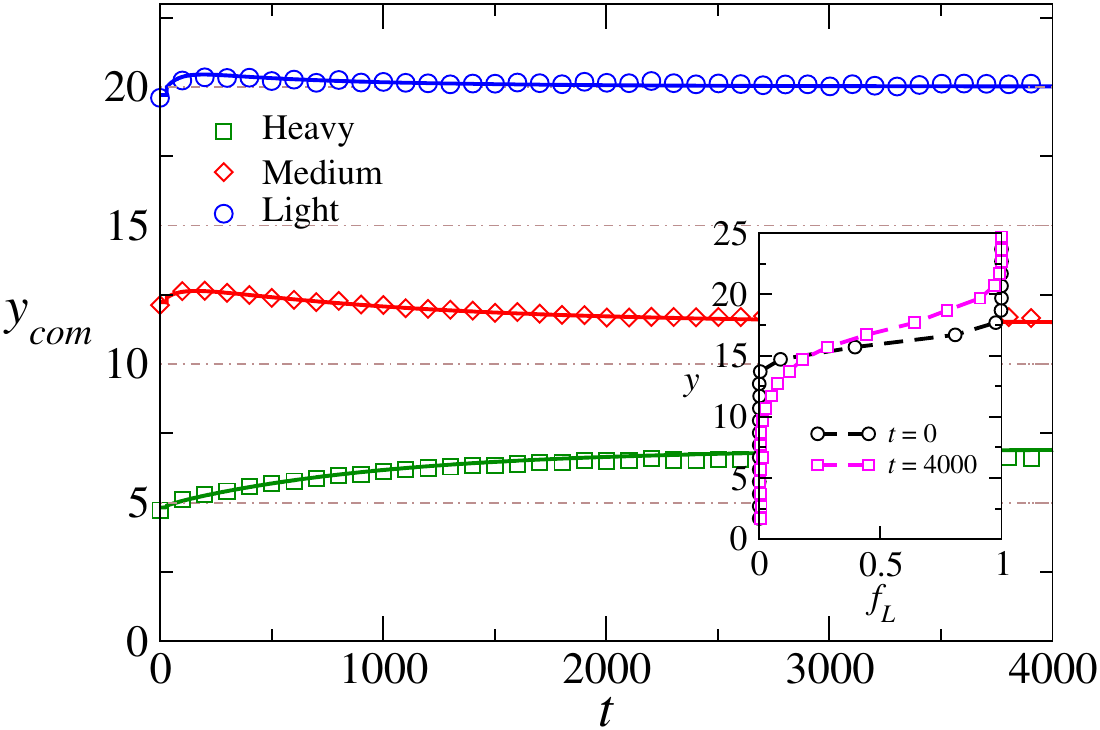}\put(-180,125){(d)}\put(-125,140){Ternary mixture}\hfill
    \includegraphics[scale=0.25]{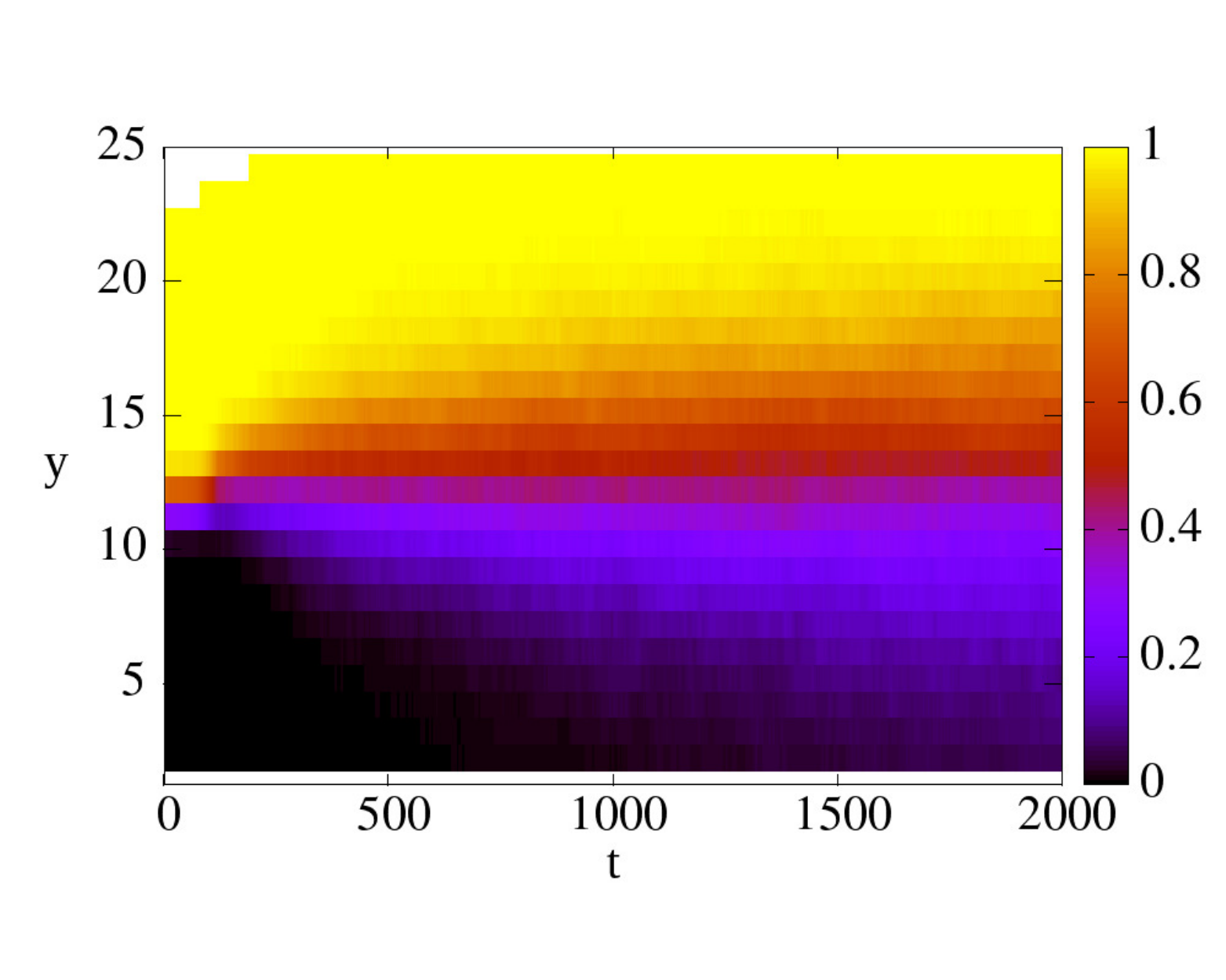}\put(-190,135){(b)} \put(-22,135){$f_L$} \put(-105,135){DEM}
    \hfill  
    \includegraphics[scale=0.25]{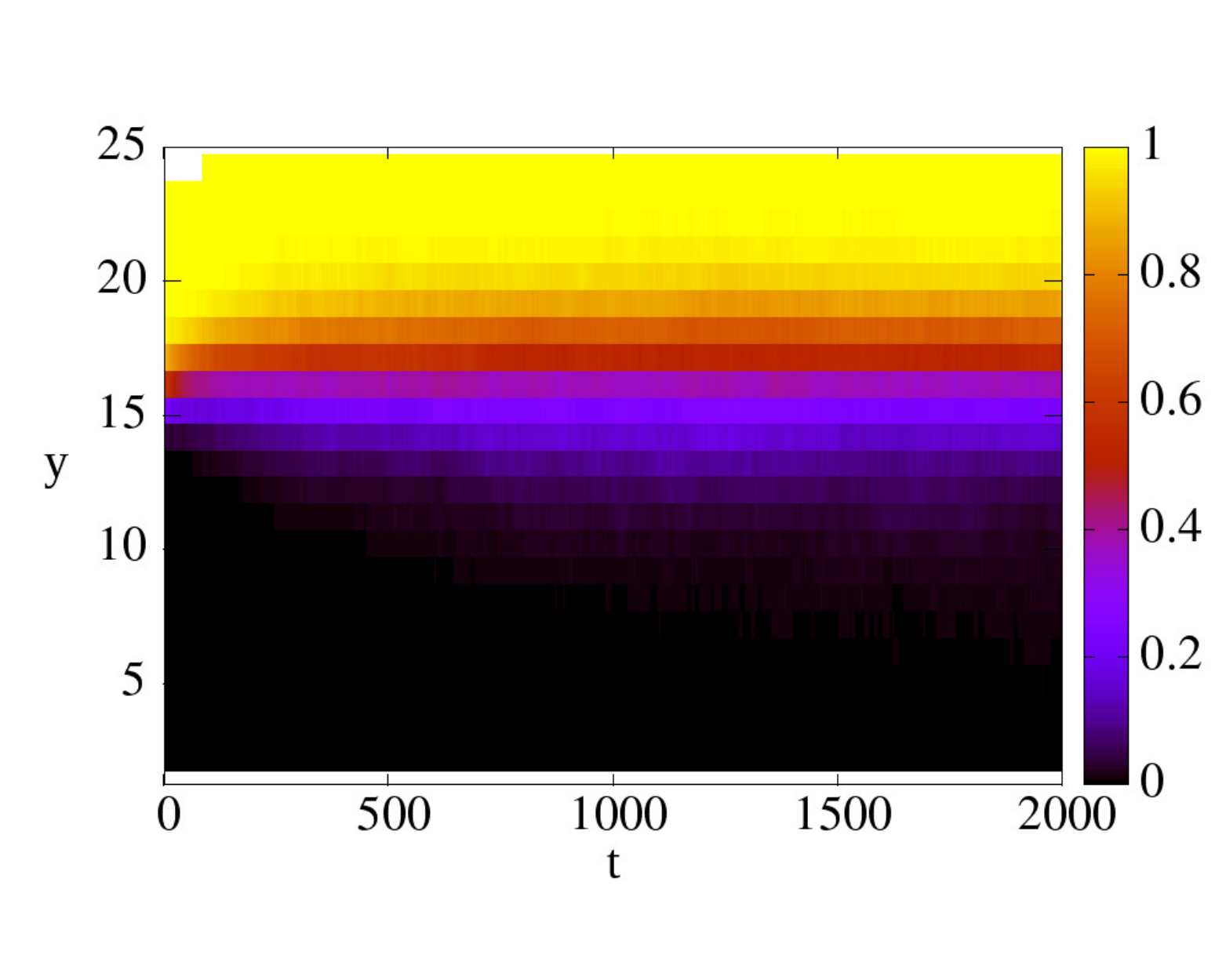}\put(-190,135){(e)} \put(-22,137){$f_L$}\put(-105,135){DEM}
    \hfill 
    \vspace{-0.5cm}
   \includegraphics[scale=0.25]{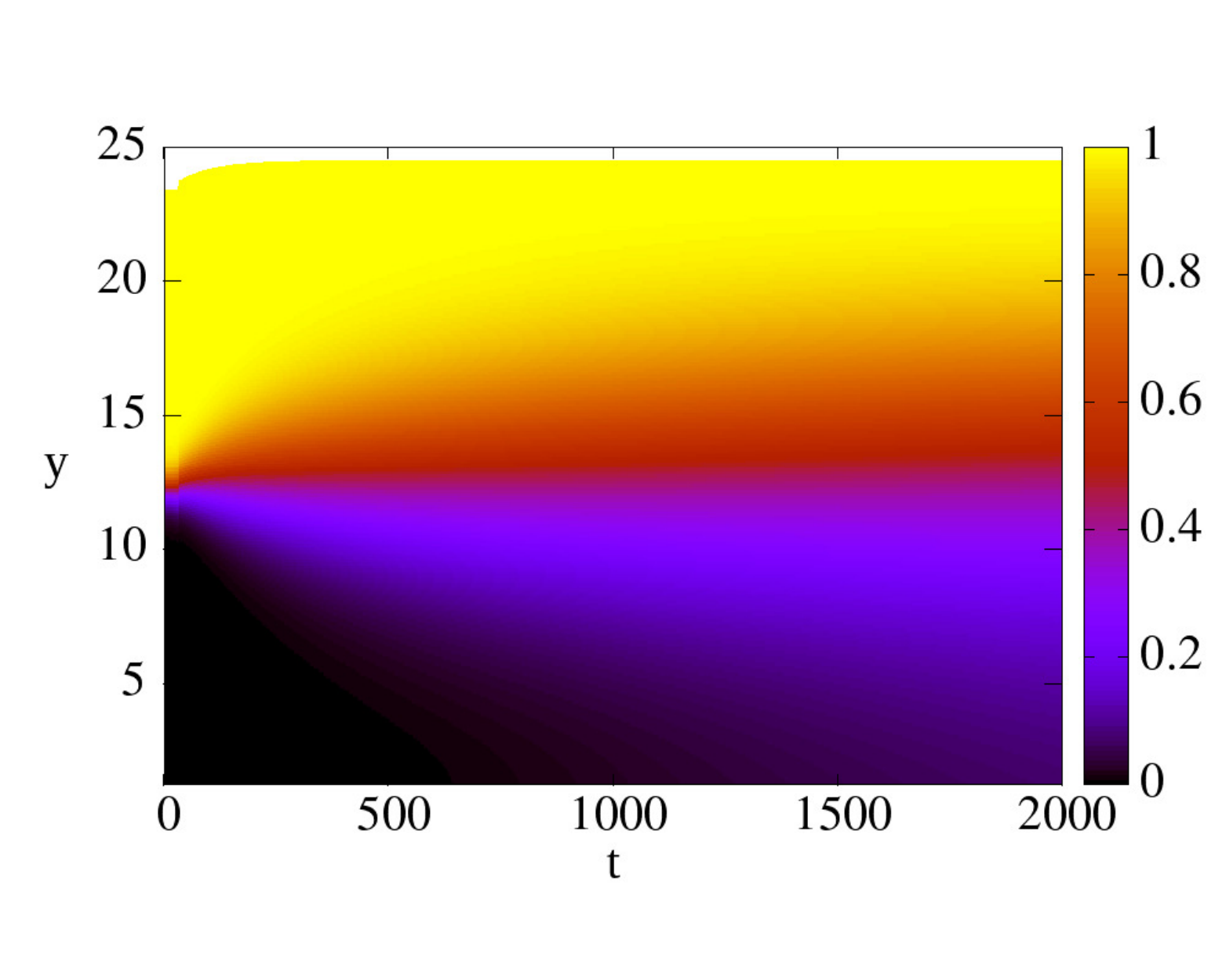}\put(-190,135){(c)} \put(-22,135){$f_L$}\put(-115,135){Continuum}\hfill
  % \hspace{0.01cm}
   \includegraphics[scale=0.25]{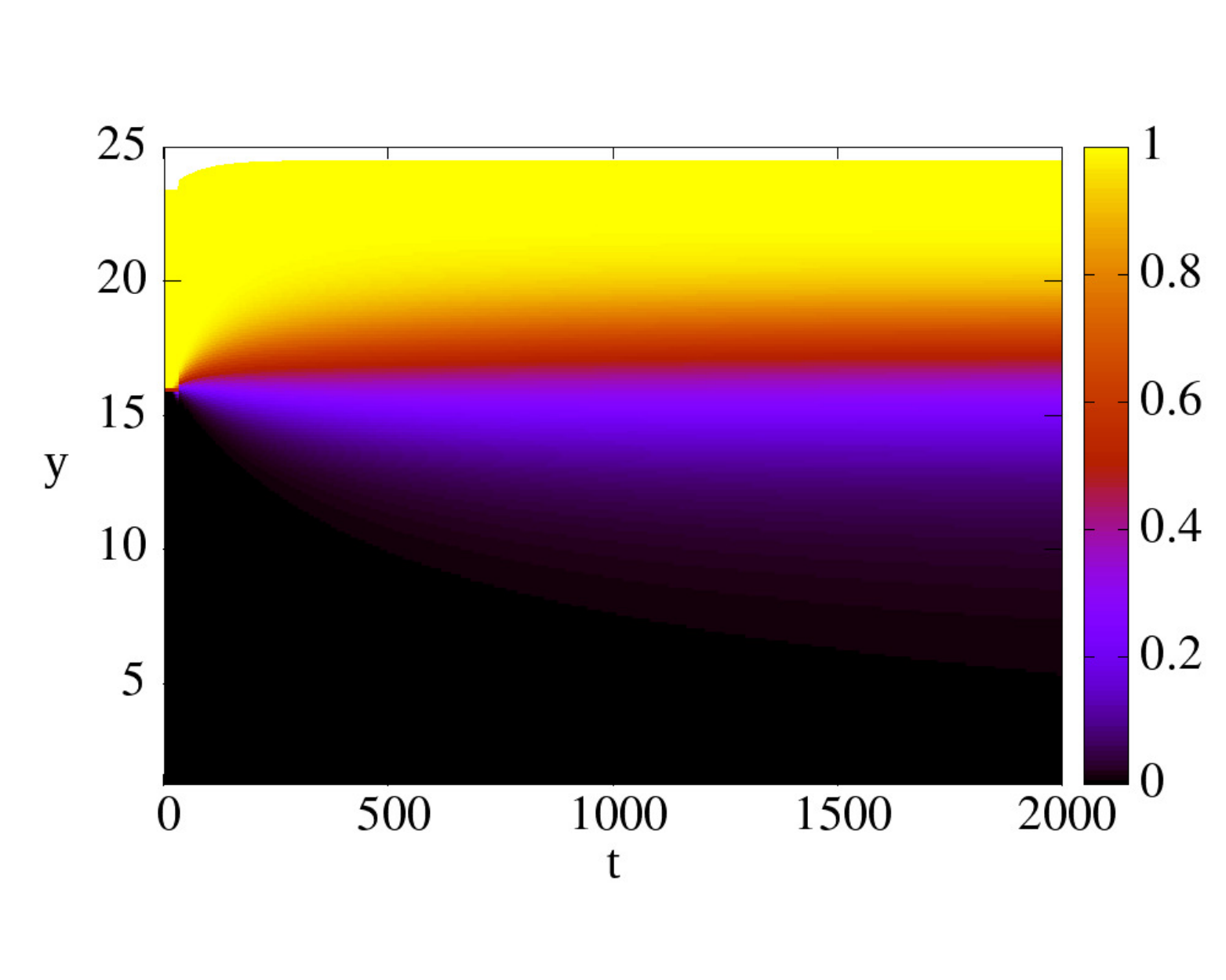}\put(-190,135){(f)} \put(-22,135){$f_L$}\put(-115,135){Continuum}\hfill 
    \caption{\textcolor{black}{(a) Time variation of species centre of mass ($y_{com}$) starting from initial concentration profile ($t = 0$) shown in inset. Variation of concentration of light species across the layer height with time for binary mixture having density ratio $\rho = 1.5$ starting from heavy-near-base initial configuration at an inclination angle $\theta = 25^\circ$ from (b) DEM simulations, and (c) Continuum model predictions. (d), (e) and (f) show the same data for ternary mixture with $\rho_H: \rho_M: \rho_L = 3:2:1$.}  %The corresponding profiles of the time evolution of $y_{com}$ using DEM (symbols), as well as the theory (solid lines) for both mixtures, are shown in (c) and (f) respectively. 
    }
\label{fig:HNB_bin_tern_diff_conf}
\end{figure}
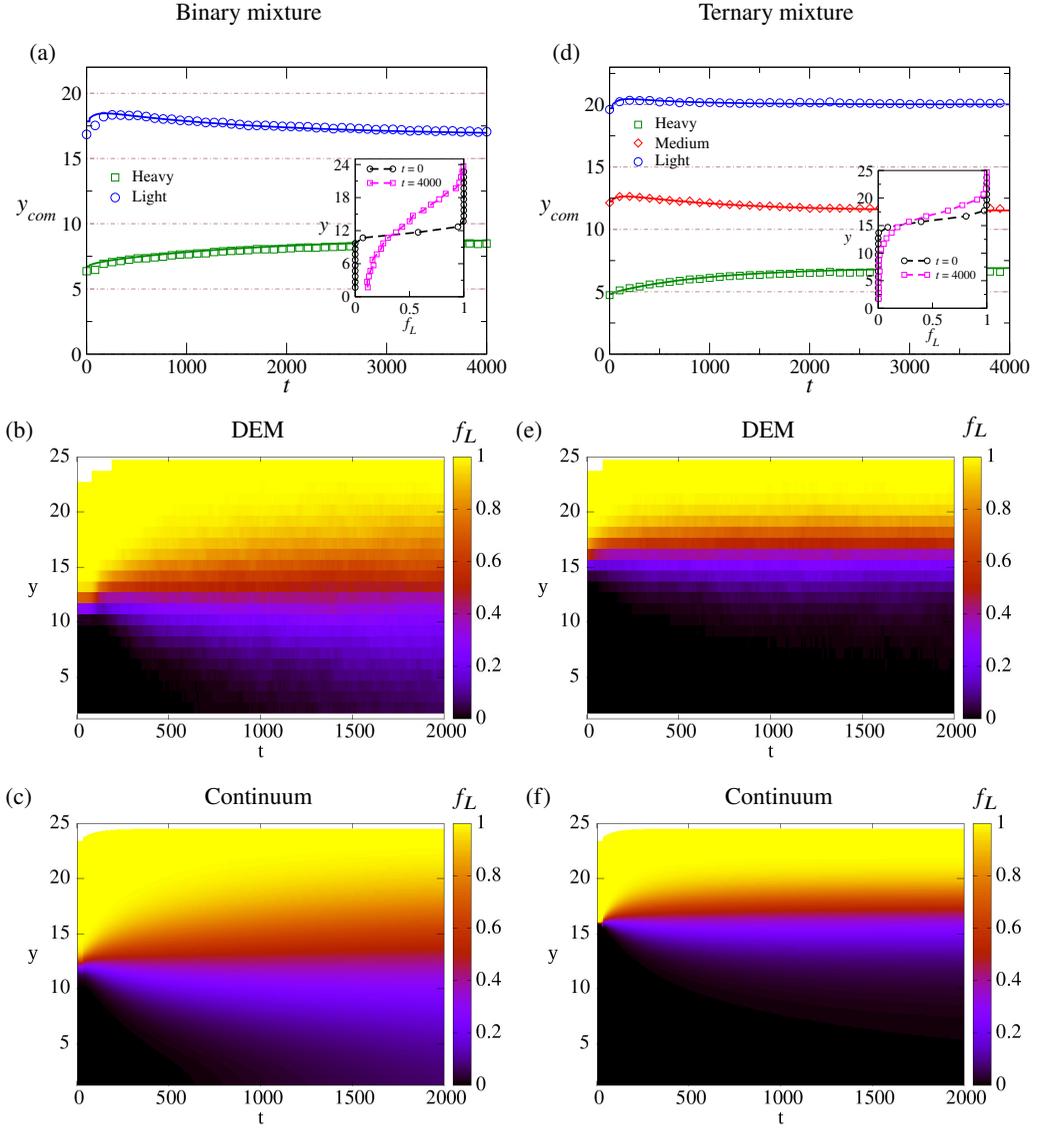

% \begin{figure}
%     \centering
%      \includegraphics[scale=0.22]{Diff_conf/Com-rho-1.5-fl-50-HNB_v1.pdf}%\put(-180,125){(a)} 
%     \includegraphics[scale=0.18]{Colormap/HNB-rho-1.5-fl-50-theta-25_DEM.png}%\put(-190,137){(a)} \put(-22,137){$f_L$}\put(-200,70){\rotatebox{90}{Binary mixture}}
%    % \hfill   
%    \includegraphics[scale=0.18]{Colormap/HNB-rho-1.5-fl-50-theta-25_th.png}%\put(-190,137){(b)} \put(-22,137){$f_L$}
%    %\hfill
%    \hfill
%    \includegraphics[scale=0.22]{Ternary_figs/rho-1-2-3-Ptcle-com-theta-25-H-M-L.pdf}%\put(-180,125){(b)}
%    \includegraphics[scale=0.18]{Colormap/Ternary_HML/H_M_L_Light_DEM.png}%\put(-190,137){(c)} \put(-22,137){$f_L$}\put(-200,70){\rotatebox{90}{Ternary mixture}}
%  %  \hfill 
%    \includegraphics[scale=0.18]{Colormap/Ternary_HML/H_M_L_Light_Theory.png}%\put(-190,137){(d)} \put(-22,137){$f_L$}
%   % \hfill 
%     \hfill
%     \caption{DEM snapshot of heavy-near-base initial configuration for equal composition (a) Binary mixture having density ratio ($\rho = 2.0$), and (b) Ternary mixture having density ratio ($\rho_H: \rho_M: \rho_L = 3:2:1$). Green, red, and blue spheres represent the heavy, medium, and light-density particles, respectively. Black spheres represent the static particles, used to form the rough bumpy base. The corresponding profiles of the time evolution of $y_{com}$ using DEM (symbols), as well as the theory (solid lines) for both cases, are shown in (b) and (d) respectively. Inclination angle $\theta = 25^o$. }
% \label{fig:HNB_bin_tern_diff_conf}
% \end{figure}

We first consider the case of a binary mixture with high density species located near the bottom half of the layer. 
Figure~\ref{fig:HNB_bin_tern_diff_conf}a shows the temporal variation of $y_{com}$ for both the species. The inset of figure ~\ref{fig:HNB_bin_tern_diff_conf}a shows the initial concentration profiles of light species at $t=0$ (black circles). Due to this nearly segregated \textcolor{black}{initial} configuration, the center of mass of the light and heavy species start from different values at $t=0$. While the heavy species center of mass exhibits slow rise from $y \simeq 6$ to $y \simeq 8$ over a duration of $4000$ time units, the light species center of mass shows a rising trend for a short time followed by slow decrease and eventually comes back to nearly the \textcolor{black}{same height at $t=4000$ due to mixing of two species (concentration profiles shown using magenta squares in inset of figure~\ref{fig:HNB_bin_tern_diff_conf}a).} 

Figures~\ref{fig:HNB_bin_tern_diff_conf}b and \ref{fig:HNB_bin_tern_diff_conf}c show the color map for the light species concentration ($f_L$) in binary mixture using DEM simulations and continuum model.
At early times, the upper half region is yellow ($f_L = 1$) and the lower half region is black ($f_L = 0$) due to the segregated initial configuration of the mixture with a very narrow region of red and purple. The width of the black as well as the yellow region keeps decreasing with time while that of the red and purple region keeps growing because of the mixing of the grains due to diffusion. 
This diffusional mixing observed in the concentration color maps obtained from the DEM simulation (figure~\ref{fig:HNB_bin_tern_diff_conf}b) is very well captured by the \textcolor{black}{continuum model} as well and two concentration color map (shown in figure~\ref{fig:HNB_bin_tern_diff_conf}b and \ref{fig:HNB_bin_tern_diff_conf}c) are found to be nearly identical to each other. The concentration using DEM data in figure~\ref{fig:HNB_bin_tern_diff_conf}b shows more discrete variations compared to continuum model because of the large grid size \textcolor{black}{used in case of DEM}. 

Figures~\ref{fig:HNB_bin_tern_diff_conf}d-\ref{fig:HNB_bin_tern_diff_conf}f show the results  
%variation of the light species concentration ($f_L$) across the space-time plane using color maps 
for an equal composition ternary mixture having density ratios $\rho_H: \rho_M: \rho_L = 3:2:1$ and flowing at $\theta = 25^\circ$.
The flow starts from a heavy-near-base configuration in which high density particles are kept near the bottom and light density particles are kept near the surface with the medium density particles sandwiched in between the two. Time variation of $y_{com}$ for all the three species are shown in figure~\ref{fig:HNB_bin_tern_diff_conf}d. Symbols represent the DEM data and solid lines represent continuum predictions. A very good match between two is observed. 
Figures~\ref{fig:HNB_bin_tern_diff_conf}e and \ref{fig:HNB_bin_tern_diff_conf}f show concentration color maps of the light species  obtained from the DEM and continuum simulations, respectively.
Since the initial concentration of light species is present only near the upper one third of the layer in this case, the yellow region with $f_L = 1$ is observed near the top portion of the layer at early times along with narrow regions of red and purple (corresponding to the linear part of the concentration data shown in the inset of figure~\ref{fig:HNB_bin_tern_diff_conf}d). As time increases, the span of the yellow region across the layer appears to decrease while that of the red and purple regions increase due to the diffusional mixing. Again the agreement between DEM simulations and continuum predictions in figure~\ref{fig:HNB_bin_tern_diff_conf}e and \ref{fig:HNB_bin_tern_diff_conf}f is found to be pretty good. Similar results are observed for heavy $f_H$ and medium species $f_M$ (not shown here) as well.
%The color scheme of green, red, and blue signifies the species with high, medium, and low densities, respectively. A similar behavior is also observed in ternary mixtures.  
\begin{figure}
    \centering
    \includegraphics[scale=0.15]{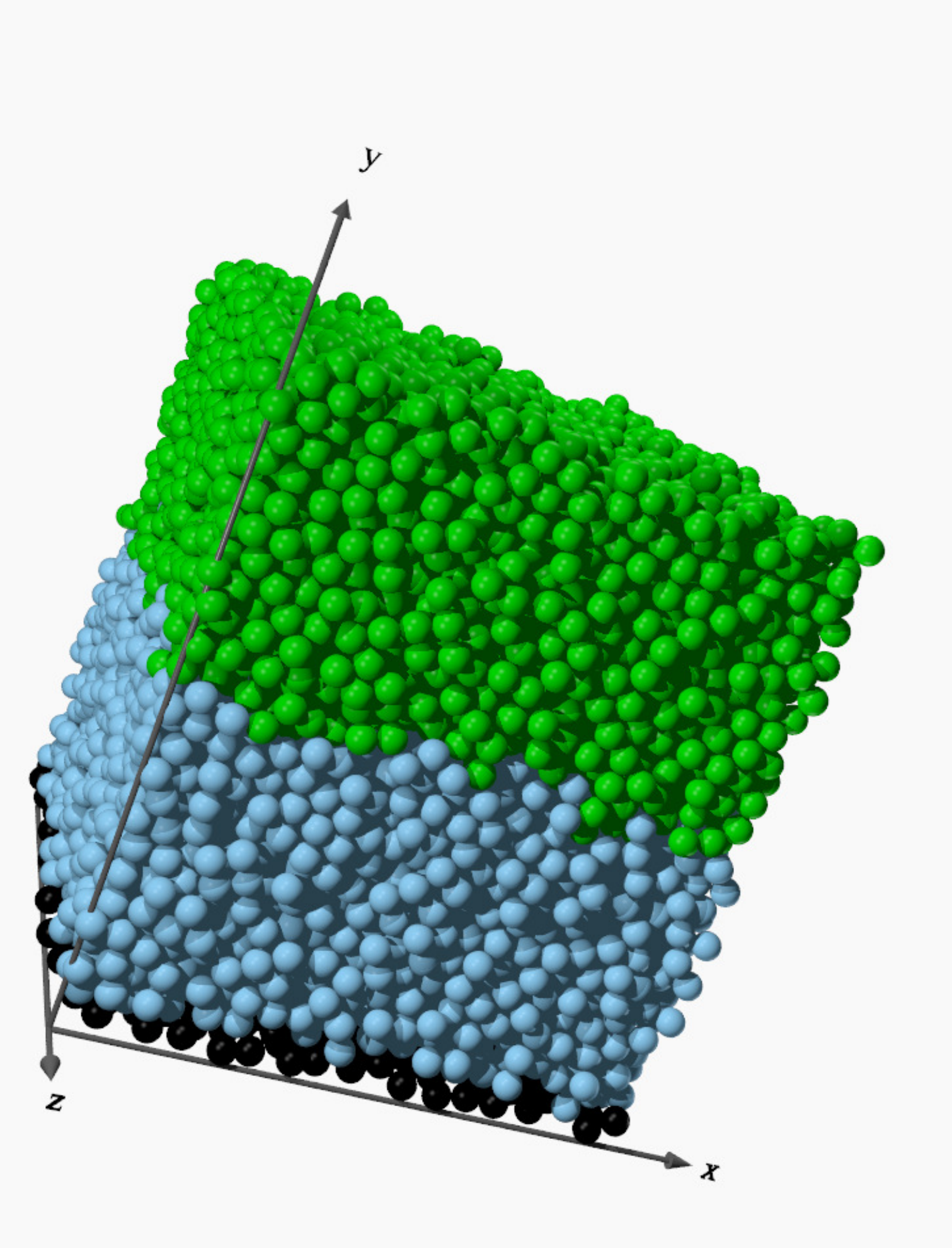}\put(-125,140){(a)} \quad \quad  
    \includegraphics[scale=0.4]{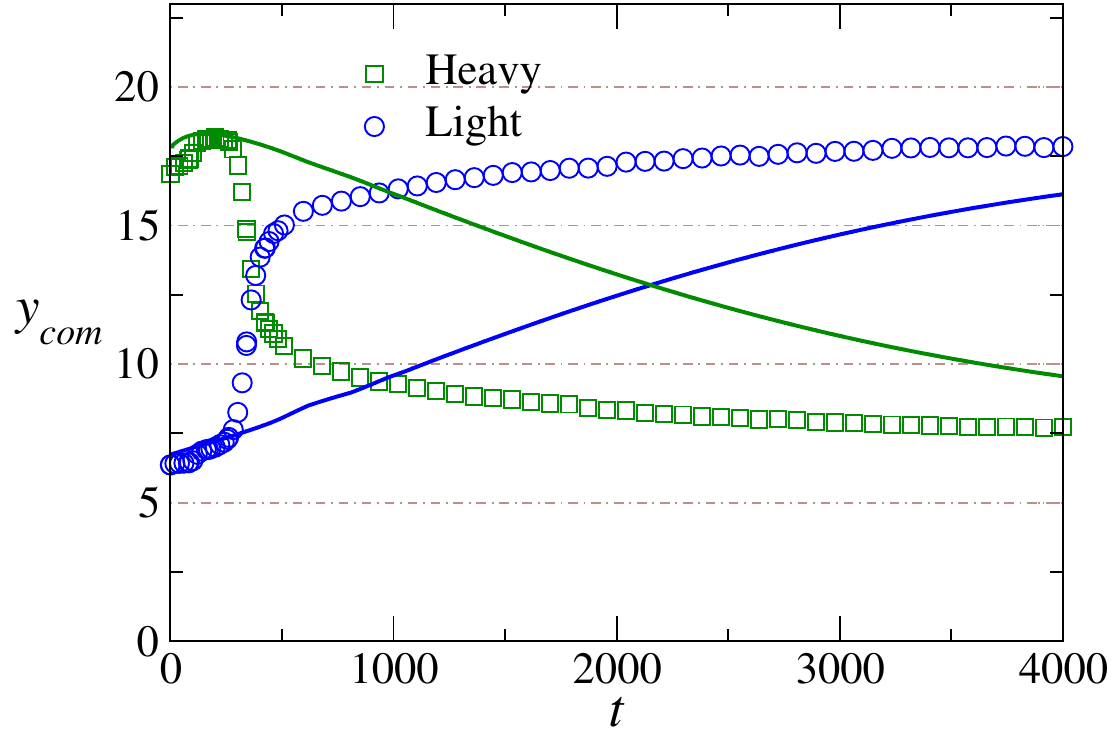}\put(-200,140){(b)}\hfill
  %   \vspace{0.5cm}
    \includegraphics[scale=0.25]{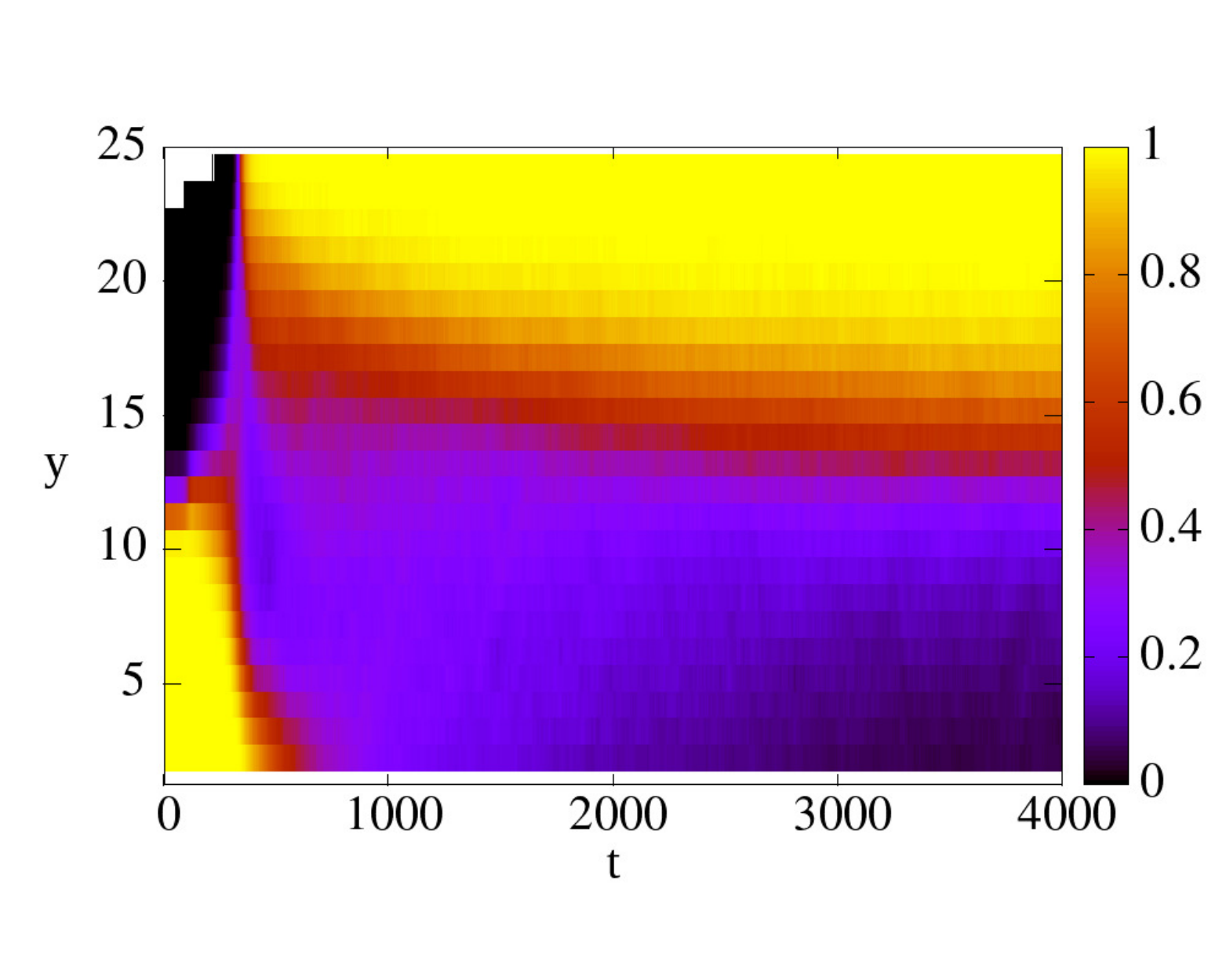}\put(-190,135){(c)}\put(-22,135){$f_L$}\put(-105,135){DEM}
    \includegraphics[scale=0.25]{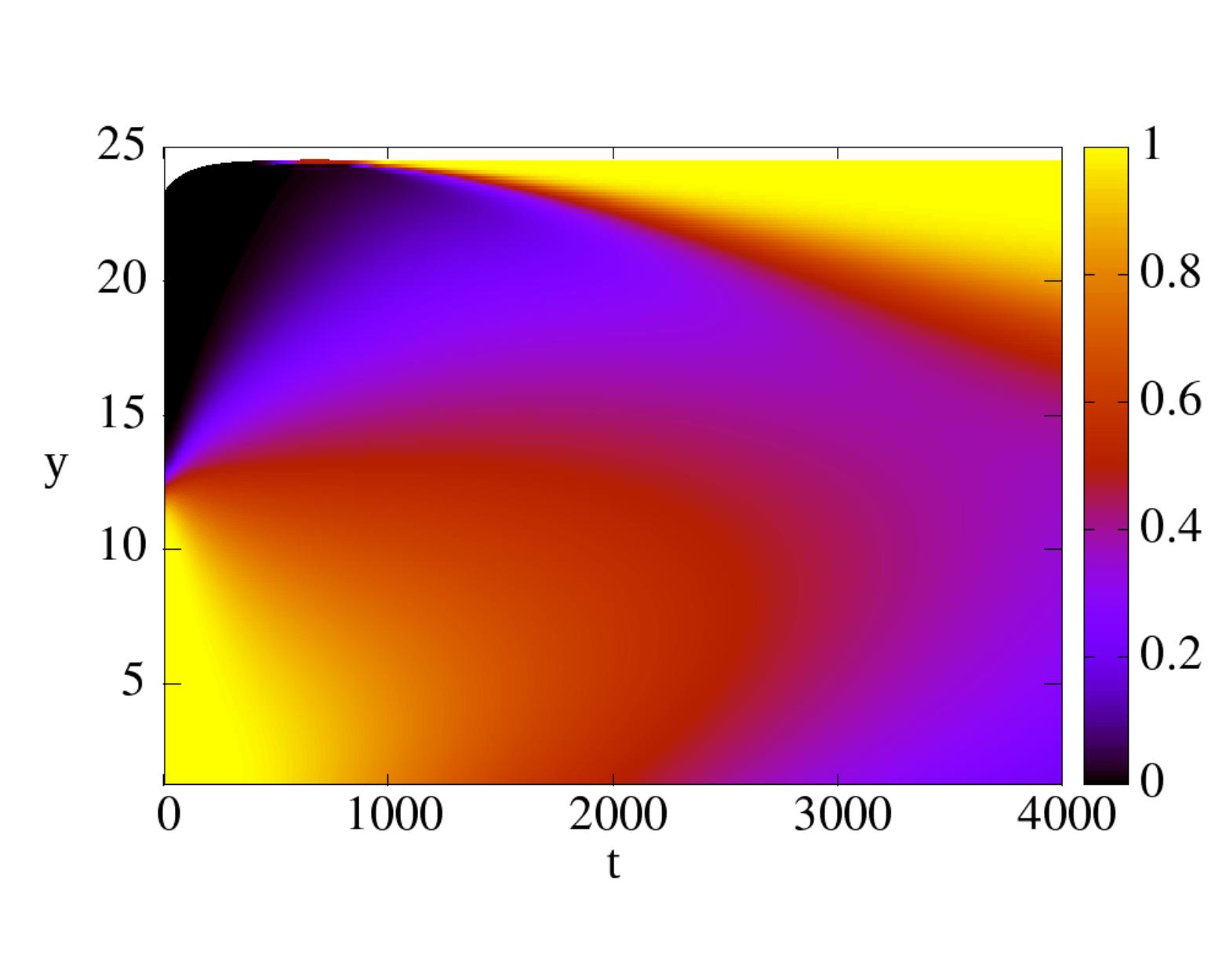}\put(-190,135){(d)}\put(-22,135){$f_L$}\put(-115,135){Continuum}\hfill
    \caption{(a) DEM snapshot for equal composition binary mixture of density ratio ($\rho = 2.0$) for light-near-base configuration. Green spheres represent the heavy-density particles while blue spheres represent the light-density particles. (b) Time evolution of $y_{com}$ for both light (blue) and heavy (green) species using DEM data (symbol) and model predictions (solid lines). Color map of light species concentration in the flowing layer over time range using (c) DEM, and (d) \textcolor{black}{Continuum model} at inclination angle $\theta = 25^o$. }
\label{fig:LNB_binary_diff_conf}
\end{figure}
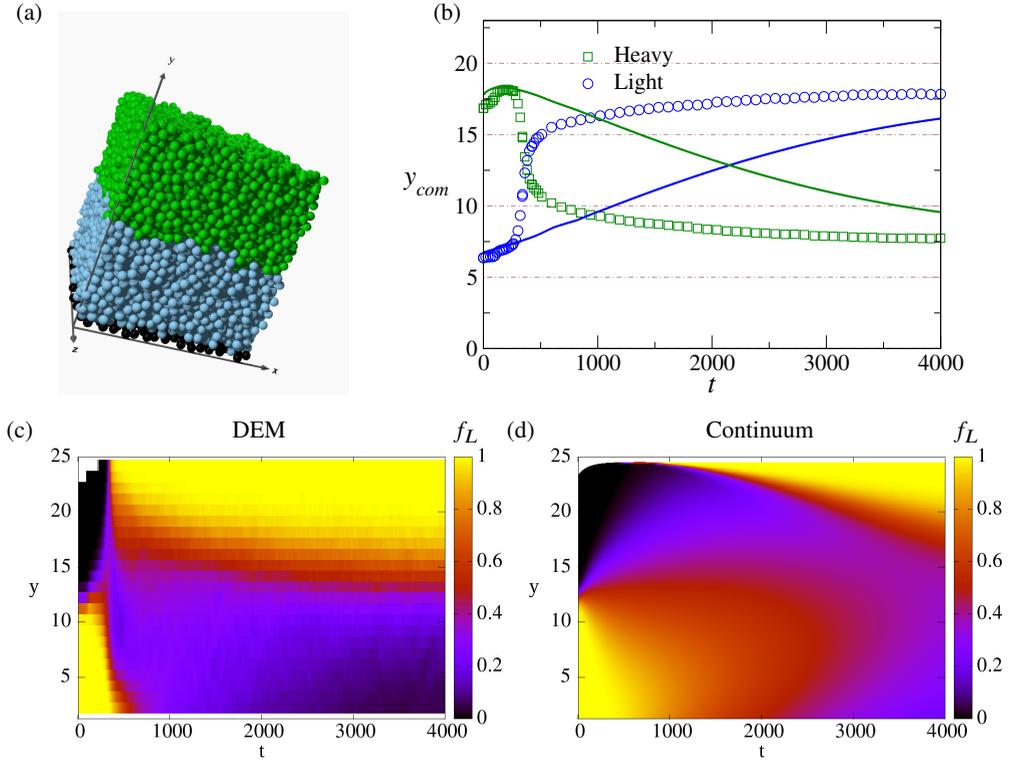

\subsection{Light-Near-Base configuration}

We next report the results for the cases where the heavy species is near the free surface and the light species is near the base at the starting of the simulation. Figure~\ref{fig:LNB_binary_diff_conf}a shows the DEM snapshot of the initial configuration for an equal composition binary mixture having density ratio $\rho = 2.0$. The flow occurs at $\theta = 25^\circ$
% starting from the light-near-base initial configuration where light particles are concentrated near the bottom.
%The corresponding DEM snapshot is shown in figure~\ref{fig:LNB_binary_diff_conf}a. Green spheres represent the heavy-density particles while blue spheres represent the light-density particles. Black spheres represent the static particles, used to form the rough bumpy base. 
and the variation of $y_{com}$ of both light and heavy species with time is shown in figure~\ref{fig:LNB_binary_diff_conf}b. The center of masses of both the species rise for a brief initial period (up to $250$ time units) due to the layer dilation. 
After that a sharp rise in the $y_{com}$ of light species (blue circles) and sharp decline in the $y_{com}$ of heavy species (green squares) is observed from the DEM simulations (symbols) during the next $250$ time units. After $t = 500$ time units, $y_{com}$ of both the species change gradually. Contrary to all other cases considered before, the predicted center of mass from the continuum model (shown using solid lines) deviate significantly from the DEM data. It is evident that the model fails to capture the center of mass behavior observed in the case of DEM simulations. 
%do not capture this beshow the gradual change over time for this case and fail to match the DEM data.  
%For the detailed investigation of this case, we show 

\begin{figure}
    \centering
      \includegraphics[scale=0.14]{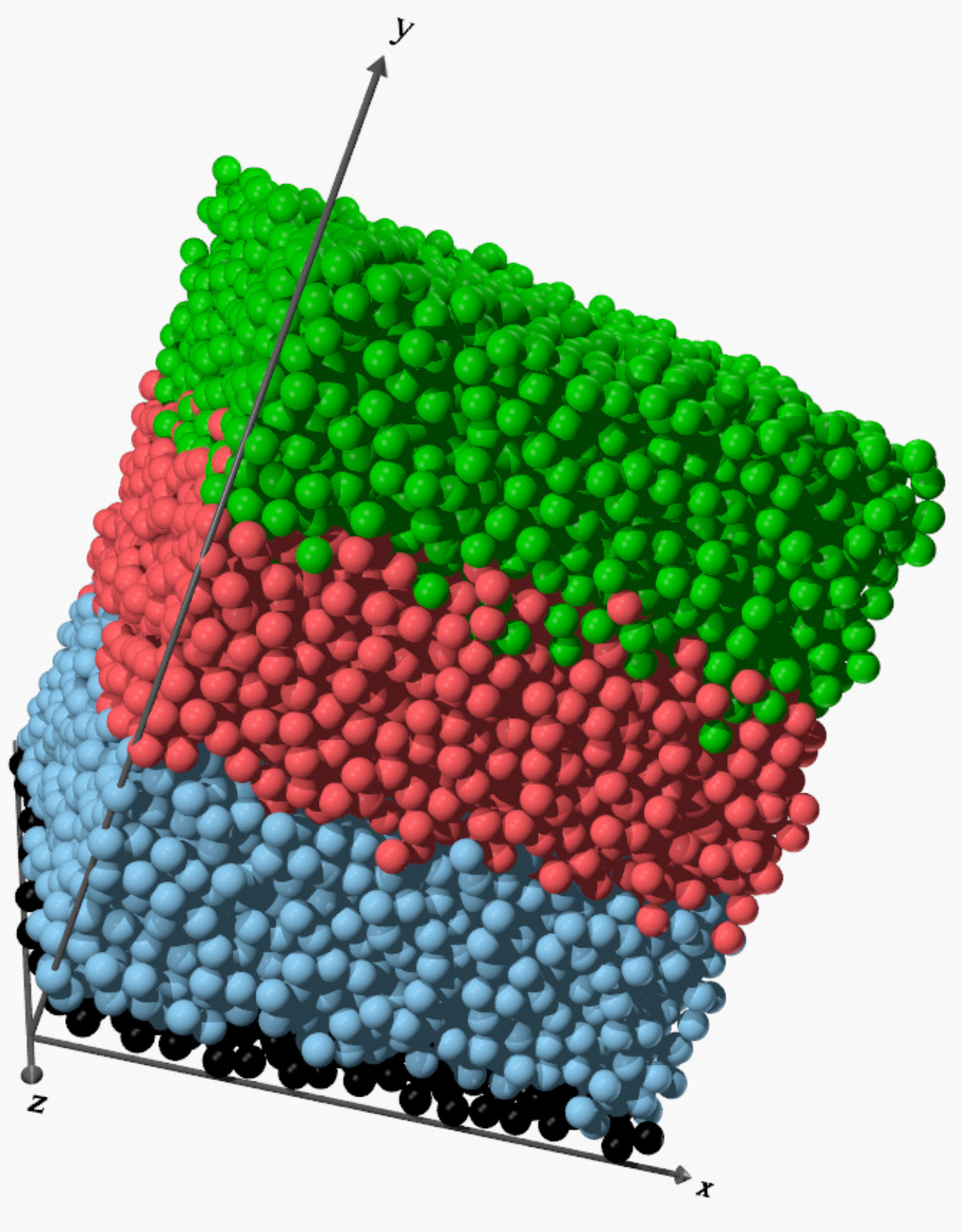}\put(-120,120){(a)} \quad \quad \quad 
     \includegraphics[scale=0.35]{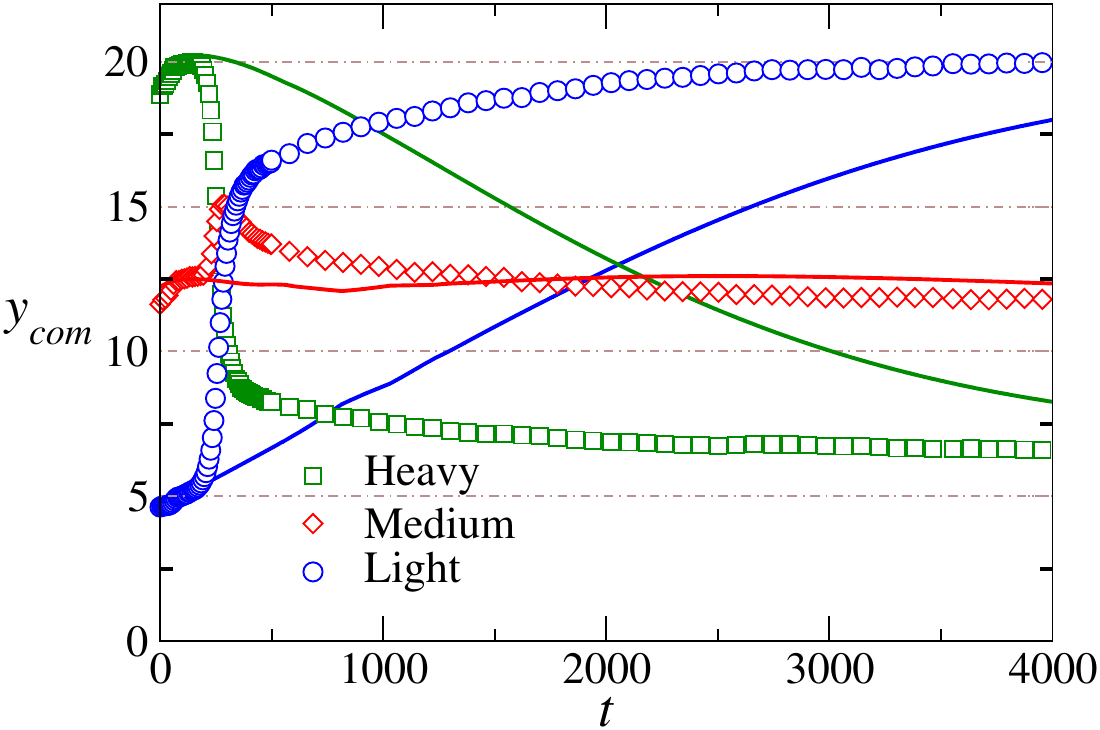}\put(-190,120){(b)} \quad \quad \quad
 \includegraphics[scale=0.25]{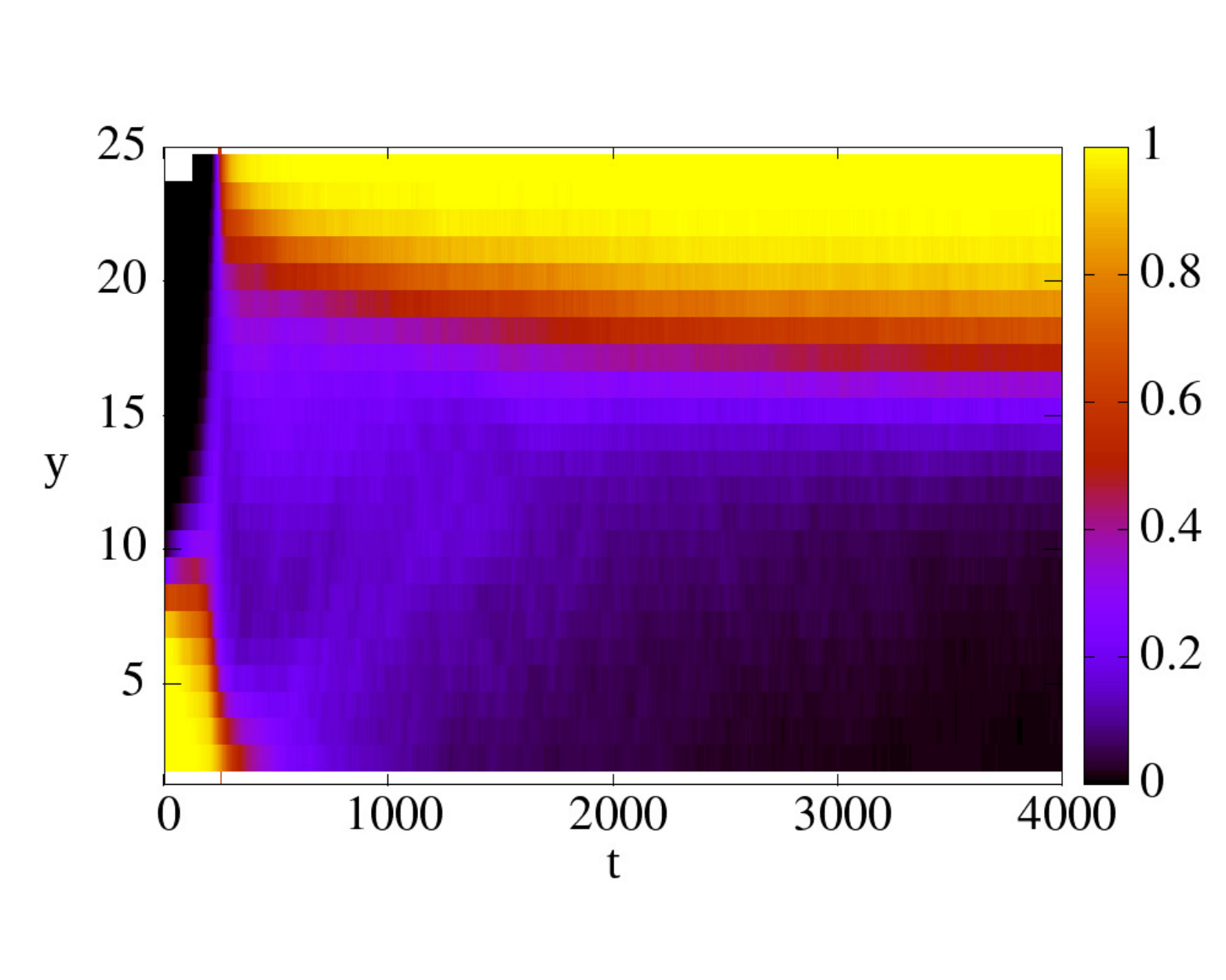}\put(-190,135){(c)}\put(-22,135){$f_L$}\put(-105,135){DEM}\hfill
 \includegraphics[scale=0.25]{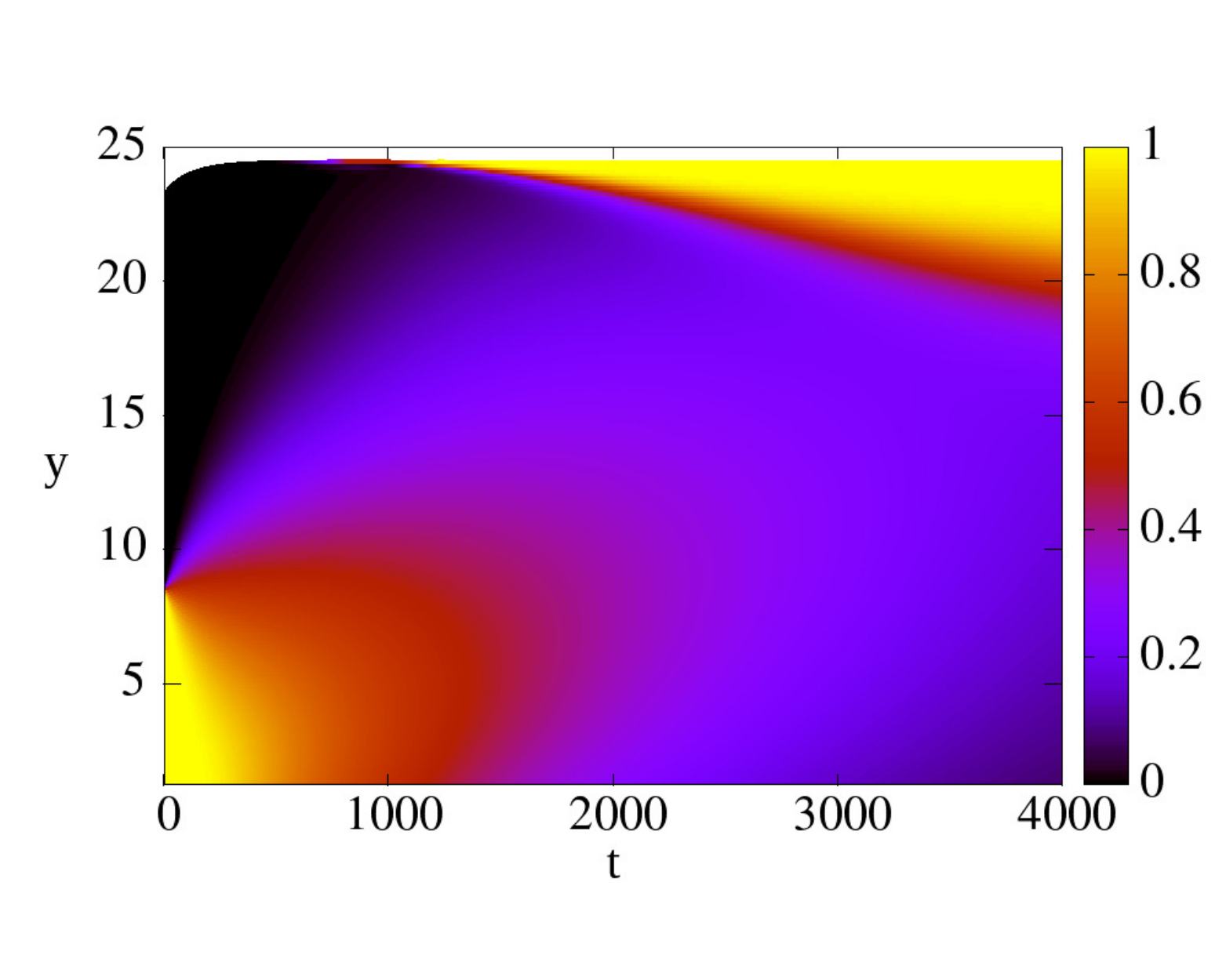}\put(-190,135){(d)}\put(-22,135){$f_L$}\put(-115,135){Continuum}\hfill
 %\vspace{-0.5cm}
 %\includegraphics[scale=0.24]{Colormap/Ternary_v1/LMH-DEM-Medium_conc.png}\put(-185,125){(d)}\put(-22,130){$f_M$}\hfill
% \includegraphics[scale=0.24]{Colormap/Ternary_v1/LMH-Theory-Medium_conc.png}\put(-185,125){(g)}\put(-22,130){$f_M$}\hfill
 \vspace{-0.5cm}
  \includegraphics[scale=0.25]{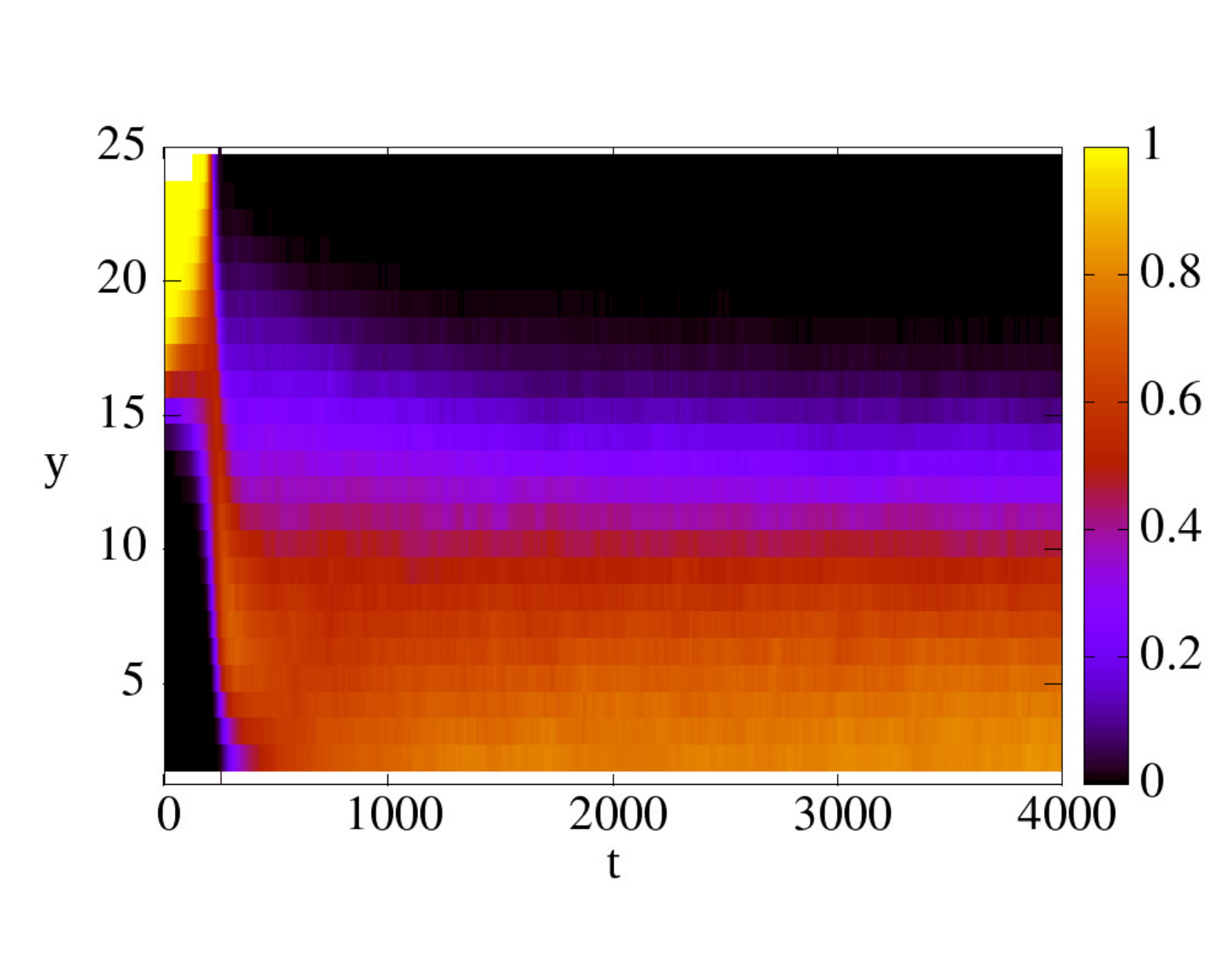}\put(-190,135){(e)}\put(-22,135){$f_H$}\put(-105,135){DEM}\hfill
 \includegraphics[scale=0.25]{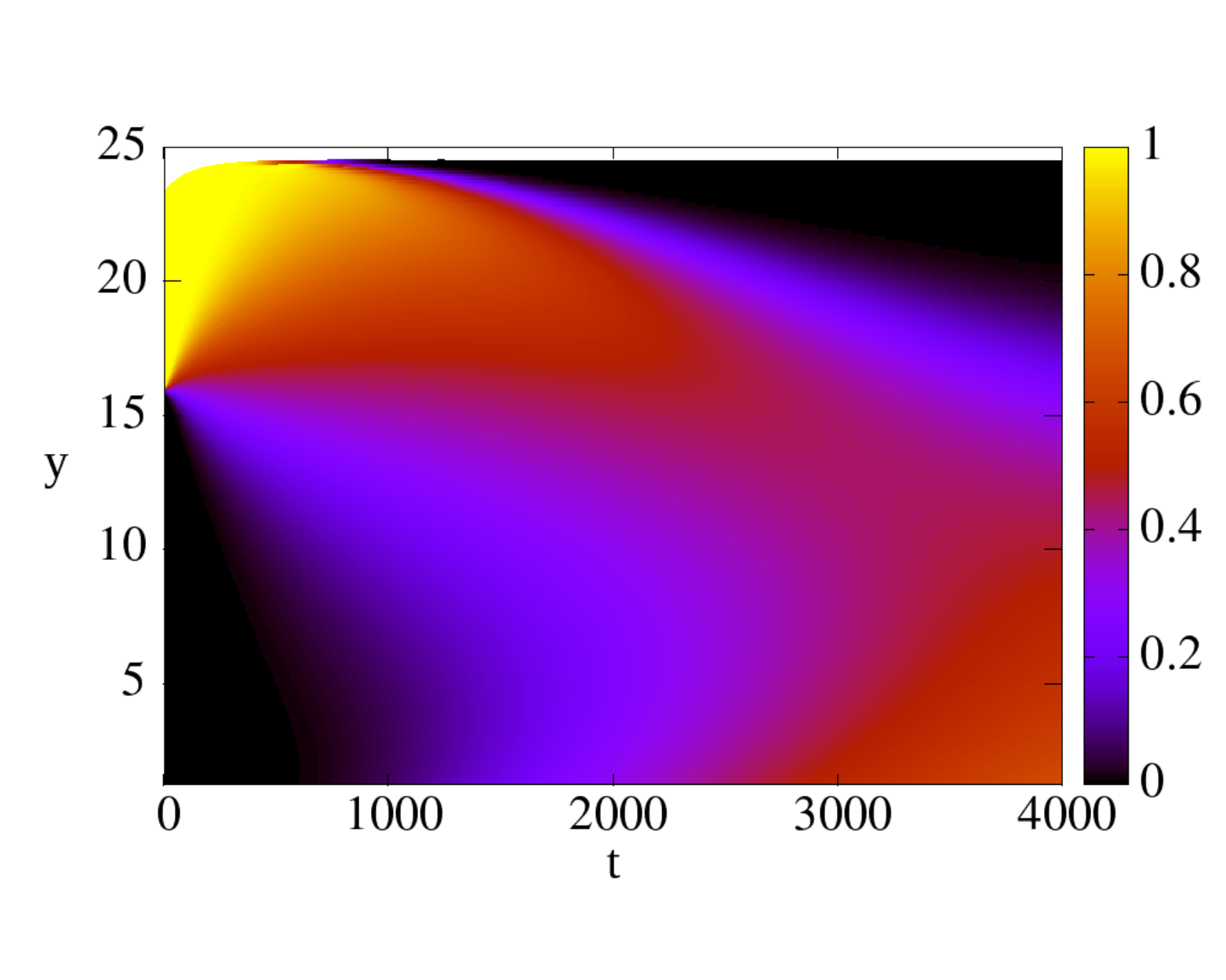}\put(-190,135){(f)}\put(-22,135){$f_H$}\put(-115,135){Continuum}\hfill
 \vspace{-0.3cm}
    \caption{DEM snapshot of ternary mixture having density ratio $\rho_H: \rho_M: \rho_L = 3:2:1$ for light-near-base initial configuration at inclination angle $\theta = 25^\circ$. Green, red, and blue particles represent the heavy, medium, and light-density particles, respectively. Black particles represent the rough bumpy base of thickness $1.2d$. (b) Variation of $y_{com}$ with time for all species. Color map for the concentration of light species using (c) DEM, and (d) Continuum model. \textcolor{black}{(e) and (f) show the same for heavy species.}}
\label{fig:LNB_ternary_diff_conf}
\end{figure}
The concentration color map of the light species, as obtained from the DEM simulations, is shown in figure~\ref{fig:LNB_binary_diff_conf}c. The predictions from the continuum model are shown in figure~\ref{fig:LNB_binary_diff_conf}d. 
Evidently, the predicted behavior from the model is qualitatively different from that observed in the DEM simulations. The concentration color map from the DEM simulations (figure~\ref{fig:LNB_binary_diff_conf}c) show that the yellow region (having pure light species) in the lower part of the layer disappears around $500$ time units and very quickly appears near the top part of the layer. In addition, the concentration of heavy species in the lower part of the layer remains high (purple region) for all times larger than $750$ time units except for a very brief duration, (shown by red colored region in lower part of the layer) where both the species appear to be well-mixed. 
The theoretical predictions in figure~\ref{fig:LNB_binary_diff_conf}d, however, show a gradual decrease in the light species concentration in the lower part of the layer and high concentration of heavy species (purple region) is observable only after $3000$ time units. 
Similar qualitative differences are observable in the upper part of the layer as well.

%Theoretical predictions perfectly match the DEM data for both binary and ternary mixtures after the instability disappears.
% \begin{figure}
%     \centering
%     \includegraphics[scale=0.24]{Colormap/Ternary_LMH/LMH_rho_mix_1_2_3_DEM.png}\hfill
%     \includegraphics[scale=0.24]{Colormap/Ternary_LMH/LMH_rhomix_1_2_3_theory.png}\hfill
%     \caption{$\rho_{mix}$ for ternary mixture}
%     \label{fig:enter-label}
% \end{figure}
We further investigate this configuration for a ternary mixture, in which particles with low density are concentrated near the base, medium density particles are positioned in the middle, and particles with high density are located at the top. Figure~\ref{fig:LNB_ternary_diff_conf}a shows the DEM snapshot of such an initial configuration where green, red, and blue spheres represent the high, medium, and low density particles respectively and black spheres represent the static particles that form the rough, bumpy base. Figure~\ref{fig:LNB_ternary_diff_conf}b shows the variation of $y_{com}$ for all three species in the mixture. As before, symbols represent the DEM data and lines correspond to predictions obtained from continuum model. 
As in the case of binary mixture, DEM data show a very quick change in the center of mass of heavy (green) and light (blue) species while the theoretical lines show a gradual variation. Evidently, the quick
segregation observed in the initial $500$ time units is not captured by the model. Visible qualitative differences are observable in the concentration color maps of the light species obtained from the DEM simulations (figure~\ref{fig:LNB_ternary_diff_conf}c) and the continuum model (figure~\ref{fig:LNB_ternary_diff_conf}d). Figures~\ref{fig:LNB_ternary_diff_conf}e and \ref{fig:LNB_ternary_diff_conf}f show the concentration color map for the heavy species and notable qualitative differences are observable between the model and the DEM simulations. Similar differences are also observed in the concentration color maps of medium density species as well (not shown here).

\section{Rayleigh-Taylor instability in granular mixture flows}
\label{sec:RT_instability}
\begin{figure}
    \centering
\includegraphics[scale=0.14]{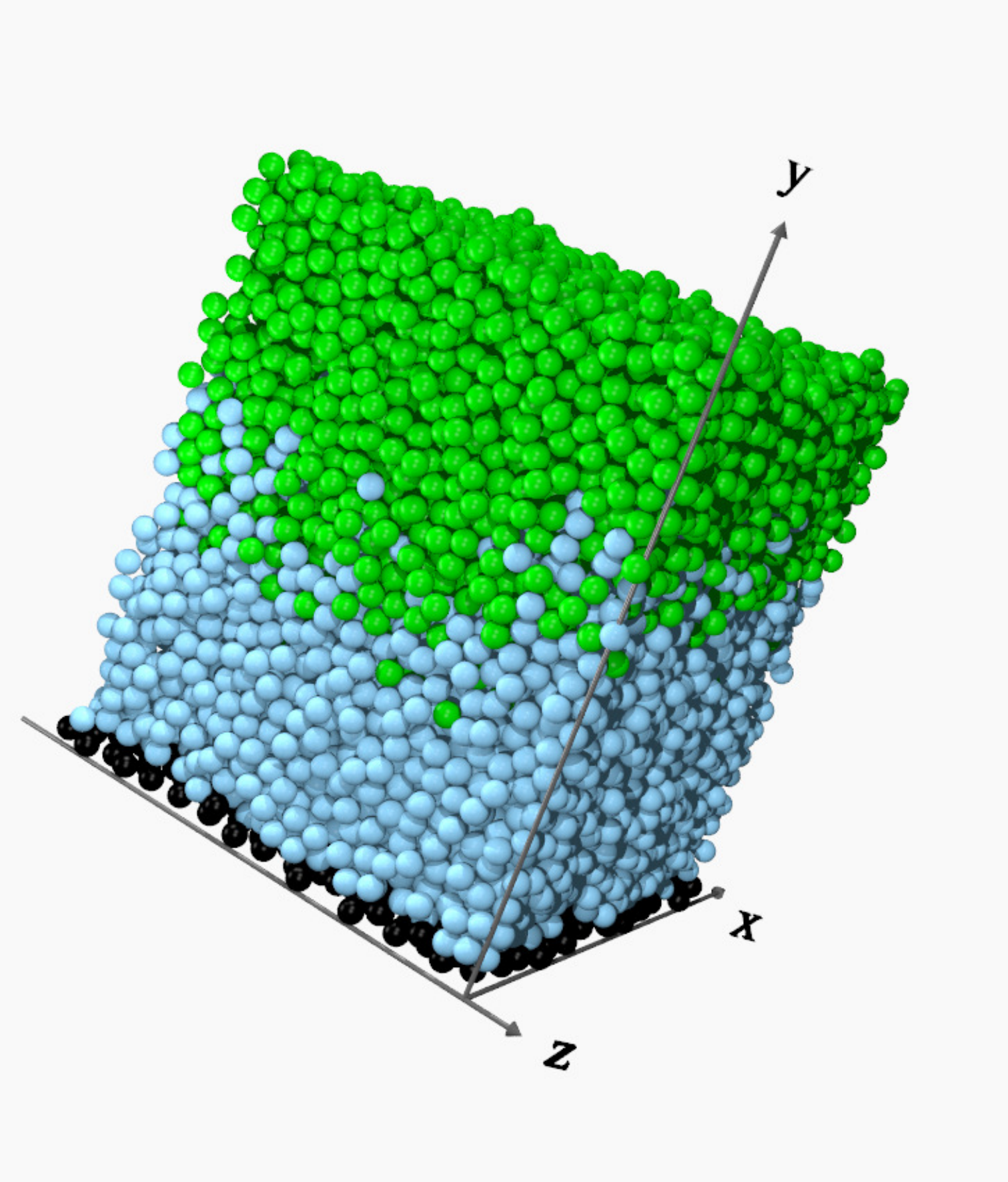}\put(-95,107){(b)}\put(-93,10){$t = 250$}\hfill
\includegraphics[scale=0.31]{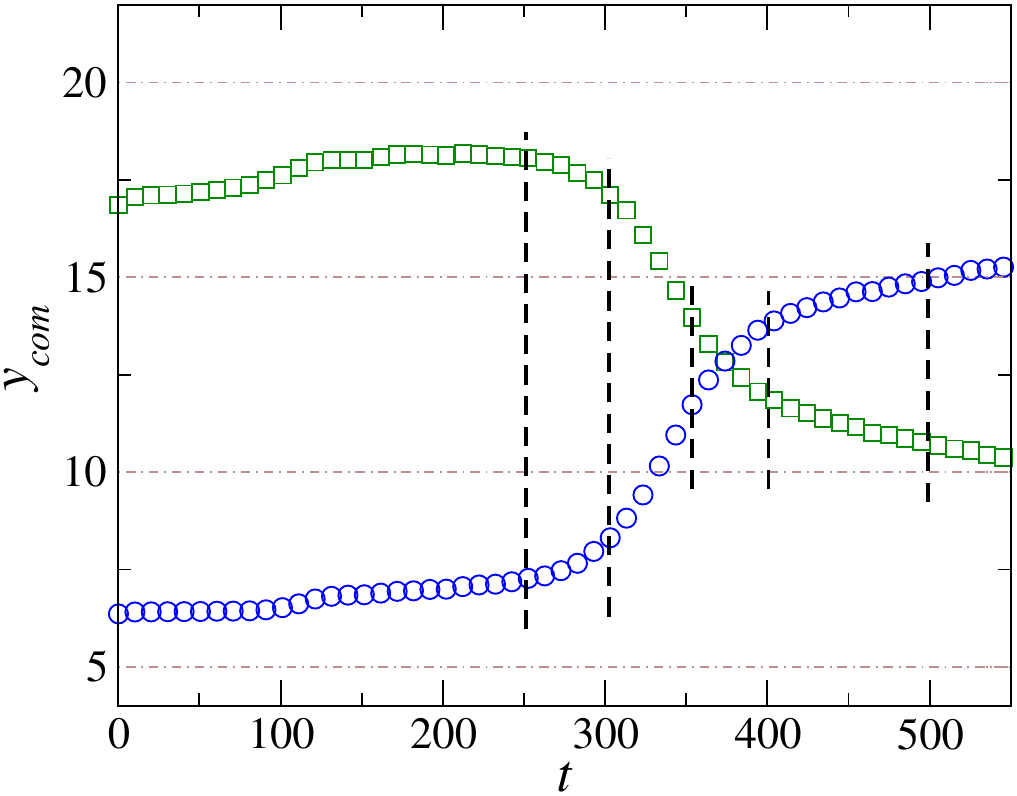}\put(-130,108){(a)}\put(-74,55){\vector(-1,0){90}}\put(-62,51){\vector(-2,-1){130}}\put(-48.5,45){\vector(0,-1){60}}\put(-38,52){\vector(1.1,-1){70}}\hfill \quad  \quad
%\put(-56,45){\vector(-1,-2){28}}
\includegraphics[scale=0.14]{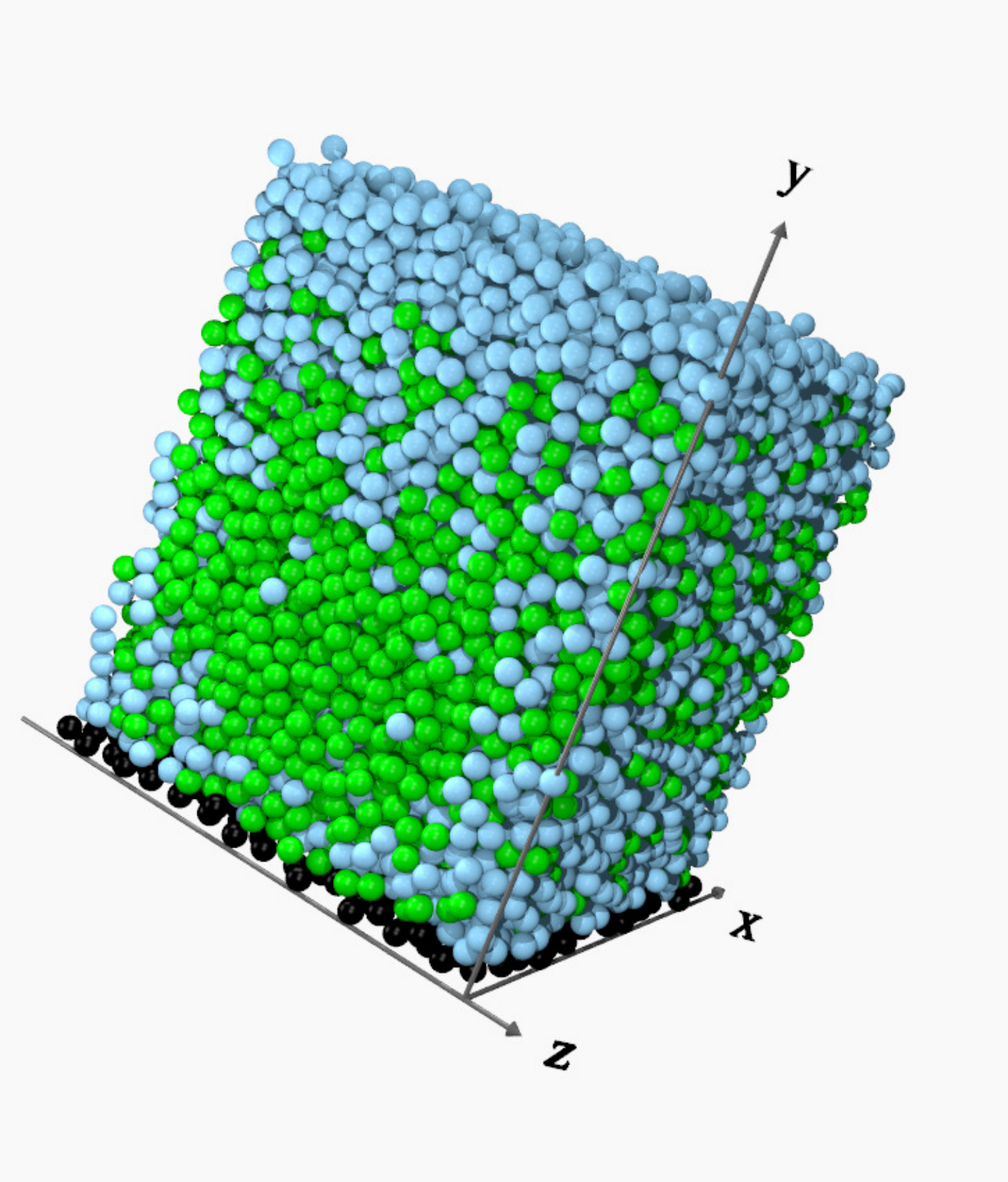}\put(-95,107){(f)}\put(-93,10){$t = 500$}\put(-140,65){\vector(1,0){50}}\hfill
%\hspace{-3cm}
\includegraphics[scale=0.14]{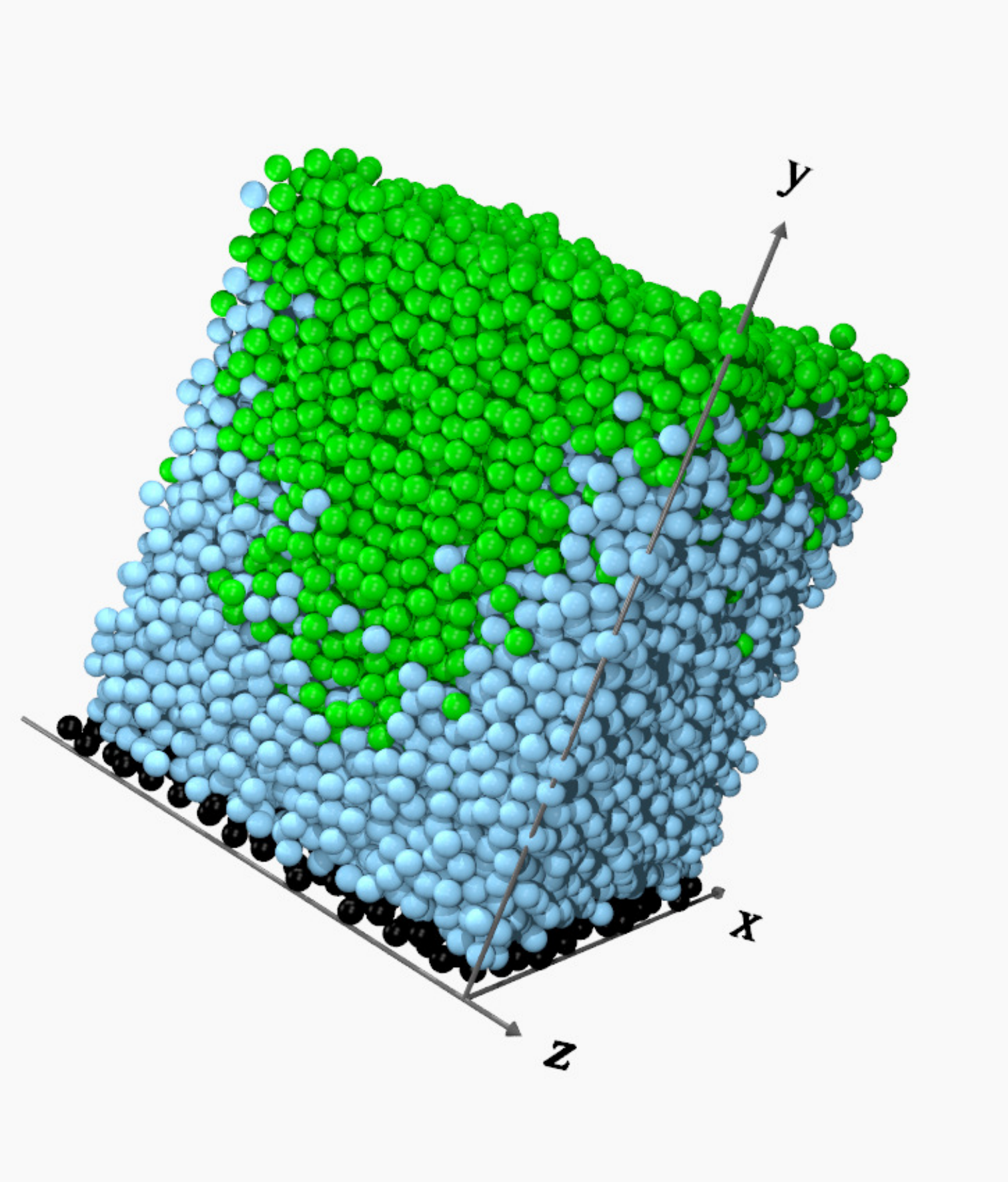}\put(-95,107){(c)}\put(-93,10){$t = 300$} \quad \quad \quad \quad
\includegraphics[scale=0.14]{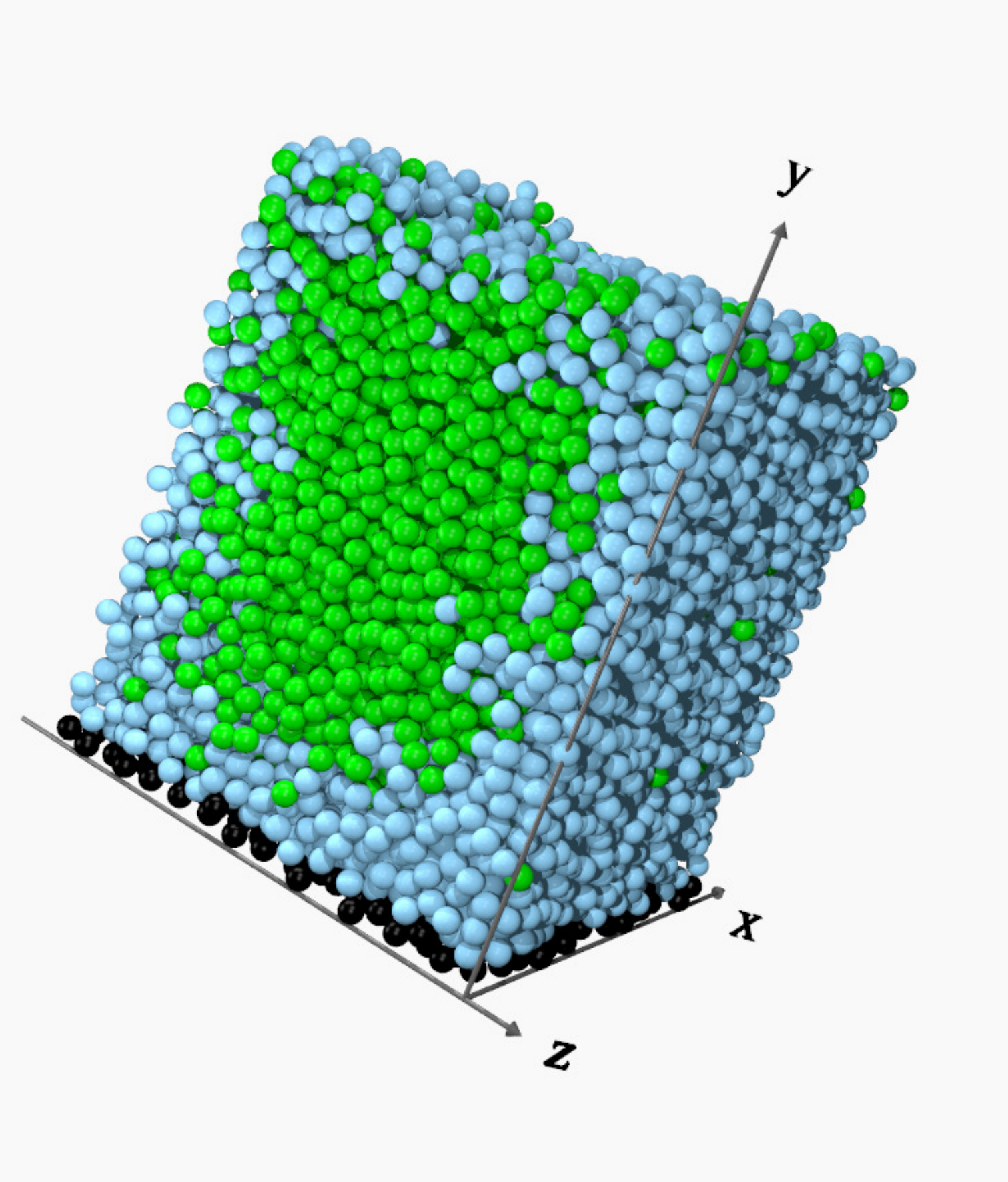}\put(-95,107){(d)}\put(-93,10){$t = 350$} \quad \quad \quad \quad
\includegraphics[scale=0.14]{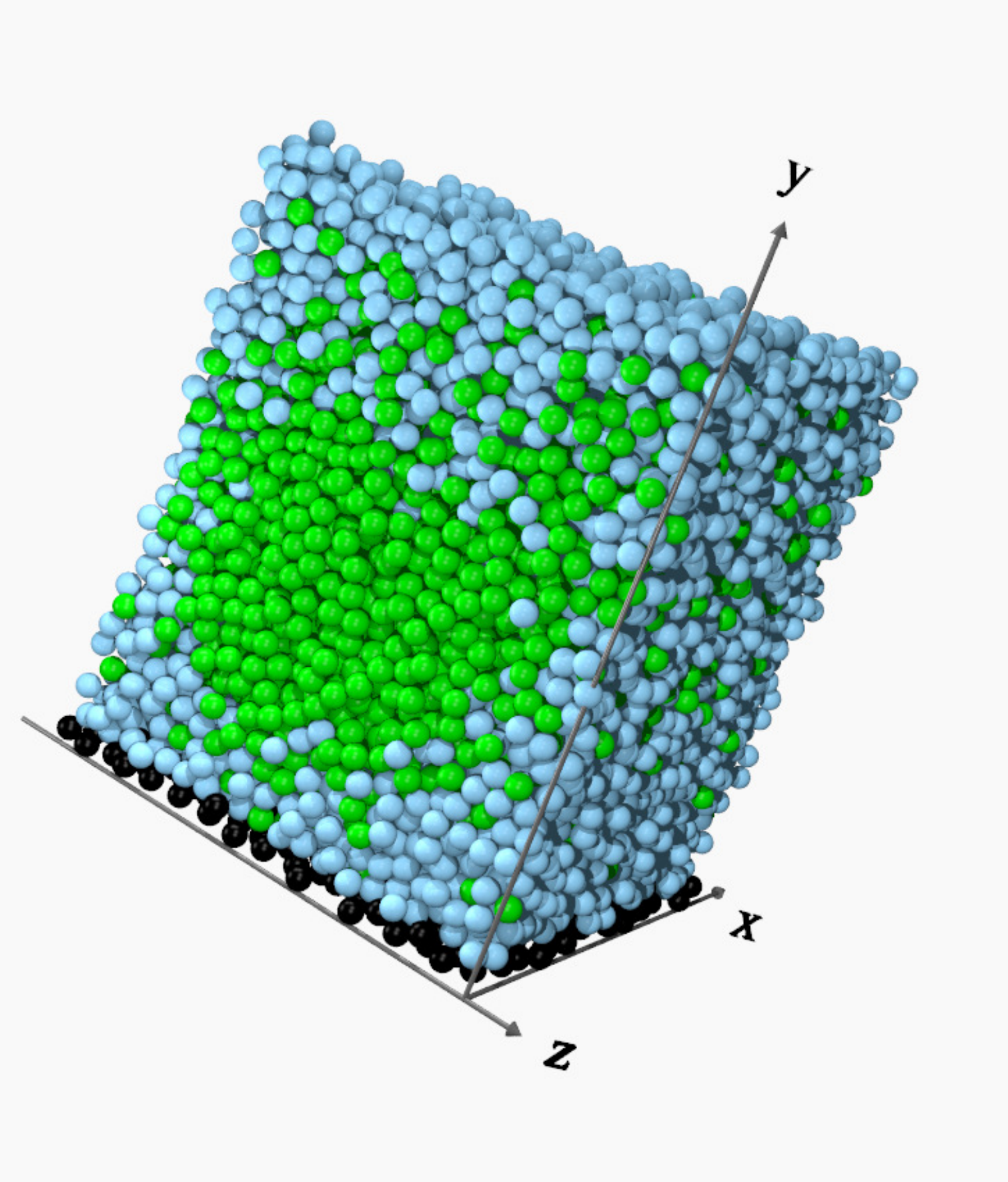}\put(-95,107){(e)}\put(-93,10){$t = 400$}
\caption{(a) Variation of $y_{com}$ with time for binary mixture at early times in case of light-near-base initial configuration. DEM simulation snapshots for equal composition binary mixture of density ratio $\rho = 2.0$ at different times (b) $t = 250$, (c) $t = 300$, (d) $t = 350$, (e) $t = 400$, and (f) $t = 500$ staring from light-near-base initial configuration at inclination angle $\theta = 25^o$.}
    \label{fig:binary_LNB_snap}
\end{figure}

The results shown in figure~\ref{fig:LNB_binary_diff_conf} and \ref{fig:LNB_ternary_diff_conf} confirm that the model is unable to capture the evolution of segregation for the cases when the heavy species is situated at the top of the layer. The sudden changes in the species center of mass as well as the light and heavy species concentration color maps observed in the DEM simulations indicate that the mechanism of segregation operative in this case is not captured using our one-dimensional continuum model. 
In order to identify the reason for the quick segregation observed in these cases, \textcolor{black}{we take a more careful look at our DEM simulation results in figure~\ref{fig:binary_LNB_snap}}. 

\begin{figure}
\centering
\includegraphics[scale=0.14]{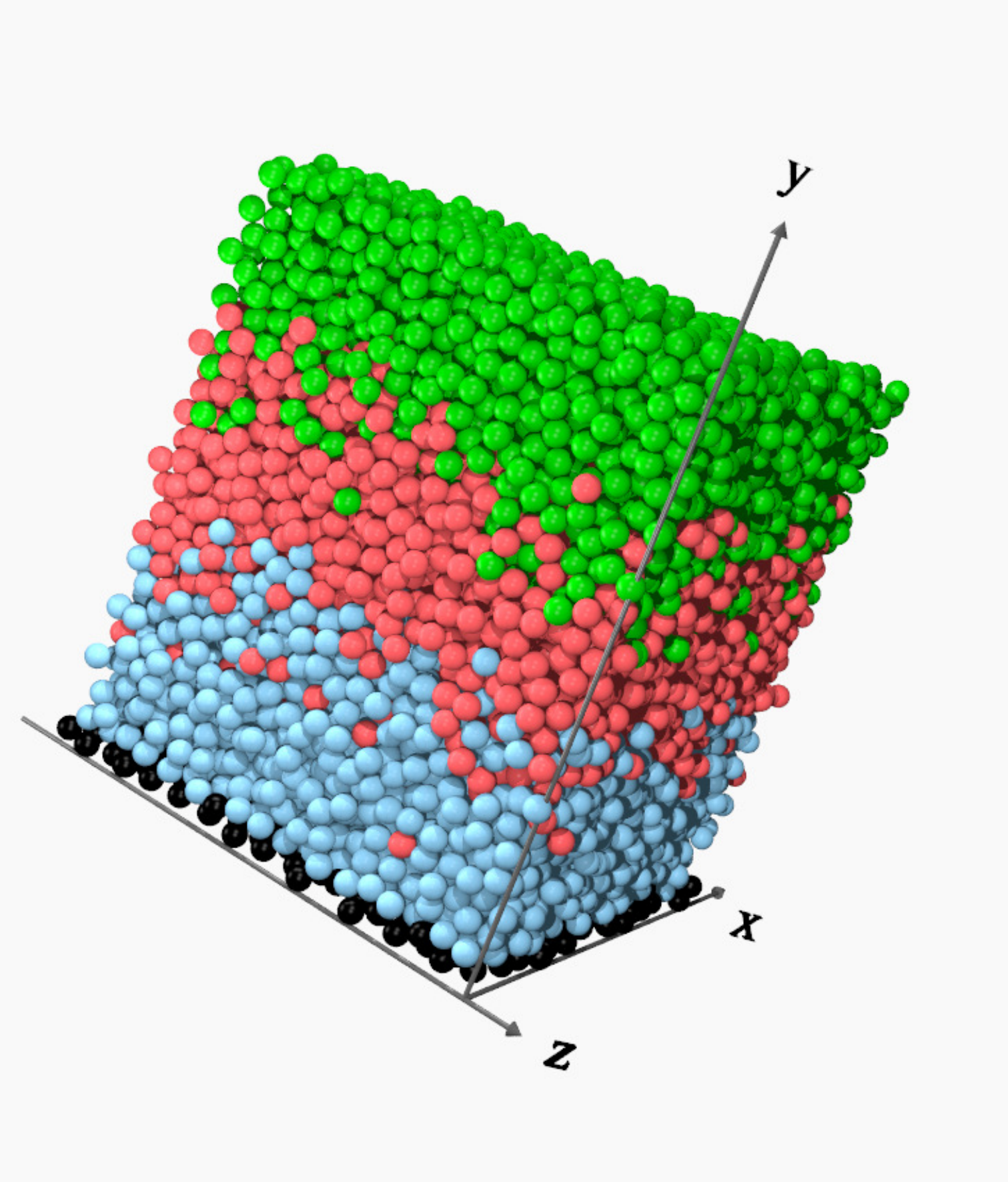}\put(-95,107){(b)}\put(-93,10){$t = 150$} \hfill
\includegraphics[scale=0.31]{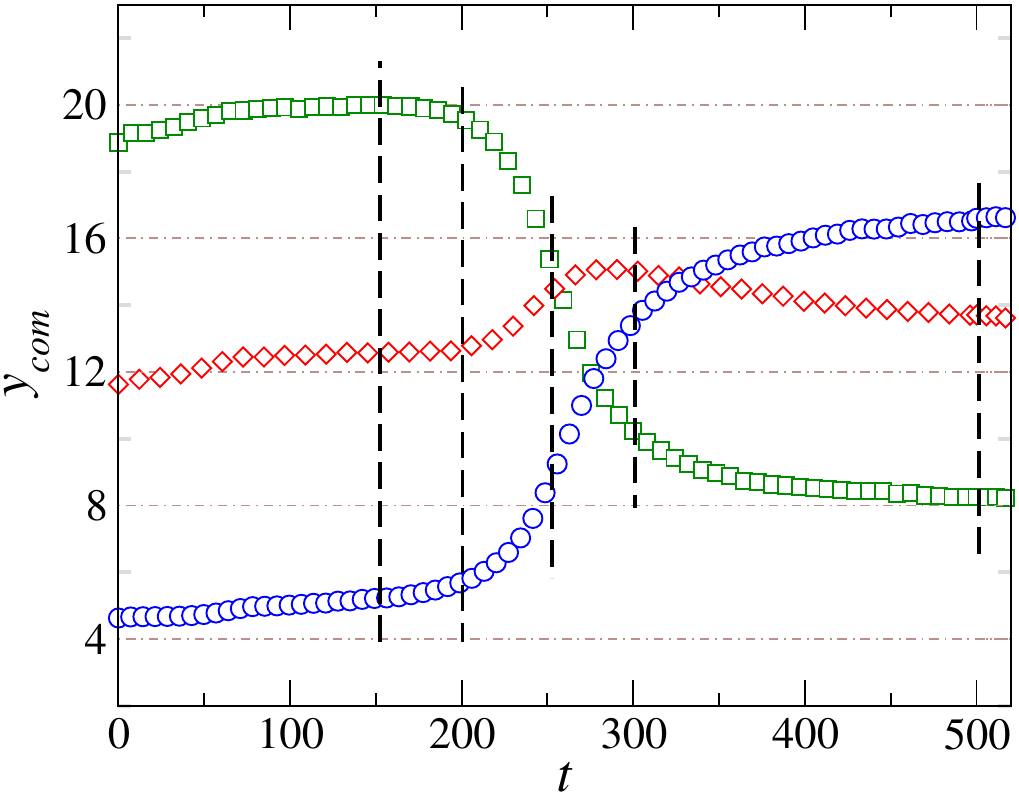}\put(-130,108){(a)}\put(-95,78){\vector(-1,0){75}} \put(-85,51){\vector(-1.8,-1.3){100}}\put(-70,35){\vector(0,-1){55}}\put(-58,58){\vector(1.3,-1){100}}\hfill \quad
%\put(-78,32){\vector(-1,-2){28}}
\includegraphics[scale=0.14]{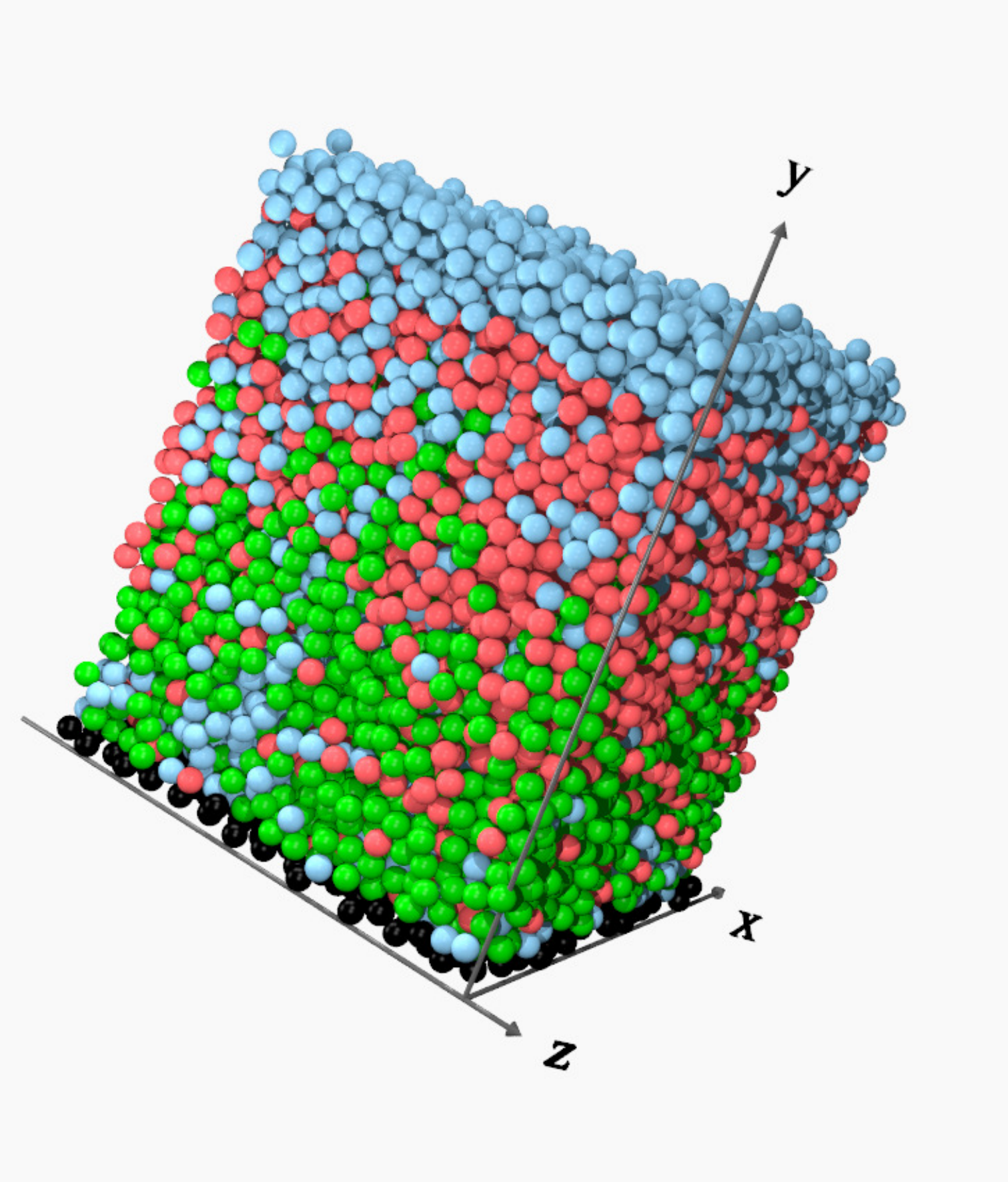}\put(-95,107){(f)}\put(-93,10){$t = 500$}\put(-125,65){\vector(1,0){37}}\hfill
\includegraphics[scale=0.14]{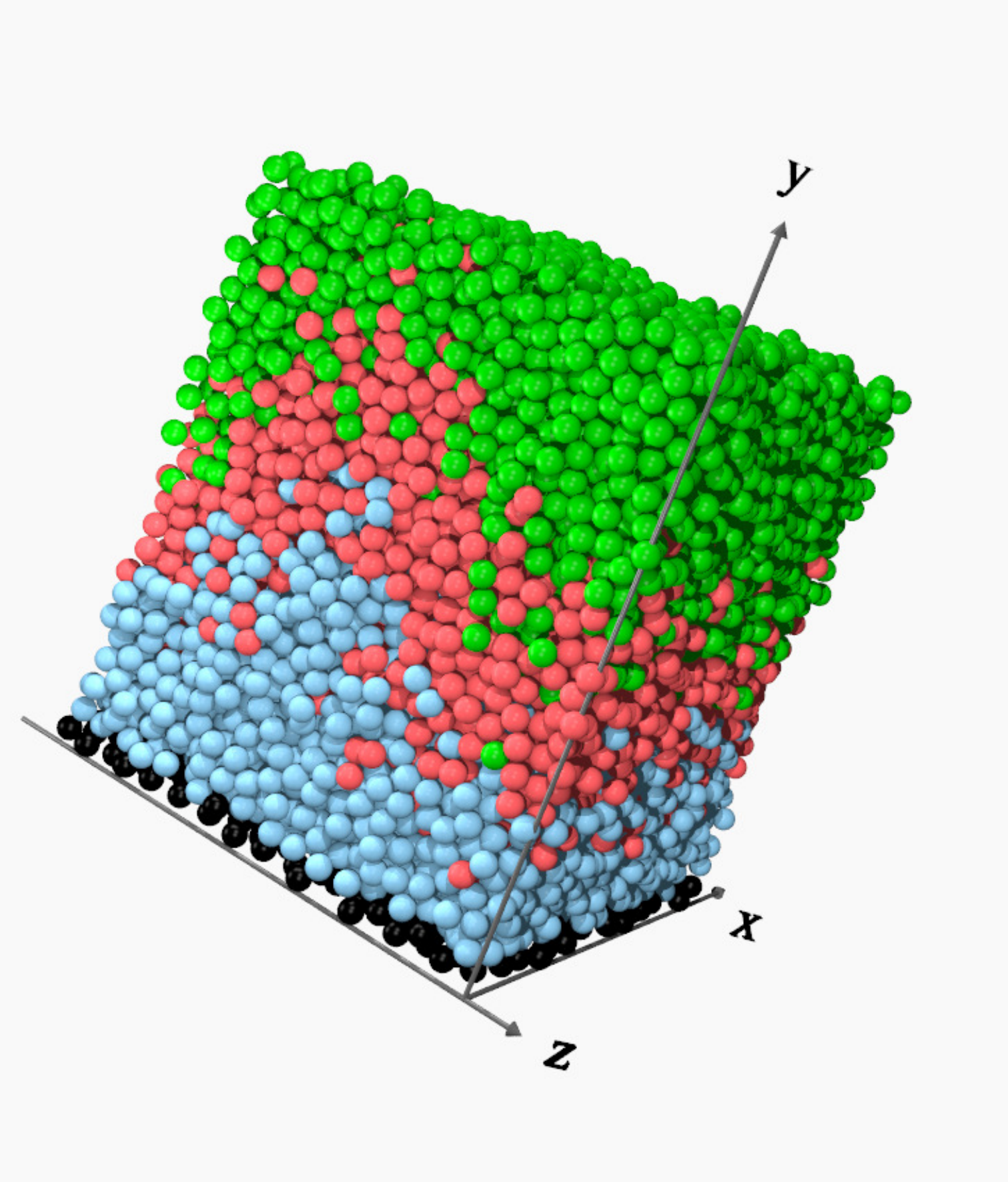}\put(-95,107){(c)}\put(-93,10){$t = 200$} \quad \quad \quad \quad
\includegraphics[scale=0.14]{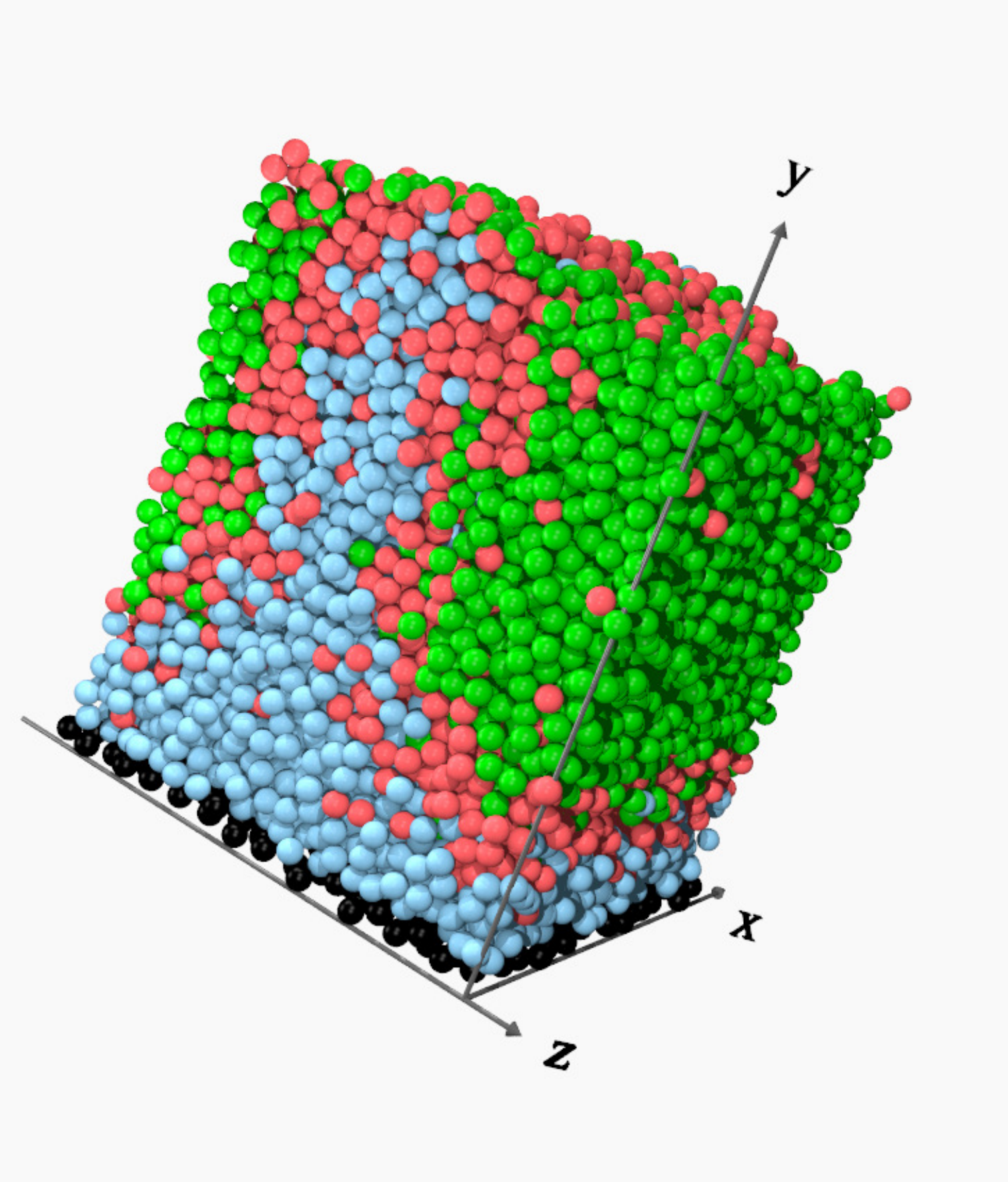}\put(-95,107){(d)}\put(-93,10){$t = 250$}\quad \quad \quad \quad
\includegraphics[scale=0.14]{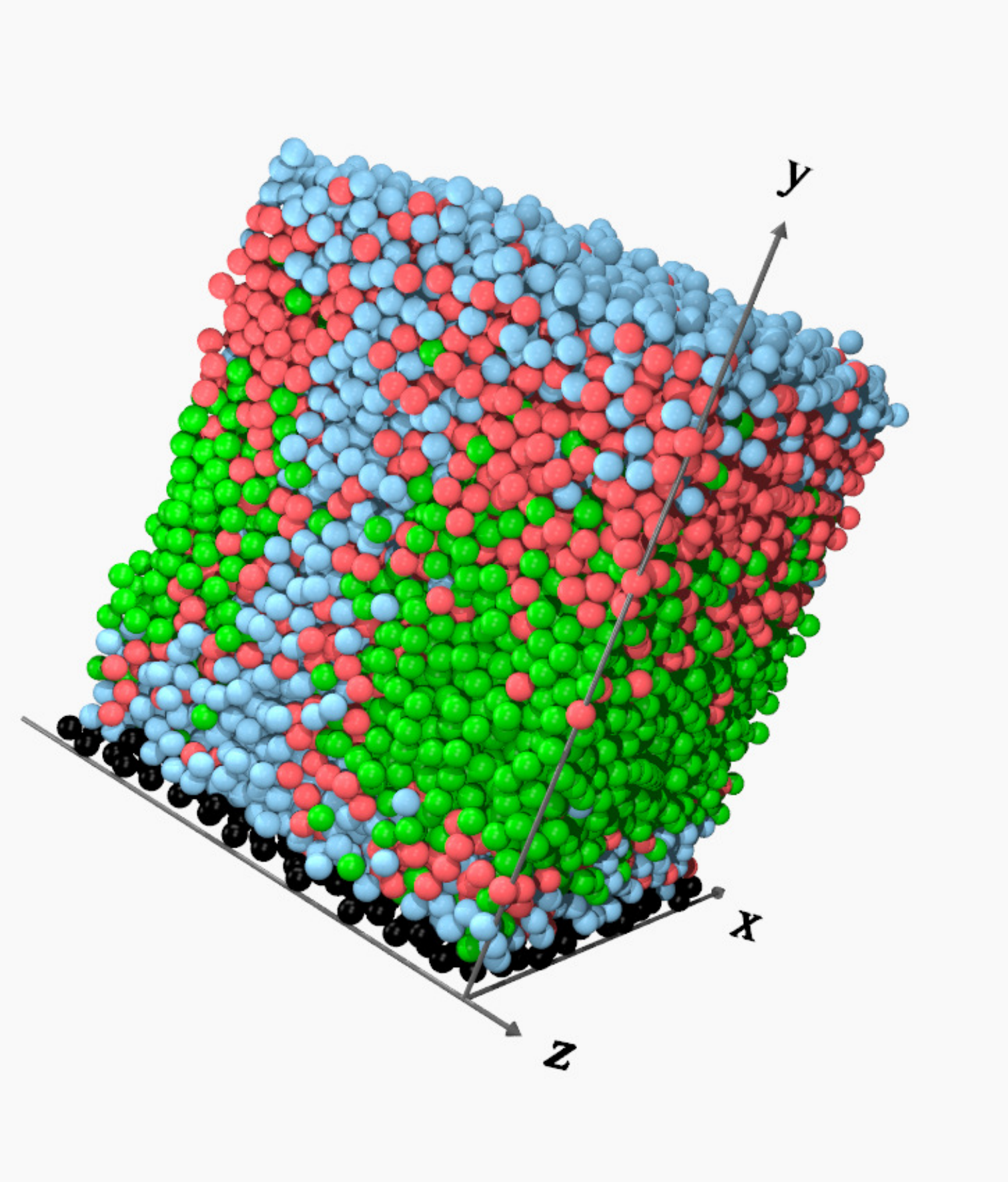}\put(-95,107){(e)}\put(-93,10){$t = 300$}
\caption{(a) Variation of $y_{com}$ with time for ternary mixture at early times in case of light-near-base initial configuration. DEM snapshots for equal composition ternary mixture at different times (b) $t = 150$, (c) $t = 200$, (d) $t = 250$, (e) $t = 300$, and (f) $t = 500$ starting from light-near-base initial configuration at inclination angle $\theta = 25^o$ with density ratios $\rho_H : \rho_M : \rho_L = 3:2:1$.}
    \label{fig:Ternary_LNB_snap}
\end{figure}

\begin{figure}
    \centering
\includegraphics[scale=0.17]{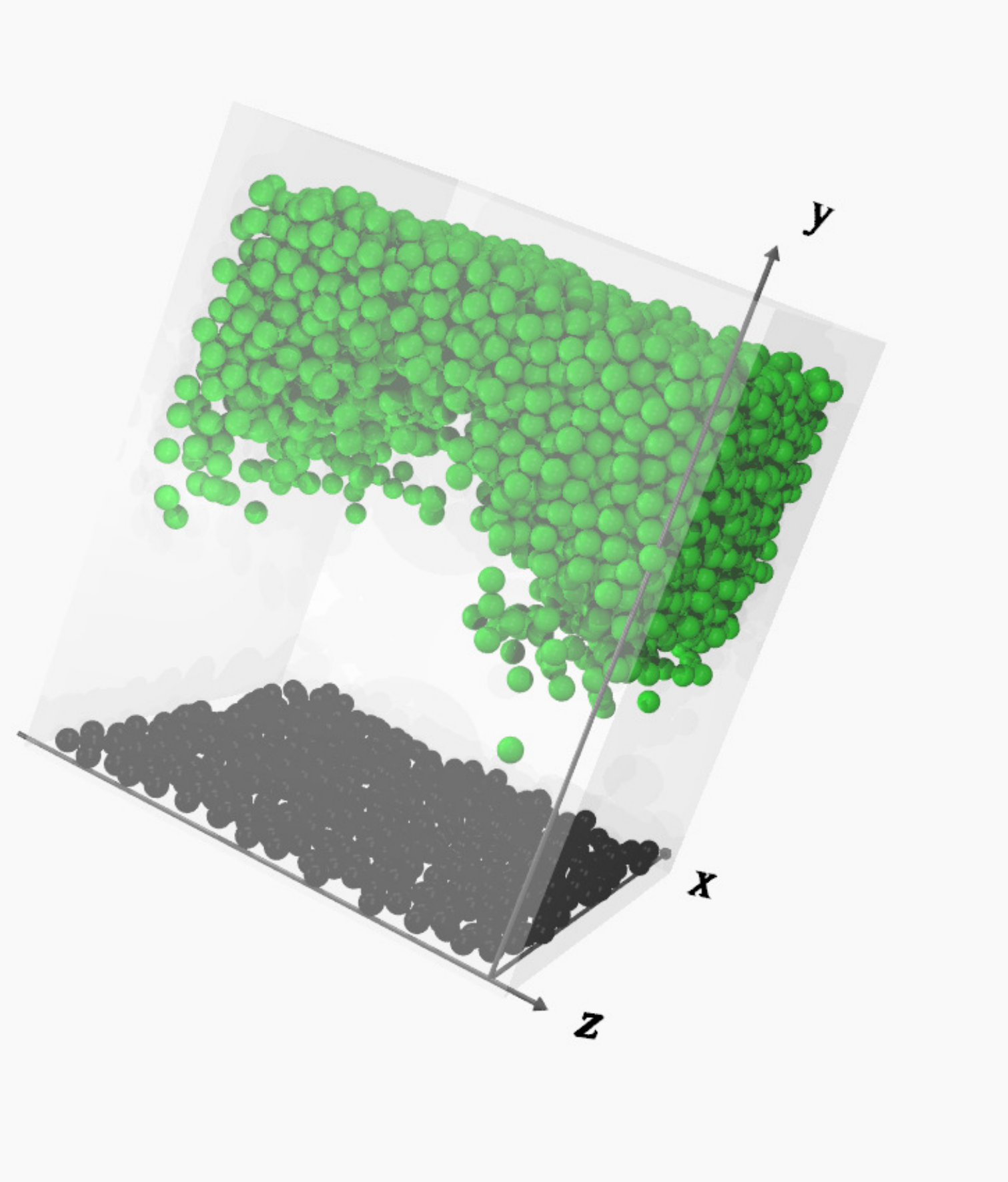}\put(-115,125){(a)}\put(-73,137){Heavy} \put(-135,60){\rotatebox{90}{$t = 200$}}\hfill
\includegraphics[scale=0.17]{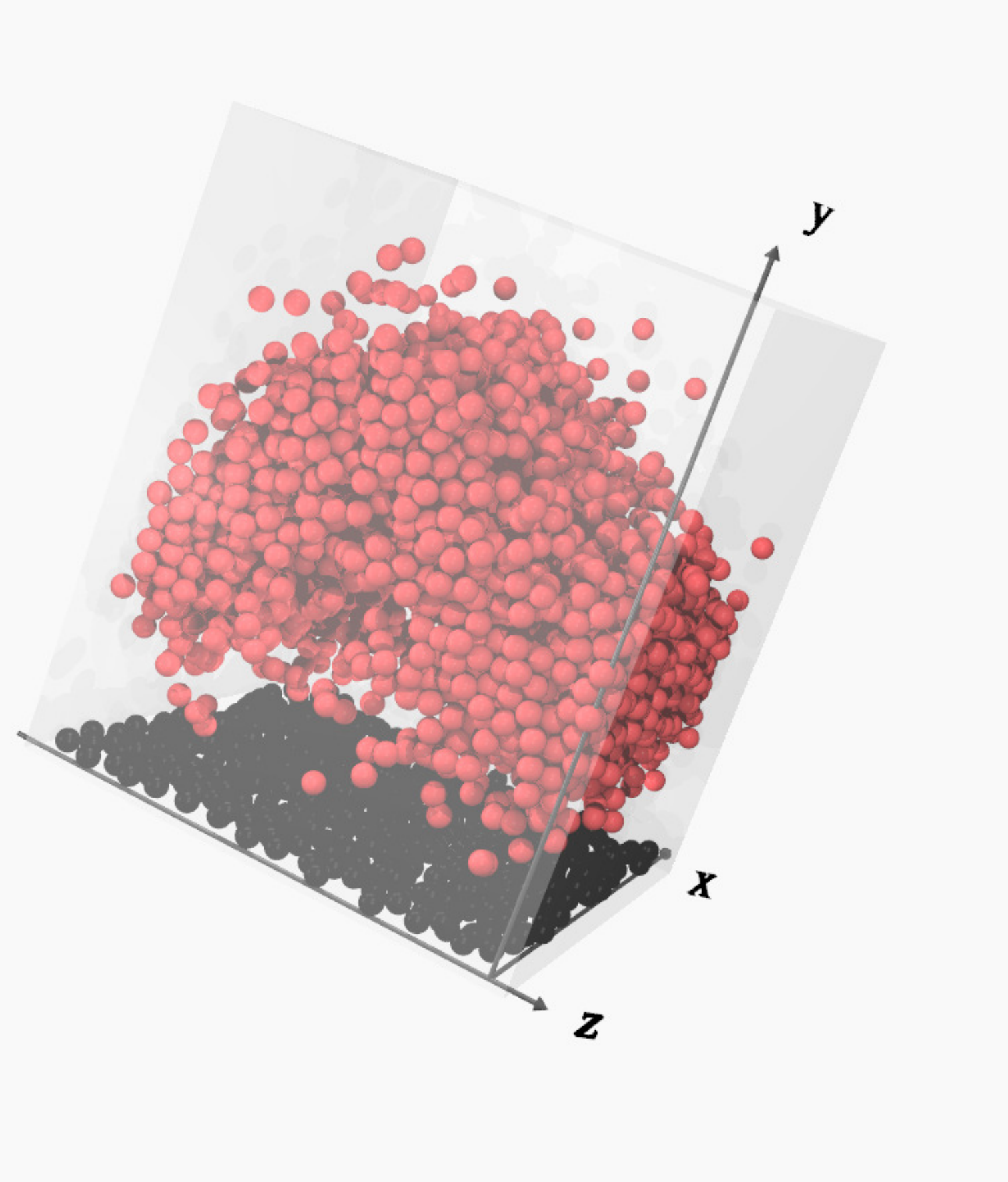}\put(-115,125){(b)}\put(-73,137){Medium}\hfill
\includegraphics[scale=0.17]{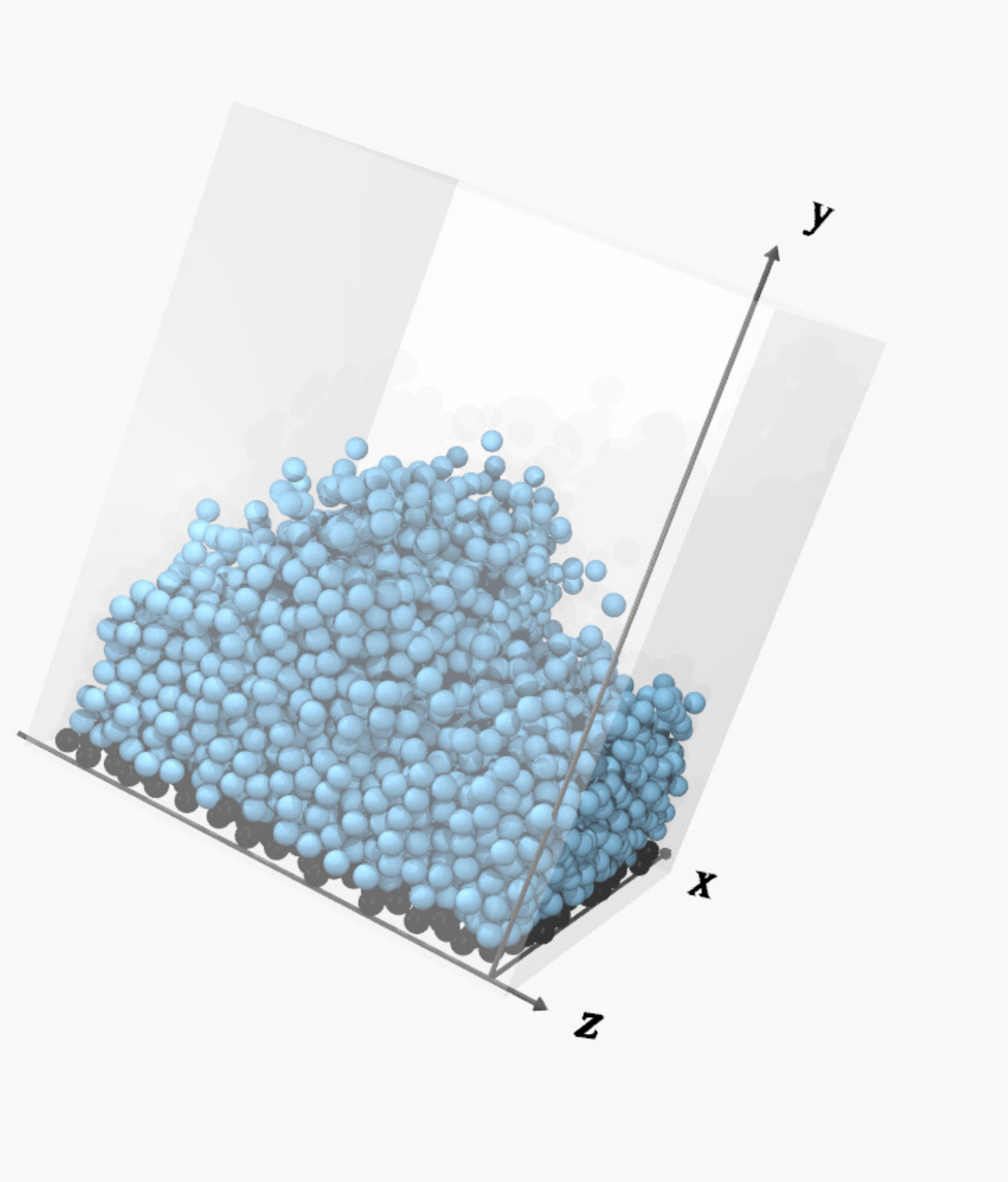}\put(-115,125){(c)}\put(-73,137){Light}\hfill
\includegraphics[scale=0.17]{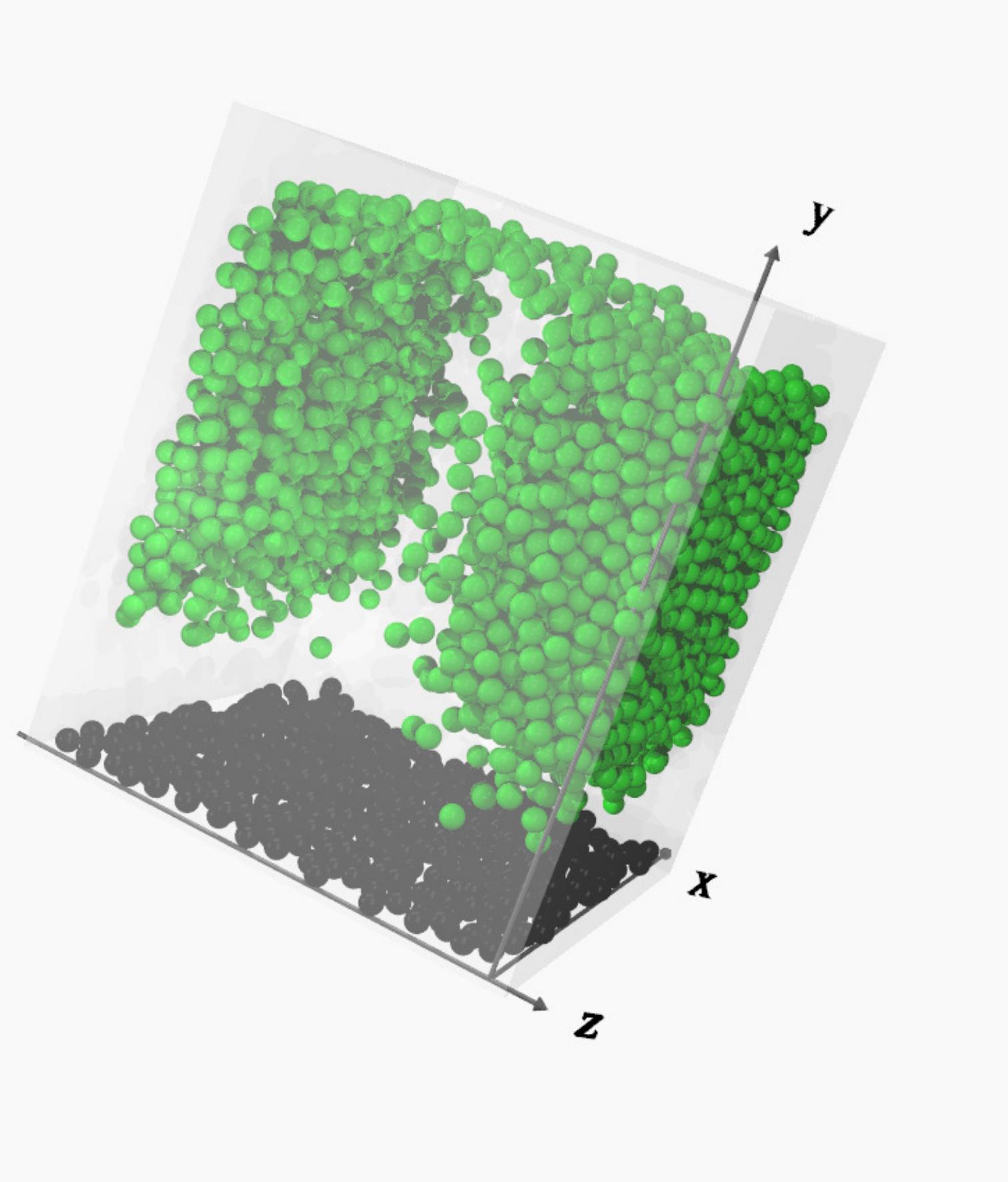}\put(-115,125){(d)}\put(-135,60){\rotatebox{90}{$t = 250$}}\hfill
\includegraphics[scale=0.17]{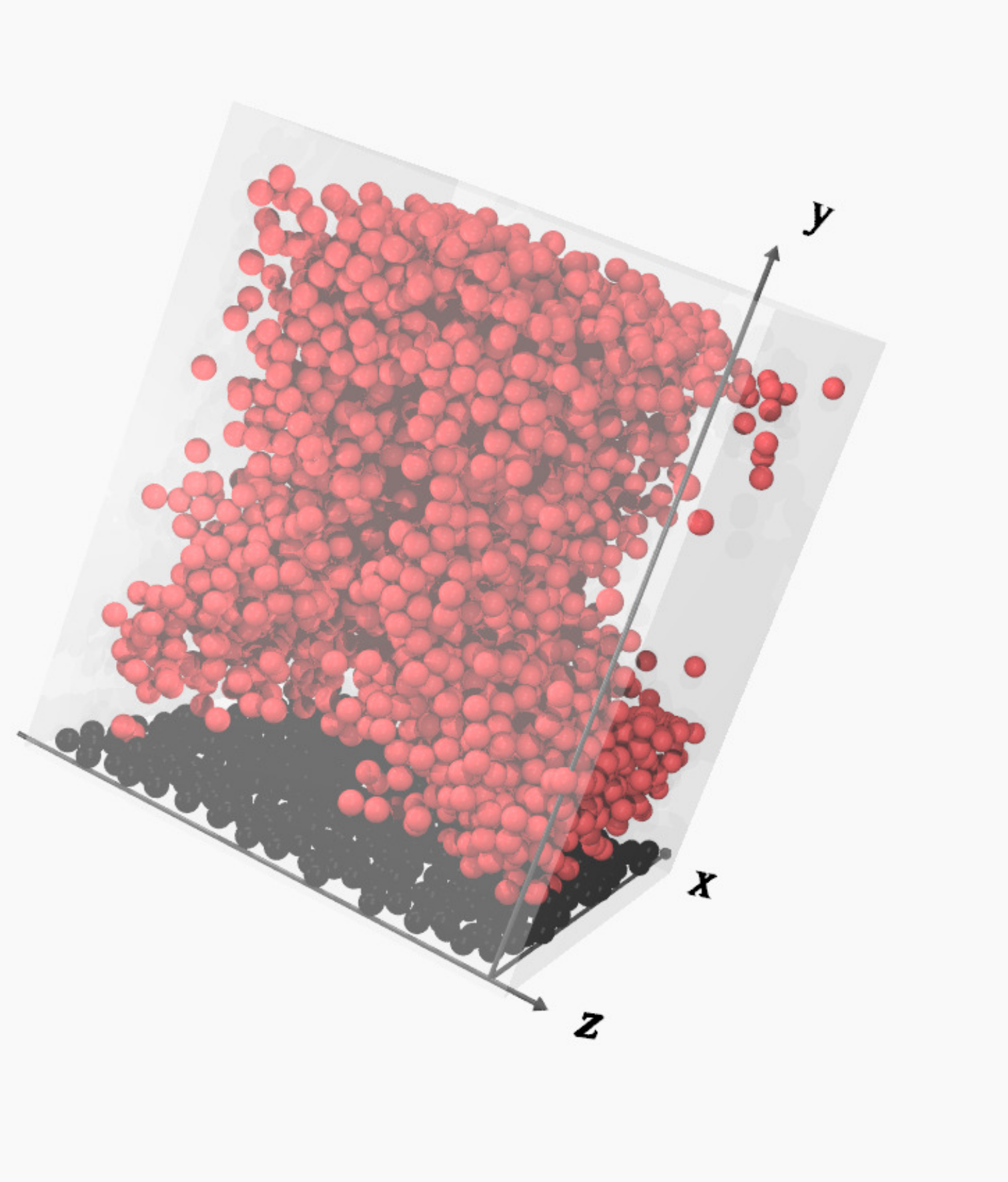}\put(-115,125){(e)}\hfill
\includegraphics[scale=0.17]{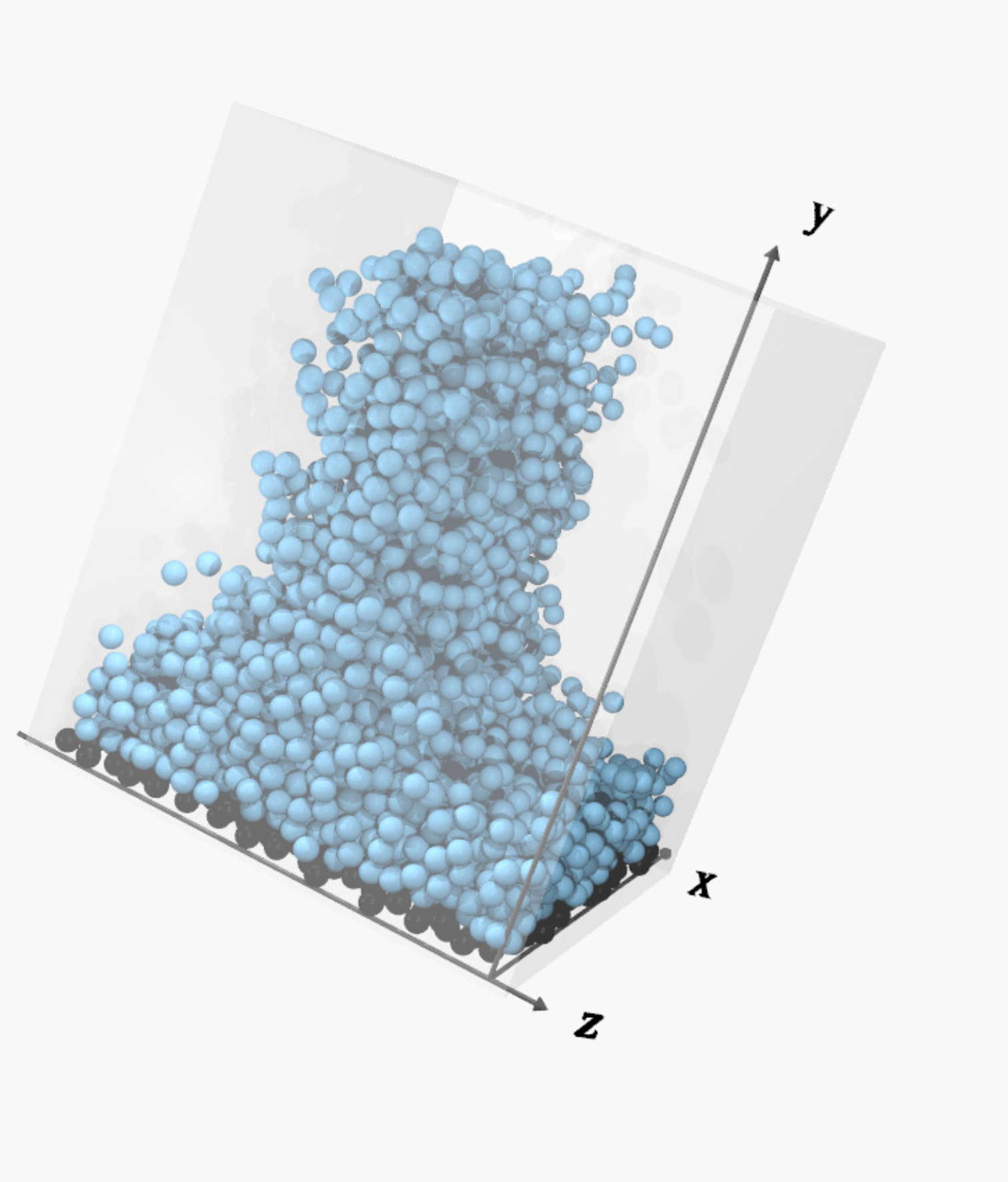}\put(-115,125){(f)}\hfill
    \caption{Arrangement of particles of (a) heavy, (b) medium, and (c) light species  at $t = 200$ corresponding to snapshot shown in figure~\ref{fig:Ternary_LNB_snap}c. 
    Arrangement of particles of (d) heavy, (e) medium, and (f) light species at  $t = 250$ corresponding to snapshot shown in figure~\ref{fig:Ternary_LNB_snap}d.}
    \label{fig:Ternary_species_DEM_snap}
\end{figure}

% \begin{figure}
%     \centering
%     \includegraphics[scale=0.251]{Ternary_rhomix_vec/snap-76.png}\put(-180,140){(a)}\hfill
%     \includegraphics[scale=0.251]{Ternary_rhomix_vec/snap-101.png}\put(-180,140){(b)}\hfill
%     \includegraphics[scale=0.251]{Ternary_rhomix_vec/snap-126.png}\put(-180,140){(c)}\hfill
%       \includegraphics[scale=0.251]{Ternary_rhomix_vec/snap-151.png}\put(-180,140){(d)}\hfill
%     \caption{Color maps for $\rho_{mix}$ in case of ternary mixture having density ratio $1:2:3$ with velocity vector in $x-z$ plane at different times, (a) $t = 150$, (b) $t = 200$, (c) $t = 250$, and (d) $t = 300$. }
%     \label{fig:ternary_rho_mix_with_vectors}
% \end{figure}

Figure~\ref{fig:binary_LNB_snap}a (in the center) shows the variation of the center of mass of the heavy and light species for the first $500$ time units after starting the flow. Figures~\ref{fig:binary_LNB_snap}b, \ref{fig:binary_LNB_snap}c, \ref{fig:binary_LNB_snap}d, \ref{fig:binary_LNB_snap}e, and \ref{fig:binary_LNB_snap}f show the DEM simulation snapshots corresponding to $250$, $300$, $350$, $400$, and $500$ time units respectively. 
The arrangement of the particles at $t=250$ (figure~\ref{fig:binary_LNB_snap}b) shows that the heavy (green) particles occupy the upper part of the layer while the light (blue) particles occupy the lower part of the layer. 
At time $t = 300$ (figure~\ref{fig:binary_LNB_snap}c), heavy particles seem to displace light particles in the central part of the simulation box pushing the light (blue) particles towards the edges leading to the rise of light particles in the upper part near the edges. This situation is reminiscent of the Rayleigh-Taylor instability observed in case of classical fluids when a high density fluid displaces a low density fluid. We note that this instability is observable in the $yz$ plane and forms a plume pattern at the interface of the light and heavy species in the $y$ direction. Similar instabilities have been observed by \cite{Instability2020PRL} in the case of different densities and sizes particle mixture flowing over periodic chute. 
%As the flow evolves in $x$ direction, the plume pattern is observable more clearly at time unit $t = 232$, shown in figure~\ref{fig:binary_LNB_snap}b. 
At time $t = 350$, the heavy (green) particles seem to be almost completely surrounded by the light (blue) particles and form a heavy core in the middle part of the layer (see figure~\ref{fig:binary_LNB_snap}d). The light particles quickly appear at the top and hence the center of mass of the heavy and light species come close to each other at $t=350$ (as shown in figure~\ref{fig:binary_LNB_snap}a). 
At time $t=400$, the core of the heavy particles consolidates more, leading to further decrease in the heavy species center of mass height while the light particles move upward and the number of the light particles at top surface (figure~\ref{fig:binary_LNB_snap}e) increases further. The heavy particles concentration slowly increases near the base and a good number of heavy (green) particles appear at the base at $t = 500$ (figure~\ref{fig:binary_LNB_snap}f).  

The onset of the instability and plume formation is much more clearly observable in the snapshots shown in figure~\ref{fig:Ternary_LNB_snap} for the case of ternary mixture. Figure~\ref{fig:Ternary_LNB_snap}a shows the variation of the center of the mass of the three species for first $500$ time units after the initiation of the flow. Figures~\ref{fig:Ternary_LNB_snap}b-\ref{fig:Ternary_LNB_snap}f show the DEM snapshots corresponding to different time instants to understand the flow pattern of different species during this time. 
Figure~\ref{fig:Ternary_LNB_snap}b shows the existence of a small perturbation of the light-medium as well as medium-heavy interface at $t=150$ which grows significantly by time $t=200$ (figure~\ref{fig:Ternary_LNB_snap}c) indicating the onset of formation of medium and light species plume. 
This can be more clearly observed in the snapshots shown in figure~\ref{fig:Ternary_species_DEM_snap} where particles of only one species are shown. Figures~\ref{fig:Ternary_species_DEM_snap}a, \ref{fig:Ternary_species_DEM_snap}b and \ref{fig:Ternary_species_DEM_snap}c to show the arrangement of particles of heavy, medium and light species at $t=200$. Figure~\ref{fig:Ternary_species_DEM_snap}a clearly shows the slightly perturbed interface of the heavy particles. The light particles also occupy a larger height in the center of the simulation box compared to the sides at $t=200$ (figure~\ref{fig:Ternary_species_DEM_snap}c) while the medium species particles appear to be in the form a mushroom-cap (figure~\ref{fig:Ternary_species_DEM_snap}b). 

\textcolor{black}{Returning back to figure~\ref{fig:Ternary_LNB_snap}d, we see} that by the time $t=250$, a plume of light (blue) and medium (red) species rises to the free surface due to the settling of the heavy particles from the sides. 
The arrangement of the particles of individual species for this particular instant is shown in figures~\ref{fig:Ternary_species_DEM_snap}d, \ref{fig:Ternary_species_DEM_snap}e and \ref{fig:Ternary_species_DEM_snap}f. 
A comparison of the heavy species particles shown in figure~\ref{fig:Ternary_species_DEM_snap}a and \ref{fig:Ternary_species_DEM_snap}d confirms that the interface between the heavy and medium species deforms to a large extent and almost all the heavy particles concentrate in form of two columns at the z-boundaries of the simulation box. This settling of the heavy species from the sides pushes the medium and light density particles concentrated in the center portion of the box upwards. The formation of the plume of light and medium species particles is clear by the column-like structural arrangement of the light species (figure~\ref{fig:Ternary_species_DEM_snap}f) which \textcolor{black}{are surrounded} by the medium density particles (figure~\ref{fig:Ternary_species_DEM_snap}e). The rising plume of the light and medium species first brings the medium density particles at $t=250$ (see figure~\ref{fig:Ternary_LNB_snap}d) and then light density particles to the top at $t=300$ (figure~\ref{fig:Ternary_LNB_snap}e). The medium density particles quickly settle downwards leaving a pure layer of the light species at the top which grows further in thickness as time progresses to $t=500$ (figure~\ref{fig:Ternary_LNB_snap}f).

\begin{figure}
    \centering
     \includegraphics[scale=0.35]{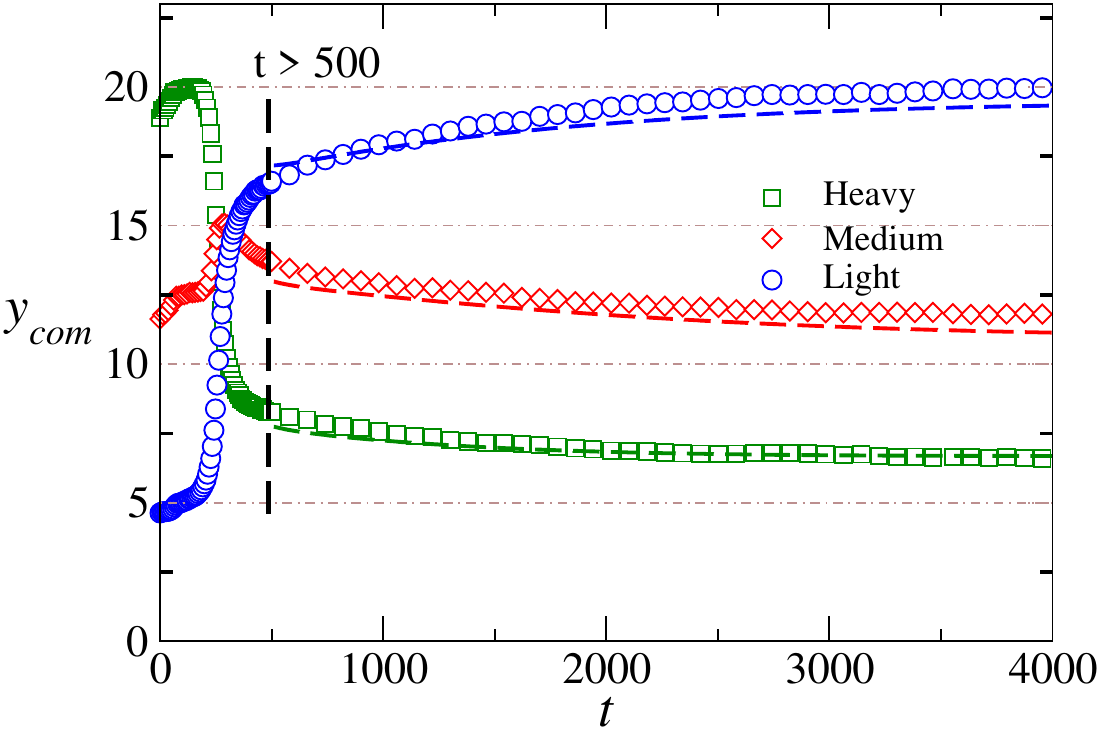}\put(-180,125){(a)}
      \includegraphics[scale=0.35]{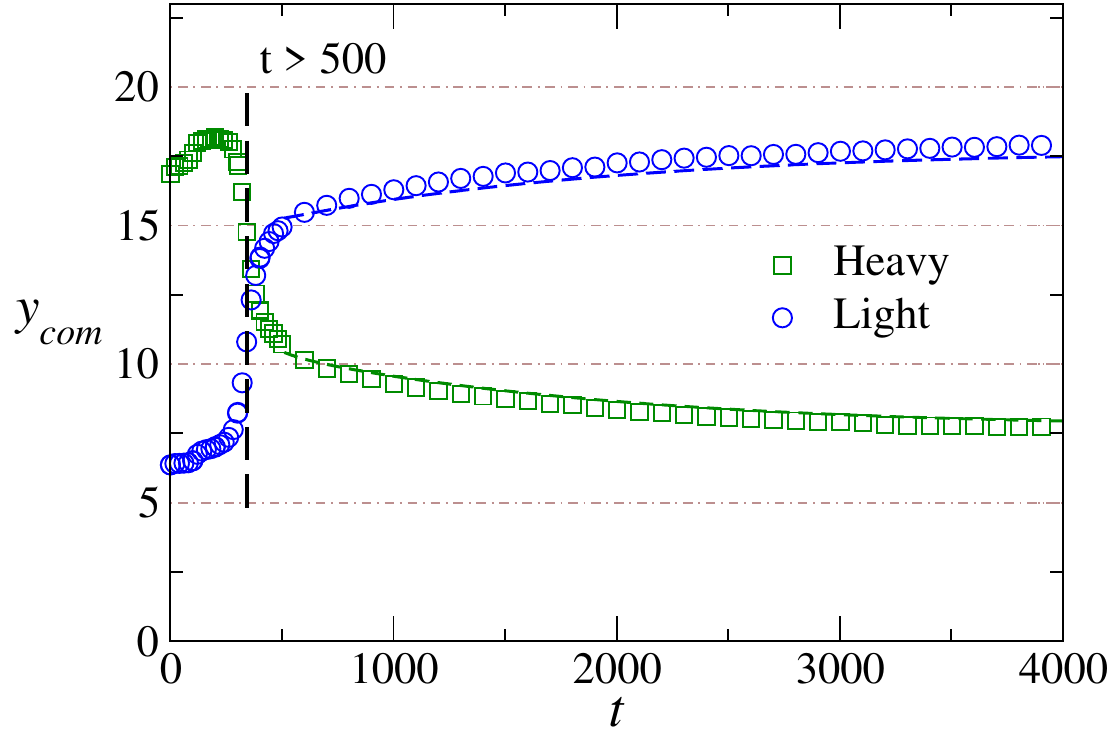}\put(-180,125){(b)}
    \caption{Time evolution of $y_{com}$ for light-near-base configuration after $t = 500$ time \textcolor{black}{units when the quick segregation due to instability is over}. (a) Equal composition ternary mixture having density ratios $\rho_H : \rho_M : \rho_L = 3:2:1$ and (b) Equal composition binary mixture having density ratio $\rho = 2.0$ for inclination angle $\theta = 25^o$. }
    \label{fig:AfterInstabilityPrediction}
\end{figure}

\textcolor{black}{The arrangement of particles depicted in figure~\ref{fig:Ternary_LNB_snap}f shows that the instability quickly drives the light near base and heavy at the top arrangement of particles into a configuration where the heavy species concentrates near the base and the light species gets concentrated near the free surface with the medium species occupying the space between these two species.} This configuration is very similar to the configuration studied in figure~\ref{fig:HNB_bin_tern_diff_conf}d-\ref{fig:HNB_bin_tern_diff_conf}f. Given the successful prediction of the center of mass for the configuration in figure~\ref{fig:HNB_bin_tern_diff_conf}d from our model, we expect that the model should be able to capture the evolution of the species concentrations and \textcolor{black}{species center of masses} reasonably well from this time onwards. 
Hence, we use the DEM data at $t = 500$  to obtain the concentration profile $f_i(y)$ of the three species at \textcolor{black}{this particular} instant (snapshot shown in figure~\ref{fig:Ternary_LNB_snap}f) as the initial condition in the continuum model and predict the evolution of the species center of mass for the ternary mixture. \textcolor{black}{These model predictions are shown as dashed lines in figure~\ref{fig:AfterInstabilityPrediction}a while symbols represent the DEM data.} Interestingly, the theory can accurately predict the segregation behavior for $t > 500$ time units by using the concentration profiles at $t=500$ as the initial condition. The concentration color maps from the theoretical predictions are also found to be qualitatively as well as quantitatively very similar to those obtained from the DEM simulations for $t>500$ (shown in figure~\ref{fig:LNB_ternary_diff_conf}c and \ref{fig:LNB_ternary_diff_conf}e) \textcolor{black}{and hence are not shown the sake of brevity.} Similarly, figure~\ref{fig:AfterInstabilityPrediction}b shows the center of mass evolution for the binary mixture case with heavy species on top (see figure~\ref{fig:LNB_binary_diff_conf}a) using the concentration profiles of heavy and light species at $t=500$ for the snapshot shown in figure~\ref{fig:binary_LNB_snap}f. Again, the continuum model predictions for the center of mass are in very good agreement with the DEM data for $t>500$. 

The results shown in figures~\ref{fig:AfterInstabilityPrediction}a and \ref{fig:AfterInstabilityPrediction}b confirm that the continuum model fails to predict the center of mass evolution only for the brief time interval during which the instability occurs for the \textcolor{black}{higher density species on top of lower density species} configurations. 
This is \textcolor{black}{not unexpected since the continuum model accounts} for property variations only along the $y$ direction and ignores the variations along the other two directions. 
While \textcolor{black}{this assumption is reasonable} for the well-mixed and heavy-near-base configurations studied in this work, it is not valid for the duration over which the instability is observed for the configurations with high density species at the top. The snapshots shown in figures~\ref{fig:binary_LNB_snap} and \ref{fig:Ternary_LNB_snap} confirm that \textcolor{black}{the velocity as well as the species concentrations vary in all three directions. Hence} the momentum balance equations along with the rheological and segregation model must be solved in the three dimensions to be able to capture the flow and segregation due to the onset of the instability and the resulting quick segregation for heavy on top configurations. 
The snapshots depicting the position of the particles of the individual species (shown in  figure~\ref{fig:Ternary_species_DEM_snap}) suggest that a two-dimensional model accounting for the variations in $y$ and $z$ direction may also be sufficient to capture the essential features since the variations along the flow direction don't seem to be \textcolor{black}{very significant}.
We note that \textcolor{black}{while} our simulations utilize periodic boundary conditions in the flow ($x$) as well as the vorticity ($z$) direction, the instability is observed only in the $y-z$ plane for the cases studied in this work. Exploring the reason for this observation is beyond the scope of this work and will be investigated in future.
%The continuum model is incapable of forecasting the behavior during the initial time period when instability is present in the LNB initial configuration. Conversely, figure \ref{fig:LNB_binary_diff_conf}b presents the data for initial light-near-base (LNB) configuration. The arrangement of particles in the LNB case is such that it forms an unstable configuration due to heavy-density particles on the top and the segregation force being maximum. Note that the continuum model fails to capture this instability behavior as it is a one-dimensional model and hence, needs to solve the equations in $2D$.

%It is important to note that their theoretical model failed to accurately capture the segregation phenomena in some cases. Consequently, instability could potentially be a contributing factor to this.
%Examining this instability for this specified configuration lies beyond the scope of this study.  

%\section{Discussion}
%\label{sec:discussion}

\section{Summary}
\label{sec:summary}
In this work, we have developed a continuum model to predict the time-dependent density segregation for binary as well as multicomponent mixtures in a periodic chute flow. The model incorporates transient segregation-diffusion equation for all the species along with an inertial number based mixture rheological description to solve the momentum balance equations for same-size, different-density granular mixtures. 
%by using $PDEPE$ solver in MATLAB. 
Specifically, the segregation model uses the particle force-based \textcolor{black}{theory proposed by} ~\cite{sahu_kumawat_agrawal_tripathi_2023} to calculate the segregation velocity. 
In contrast to the existing empirical segregation models (\cite{xiao2016modelling,jones2018asymmetric}) for density segregation, the particle force-based model does not require any empirical parameter.
The generalized form of the inertial number based rheology (\cite{jop2006constitutivenature}) for granular mixtures, proposed by \cite{tripathi2011rheology}, is utilized to capture the flow evolution.
It is important to note that solving the segregation-diffusion equation requires knowledge of the flow kinematics. 
Unlike most other approaches that utilize the flow kinematics data obtained from experiments or DEM simulations, we solve the time-dependent momentum balance equations by inter-coupling the mixture rheology and the segregation model. 
Due to the inter-coupling of rheology and segregation, the predictions from our model are able to capture non-monotonic evolution of the velocity field in the case of well-mixed initial configuration. 
The model is able to predict the concentration profiles and flow properties such as velocity, shear stress, pressure, solids fraction, and inertial number reasonably well for two different initial configurations. 
In addition, the model successfully captures the segregation for different mixture compositions, density ratios, and inclination angles.
Significant differences, however, are observed in the predicted velocity profiles as compared to DEM data for a small time period after the start of the flow. 
These discrepancies are attributed to the \textcolor{black}{limitations} of the JFP model \textcolor{black}{to describe flow} at low inertial numbers. % where the flow tends to be in the slow-flow regime. %Our DEM data showed that the $\mu-I$ and $\phi-I$ relations used in the JFP model are not valid for $I \leq 0.05$. The rheology for low inertial numbers needs to be explored in the future. 

While the properties of the well-mixed initial configuration as well as heavy-near-base initial configurations are \textcolor{black}{predicted} very well, the model does not accurately capture the behavior for the light-near-base initial configuration. 
For this configuration, quick segregation is observed at early times followed by a slow evolution of the segregation at later times. The model fails to capture the quick segregation despite the inter-coupling of rheology and segregation due to its one-dimensional nature. 
By carefully observing DEM snapshots at different time instants during this quick segregation phase, we notice the presence of a Rayleigh-Taylor like instability due to the unstable initial configuration of heavy density species being on top of a low density species. The slight perturbations of the species interface at early times are followed by bulk motion where the heavy species quickly moves downwards and pushes the lighter species upwards. Once the bulk motion induced due to the onset of instability subsides, the light species concentrates near the top and the heavy species gets concentrated near the base. 
By using the species concentration profiles at the end of this quick segregation phase as the initial concentration profile, the model is able to capture the evolution of all the flow properties at later times very well. Thus the one-dimensional formulation presented in this work fails to capture the segregation and evolution of the flow properties only for the small time duration when bulk motion and segregation in more than one dimension become important. \cite{TIRAPELLE20211annular} also utilized the light-near-base initial configuration to study the density segregation in annular shear cell and found noticeable differences in their $1D$ continuum model predictions and experiments. While the authors do not discuss the role of instability, our results indicate that the discrepancies \textcolor{black}{observed} could be attributed to instability driven bulk motion during the flow. These observations indicate the need to generalize \textcolor{black}{our one dimensional} continuum model to three dimensions. %following the approach of \cite{barker2021OpenFoam}. %A similar particle force-based theory has also been successfully implemented for size segregation in a study by \cite{tripathi2021size}. 
The continuum approach presented in this work can be easily extended to different size mixtures by incorporating the particle force-based segregation theory of \cite{tripathi2021size} to predict transient size segregation in periodic chute flows. In addition, we would like to explore whether a quantitative comparison with DEM simulations can be observed using this approach in other free surface flows such as plane shear flow or heap flow.

\backsection[Acknowledgements]{SK acknowledges the help of Mr. Satyabrata Patro in performing the DEM simulations. }

\backsection[Funding]{AT gratefully acknowledges the financial support provided by the Indian Institute of Technology Kanpur via the initiation grant IITK/CHE/20130338. AT and SK gratefully acknowledge the funding support provided to SK by the Prime Minister's Research Fellowship (Government of India) grant.}

\backsection[Declaration of interests]{The authors report no conflict of interest.}

\backsection[Data availability statement]{The data that support the findings of this study are available from the corresponding author, AT, upon reasonable request.}

\backsection[Author ORCIDs]{
\newline
Soniya Kumawat  \url{https://orcid.org/0000-0003-3314-9875};\\
Vishnu Kumar Sahu \url{https://orcid.org/0000-0002-7768-0505}; \\
Anurag Tripathi \url{ https://orcid.org/0000-0001-9945-3197}.}

\bibliographystyle{jfm}
\bibliography{Denseg_v1}

\end{document}